\documentclass[twocolumn,twocolappendix]{aastex631}

\usepackage[varg]{txfonts}
\usepackage[]{graphicx}
\usepackage{enumerate}
\usepackage{subeqnarray}
\usepackage{cases}
\usepackage{ulem}
\usepackage{mathrsfs,amssymb}
\usepackage{multirow}
\usepackage{diagbox}

\usepackage{hhline}                  
\usepackage{comment}

\usepackage{tikz}
\usetikzlibrary{tikzmark}
\usetikzlibrary{arrows.meta}

\def\myspace{\vspace{2.5mm}}          
\def\mycolumnwidth{8.35mm}           
\def\myrowheight{4.5pt}               

\newcolumntype{C}[1]{>{\centering\let\newline\\\arraybackslash\hspace{0pt}}m{#1}}

\shorttitle{Re-evaluation of the cosmic-ray ionization rate...}
\shortauthors{Obolentseva et al.}

\tightenlines

\begin{document}

\title{Re-evaluation of the cosmic-ray ionization rate in diffuse clouds}

\author{M.~Obolentseva} \affiliation{Max-Planck-Institut f\"ur extraterrestrische Physik, 85748 Garching, Germany} 
\author{A.~V.~Ivlev} \affiliation{Max-Planck-Institut f\"ur extraterrestrische Physik, 85748 Garching, Germany} 
\author{K.~Silsbee} \affiliation{University of Texas at El Paso, El Paso, TX, 79968, USA} \affiliation{Max-Planck-Institut f\"ur extraterrestrische Physik, 85748 Garching, Germany} 
\author{D.~A.~Neufeld} \affiliation{Department of Physics and Astronomy, Johns Hopkins University, Baltimore, MD 21218, USA} 
\author{P.~Caselli} \affiliation{Max-Planck-Institut f\"ur extraterrestrische Physik, 85748 Garching, Germany} 
\author{G.~Edenhofer} \affiliation{Max-Planck-Institut f\"ur Astrophysik, 85748 Garching, Germany} 
\author{N.~Indriolo} \affiliation{AURA for ESA, Space Telescope Science Institute, Baltimore, MD 21218, USA} 
\author{T.~G.~Bisbas} \affiliation{Research Center for Astronomical Computing, Zhejiang Lab, Hangzhou 311100, China} 
\author{D.~Lomeli} \affiliation{University of Texas at El Paso, El Paso, TX, 79968, USA} 

\correspondingauthor{A.~V.~Ivlev} \email{ivlev@mpe.mpg.de}
\correspondingauthor{K.~Silsbee} \email{kpsilsbee@utep.edu}
       
\begin{abstract}
All current estimates of the cosmic-ray (CR) ionization rate rely on assessments of the gas density along the probed sight lines. Until now, these have been based on observations of different tracers, with C$_2$ being the most widely used in diffuse molecular clouds for this purpose. However, three-dimensional dust extinction maps have recently reached sufficient accuracy as to give an independent measurement of the gas density on parsec scales. In addition, they allow us to identify the gas clumps along each sight line, thus localizing the regions where CR ionization is probed. We re-evaluate H$_3^+$ observations, which are often considered as the most reliable method to measure the H$_2$ ionization rate $\zeta_{\rm H_2}$ in diffuse clouds. The peak density values derived from the extinction maps for 12 analyzed sight lines turn out to be, on average, an order of magnitude lower than the previous estimates, and agree with the values obtained from revised analysis of C$_2$ data. We use the extinction maps in combination with the {\sc 3d-pdr} code to self-consistently compute the H$_3^+$ and H$_2$ abundances in the identified clumps for different values of $\zeta_{\rm H_2}$. For each sight line, we obtain the optimum value by comparing the simulation results with observations. We show that $\zeta_{\rm H_2}$ is systematically reduced with respect to the earlier estimates by a factor of $\approx 9$ on average, to $\approx6\times10^{-17}$~s$^{-1}$, primarily as a result of the density reduction. We emphasize that these results have profound consequences for all available measurements of the ionization rate.
\end{abstract}

\section{Introduction}

Low-energy, viz., non-relativistic cosmic rays (CRs) play a crucial role in the evolution of molecular clouds \citep[see, e.g.,][and references therein]{Padovani2020}. Such CRs completely dominate gas ionization and heating in dense cores of the clouds and, therefore, govern dynamical and chemical processes accompanying practically all stages of star formation \citep{McKee1989, Keto2008, Dalgarno2006, Glassgold2012}. The ionization fraction determines the gas conductivity which controls the degree of coupling between the gas and magnetic field, and thus affects properties of turbulence and the timescale of cloud collapse \citep{Shu1987, Li2014, Turner2014}. These processes, in turn, determine the initial growth of dust grains and thus may critically affect the formation of pebble-sized aggregates occurring at later evolutionary stages in protoplanetary disks \citep{Ossenkopf1994, Ormel2009}. Furthermore, ionization of molecular hydrogen initiates the network of fast ion-neutral reactions dominating gas-phase chemistry \citep[e.g.,][]{Herbst1973, Wakelam2010, Hollenbach2012}, while the energy deposited by CRs upon their collisions with dust grains drives the solid-state reactions in icy mantles \citep{Shingledecker2018} that form on the surface of the grains due to gas freeze-out \citep{Bergin1995, Caselli1999, Gibb2004}. The gas-phase chemistry is inherently linked to the chemistry in ices, as the bombardment by CRs leads to desorption of chemical species back to the gas phase \citep{Leger1985, Hasegawa1993, Vasyunin2017, Sipila2021}.

The strength of all the processes mentioned above is typically parameterized in terms of the CR ionization rate, the value of which has been debated for over half a century.  Different groups have made estimates and upper limits based on the temperatures of interstellar gas clouds \citep{Hayakawa1961, Spitzer1968}, abundances of chemical species \citep{Odonnell1974, Black1977, Indriolo2012, Neufeld2017}, ro-vibrational excitation of H$_2$ \citep{Bialy2022}, as well as first-principles calculations using CR spectra measured on spacecraft \citep{Spitzer1968, Webber1998, Cummings2016}. The CR ionization rate in diffuse clouds has been derived from absorption-line observations of specific ions produced by CRs, such as OH$^+$, H$_2$O$^+$, and ArH$^+$ in atomic gas \citep{Indriolo2015, Neufeld2017, Bacalla2019}, or H$_3^+$ in molecular gas \citep{McCall2003, Indriolo2012, Albertsson2014}. The abundance ratios obtained from the absorption and emission measurements, such as HCO$^+$/CO and OH/CO, were also proposed as measures of the ionization rate in diffuse molecular clouds \citep{Luo2023a, Luo2023b, Gaches2019}. In dense cores, chemical models of higher complexity have been employed to estimate the ionization rate from emission measurements of various species, e.g., using the abundance ratios of DCO$^+$/HCO$^+$ \citep{Caselli1998}, HCO$^+$/N$_2$H$^+$ \citep{Ceccarelli2014}, and of neutral hydrocarbons \citep{Fontani2017, Favre2018}, or measuring H$_2$D$^+$ molecular ions to trace H$_3^+$ \citep{Bovino2020, Sabatini2023, Redaelli2024}. Much attention has been given to observations of the H$_3^+$ ions in diffuse clouds \citep{Geballe1996, McCall2003, Indriolo2007, Indriolo2012, Albertsson2014, Neufeld2017}, as their formation and destruction is described by a fairly simple chemical network.  

It was first noted in \citet{McCall2003}, and then confirmed in \citet{Indriolo2007} and \citet{Indriolo2012}, that the typical ionization rates inferred from H$_3^+$ abundances are approximately an order of magnitude higher than the value estimated from the Voyager and Pioneer spacecraft data for the CR spectra at distances up to 60~au from the Sun \citep{Webber1998}. The existence of this discrepancy was solidified in \citet{Cummings2016}, presenting the CR spectra measured by Voyager~1 after its passage through the heliopause \citep[into a region of space in which CRs are believed to be unaffected by the solar modulation; see, e.g.,][]{Stone2013}. The authors found an ionization rate of atomic hydrogen of approximately $1.6 \times 10^{-17}$~s$^{-1}$, corresponding to a rate of $2.4 \times 10^{-17}$~s$^{-1}$ for molecular hydrogen \citep{Glassgold73}.  In contrast, using H$_3^+$ observations, \citet{Indriolo2012} found an average value for the H$_2$ ionization rate of $3.5 \times 10^{-16}$~s$^{-1}$.

Several reasons for this discrepancy have been debated. \citet{Recchia19} discuss (and reject) a possible spike in the CR spectrum at energies $\lesssim3$~MeV, below the measurement cutoff in the Voyager data, which may contribute for much of the ionization. \citet{Phan2023} suggest that efficient attenuation of the low-energy CRs, combined with the discrete nature of CR sources, may lead to large fluctuations in their abundance.  \citet{Gloeckler2015} argue that Voyager has not really crossed the heliopause, and thus the measurements are not representative of the local interstellar CR spectrum. However, the later measurements of an effectively identical spectrum by Voyager~2 \citep{Stone2019} make this interpretation rather unlikely.

The molecular ion H$_3^+$ is formed in a two-step process \citep[see, e.g.,][]{Indriolo2012, Neufeld2017}: it starts with the ionization of a hydrogen molecule, ${\rm H}_2 + {\rm CR} \to {\rm H}_2^+ + e^-$, followed by reaction with another neutral molecule, ${\rm H_2^+} + {\rm H_2} \to {\rm H_3^+} + {\rm H}$.  As this second reaction in molecular gas happens nearly every time following the H$_2^+$ formation, the reaction rate is proportional to the product of the CR ionization rate and the H$_2$ abundance.  Therefore, if the destruction rate of H$_3^+$ is known, the ionization rate can be determined from measurements of the column densities of H$_3^+$ and H$_2$.  In diffuse gas, such as found in the envelopes of molecular clouds, H$_3^+$ is primarily destroyed by dissociative recombination with free electrons \citep{McCall2003}. To calculate the ionization rates, \citet{Indriolo2012} assumed the electron fraction to be set by the carbon abundance, and thus the destruction rate to be proportional to the gas density.  \citet{Neufeld2017} went beyond this approximation and analyzed the chemistry of a smaller set of sight lines using a one-dimensional (1D) model of photodissociation region (PDR), arriving at a similar mean ionization rate as \citet{Indriolo2012}. These results suggest that the destruction rate of H$_3^+$ is roughly proportional to the gas density, and therefore the inferred ionization rate is directly proportional to the assumed gas density. Thus, uncertainty in the gas density is a substantial impediment to correct determination of the ionization rate from H$_3^+$ observations.  

Gas density in diffuse molecular clouds has been typically estimated by fitting the populations of excited rotational states of C$_2$ with a model including infrared pumping and collisional de-excitation \citep{Vandishoeck1982, Sonnentrucker2007}.  However, the so-called ``extinction mapping'' technique \citep{Leike2020, Lallement2022, Edenhofer2024} has recently reached sufficient accuracy as to give an independent measurement of the gas density on the scales of the clouds that the H$_3^+$ measurements are probing. The resulting density for different sight lines turns out to be, on average, an order of magnitude lower than the previous estimates. Remarkably, the density derived from the extinction maps are, on average, in good agreement with the values obtained recently from the revised analysis of available C$_2$ data \citep{Neufeld2024}.

In the present paper, we re-analyze the CR ionization rate based on available measurements of H$_3^+$, H$_2$, and H column densities for 12 lines of sight, deriving seven rates and five upper limits.  In contrast to previous work, we use the 3D gas density distribution obtained from the extinction map \citep{Edenhofer2024}, and calculate the UV field ourselves using stellar atmosphere models \citep{ATLAS9_Castelli2003} and the locations of hot stars \citep{PastelCatalog, StarHorse2021}. The use of the extinction maps makes it possible to localize the gas clumps where CR ionization is probed in individual measurements. We employ the {\sc 3d-pdr} code by \citet{Bisbas2012} to self-consistently compute the H$_3^+$ and H$_2$ abundances in these clumps for different values of the ionization rate, which is the only unconstrained parameter in our model. The derived ionization rates are dramatically reduced compared to the earlier estimates, and their values approximately scale with the new values of gas density obtained from the extinction map. In conclusion, we discuss implications of our results for all available measurements of the ionization rate.

\vspace{.5cm}
\section{Spatial gas distribution from 3D extinction maps}
\label{3D_maps}

We adopt the 3D map of interstellar dust described in \citet{Edenhofer2024}.
The map probes a larger volume at high resolution compared to previous maps \citep{Leike2020}, and achieves a higher dynamic range. The increase in dynamic range, by at least one order of magnitude, is due to the superior extinction resolution in the underlying data \citep[see][and references therein]{Zhang2023} used in \citet{Edenhofer2024} compared to the StarHorse DR2 photometry \citep{Anders2019} in \citet{Leike2020}.
The spatial distribution of interstellar dust is discretized into voxels in a spherical coordinate system.
Voxels are spaced logarithmically in distance, achieving a parsec-scale resolution throughout the map.
The map starts at $69\,\mathrm{pc}$ and extends out to a distance of $1.25\,\mathrm{kpc}$ from the Sun.
The distribution of interstellar dust is assumed to be spatially correlated a priori.
Specifically, the logarithm of the dust density is modeled via a Gaussian process.
The authors use Iterative Charted Refinement \citep{Edenhofer2022} for modeling the Gaussian process and employ the Bayesian probabilistic imaging framework \texttt{NIFTy.re} \citep{Edenhofer2024NIFTyRE} for the inference. The final map is published in the form of 12 equally likely posterior samples and, for convenience, a mean and standard deviation map based on these samples is published as well. 

\begin{figure}
\begin{center}
	\includegraphics[width=\columnwidth]{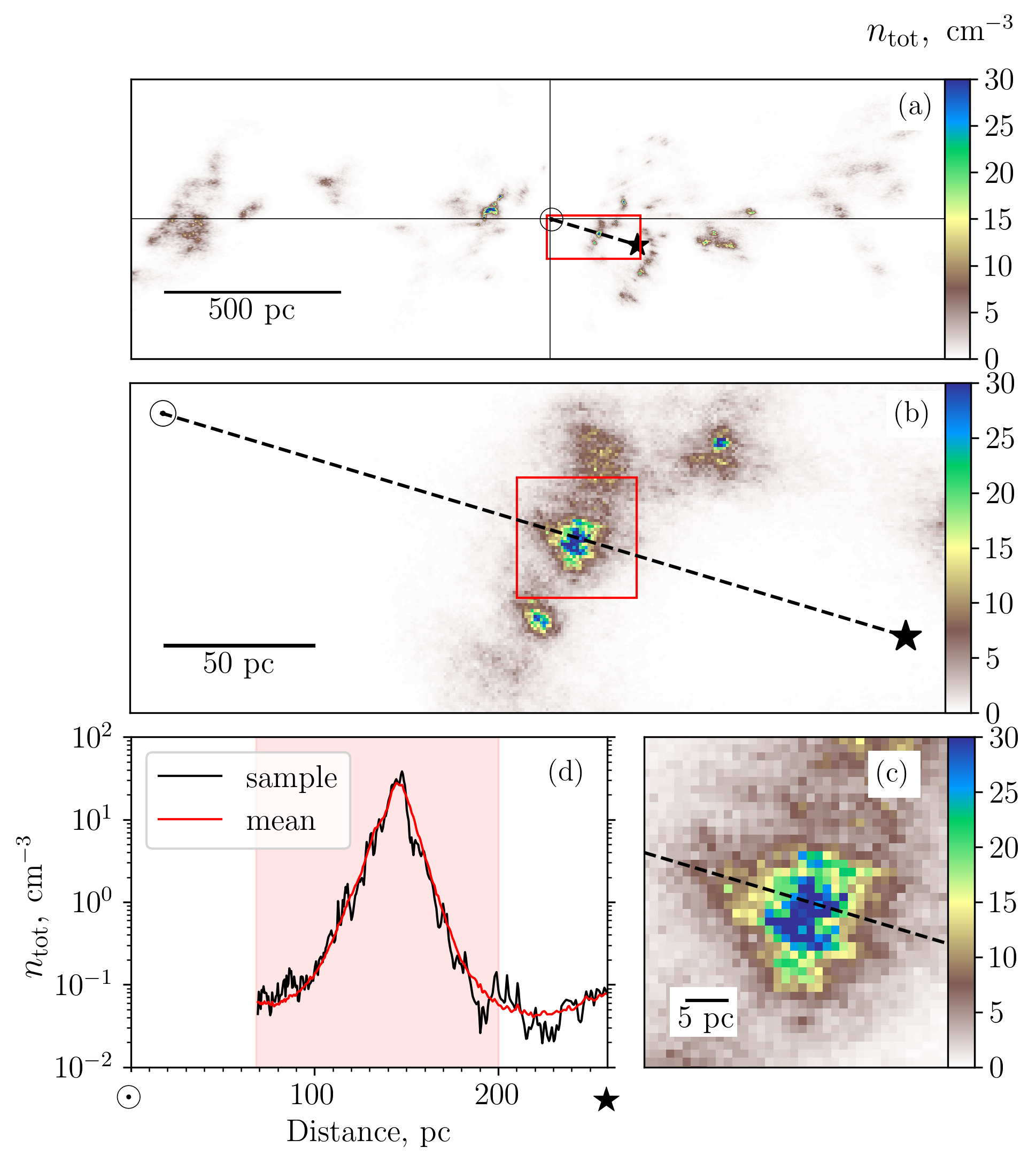}
    \caption{Gas density $n_{\rm tot}$ (total density of H nuclei, cm$^{-3}$) derived for the line of sight to HD~24398. The results represent one of the extinction map samples (see Section~\ref{3D_maps}). (a) Two-dimensional color-coded density distribution, plotted across the whole map in a vertical plane (perpendicular to the Galactic plane) which contains the sight line toward the star ($\star$), the Sun ($\odot$) is located at the origin. (b) Zoom-in to the region between the Sun and the star. (c) Zoom-in to the region containing the gas clump. (d) Density versus distance along the sight line (limited by the distance to the star): the black line corresponds to the selected map sample, the red line shows the density obtained by averaging over all samples. The shading indicates the region used for our simulations (see Section~\ref{3D_PDR}).}
    \label{f1}
\end{center}
\end{figure}

An exemplary projected view of the resulting 3D gas density reconstruction is shown in the top panel of Figure~\ref{f1}.
To derive the total volume density of hydrogen, $n_{\rm tot} = 2n_{{\rm H}_2}+ n_{\rm H}$, we adopt the column density conversion of $N_{\rm tot} = 5.3\times 10^{21}E_{\rm ZGR}$~cm$^{-2}$, where $E_{\rm ZGR}$ is the extinction\footnote{Related to the visual extinction $A(V = 540.0$~nm) via $A_V=2.8E_{\rm ZGR}$ \citep{Edenhofer2024}.} defined by \citet{Zhang2023} and utilized in \citet{Edenhofer2024}. This yields $n_{\rm tot} \approx 1700(dE_{\rm ZGR}/ds)$~cm$^{-3}$ \citep{ONeill2024}, where $s$ is the distance along the sight line in pc.

The density estimates derived from the extinction maps can be compared with previous estimates obtained from absorption-line observations of C$_2$ \citep{Sonnentrucker2007}, CO \citep{Goldsmith2013}, and CN \citep{Sheffer2008} in diffuse and translucent molecular clouds (that lie along the sight lines to nearby stars of sufficient brightness). Literature estimates based on the observed excitation of these molecules, although not entirely consistent among the different tracers, yield typical values that are almost an order of magnitude larger than those given by the extinction maps. However, in a companion paper \citep{Neufeld2024} we have revisited the excitation of C$_2$, the most widely used density tracer. Our analysis of the C$_2$ rotational populations, which are determined by a competition between radiative pumping and collisional excitation, made use of recent high-quality quantal calculations of the collisional excitation of C$_2$ by H$_2$ \citep{Najar2020}: these are considerably larger than the simple estimates assumed in previous studies, with the result that the inferred densities are considerably smaller. Therefore, in \citet{Neufeld2024} we obtained revised density estimates that are, on average, now in good agreement with those from the extinction maps. 


\section{Selected sight lines}
\label{LoS}

\begin{table*}

  \caption{Column densities for seven selected targets with H$_3^+$ detections and five targets with non-detections}
  \hspace{-1.35cm}
  \begin{tabular}{ |l || c | c || c | c || c | c || c || c |}
        \hline
        \rule[-1.5ex]{0pt}{4.5ex}
  HD~number & $N({\rm H}_3^+)^{\rm a}$ & Ref. & $N({\rm H}_2)^{\rm c}$ & Ref. & $N({\rm H})^{\rm c}$ & Ref. & $N({\rm H}_{\rm tot})^{\rm d}$ & $N({\rm H}_{\rm tot}^{\rm map})^{\rm e}$ \\
    ~(name)  & ($10^{13}$~cm$^{-2}$) & & ($10^{20}$~cm$^{-2}$) & & ($10^{20}$~cm$^{-2}$) & & ($10^{20}$~cm$^{-2}$) & ($10^{20}$~cm$^{-2}$) \\
        \hline\hline
        \rule[-.5ex]{0pt}{3.ex}
  24398 ($\zeta$~Per)   & $6.3\pm 0.5$ & (1) & $5.1^{+3.2}_{-1.6}$ & (4) & $6.4^{+0.6}_{-0.5}$ & (5) & $16.6^{+6.3}_{-3.2}$ & $15.4\pm 0.4$\\
        \hline
        \rule[-.5ex]{0pt}{3.ex}
  24534 (X~Per)         & $7.3\pm 0.9$ & (1) & $8.3^{+0.8}_{-0.7}$ & (6) & $5.5^{+0.8}_{-0.7}$ & (6) & $22.1^{+1.8}_{-1.6}$ & $24.3\pm 1.0$\\
        \hline
        \rule[-.5ex]{0pt}{3.ex}
  41117 ($\chi^2$~Ori)  & $5.3\pm 2.0$ & (2) & $5.0^{+1.4}_{-1.0}$ & (7) & $26.6^{+12.9}_{-7.2}$ & (7) & $36.6^{+12.8}_{-7.6}$ & $16.5\pm 0.3$\\
        \hline
        \rule[-.5ex]{0pt}{3.ex}
  73882      & $9.0\pm 0.5$ & (1) & $13.1^{+2.8}_{-2.1}$ & (6) & $13.7^{+6.6}_{-3.7}$ & (6) & $39.9^{+8.0}_{-5.9}$ & $31.0\pm 0.7$\\
        \hline
        \rule[-.5ex]{0pt}{3.ex}
  110432     & $5.2\pm 0.2$ & (1) & $4.4^{+0.4}_{-0.4}$ & (6) & $7.5^{+3.6}_{-2.0}$ & (6) & $16.3^{+3.6}_{-2.2}$ & $15.8\pm 0.2$\\
        \hline
        \rule[-.5ex]{0pt}{3.ex}
  154368     & $9.4\pm 1.3$ & (1) & $14.6^{+2.7}_{-2.1}$ & (6) & $10.1^{+1.3}_{-1.1}$ & (6) & $39.3^{+5.5}_{-4.3}$ & $31.8\pm 0.4$\\
        \hline
        \rule[-.5ex]{0pt}{3.ex}
  210839 ($\lambda$~Cep) & $7.6\pm 1.2$ & (1) & $6.3^{+0.8}_{-0.7}$ & (8) & $17.5^{+2.2}_{-1.8}$ & (8) & $30.2^{+2.7}_{-2.3}$ & $28.1\pm 0.4$\\
        \hline\hline
        \rule[-.5ex]{0pt}{3.ex}
  21856    & $<9.2$ & (3) & $1.2^{+0.6}_{-0.3}$ & (4) & $11.0^{+2.8}_{-1.9}$ & (5) & $13.4^{+2.9}_{-2.1}$ & $12.9\pm 0.2$\\
        \hline
        \rule[-.5ex]{0pt}{3.ex}
  22951 (40~Per)      & $<2.6$ & (3) & $3.1^{+2.0}_{-1.0}$ & (4) & $11.0^{+4.4}_{-2.6}$ & (5) & $17.3^{+5.3}_{-3.5}$ & $16.4\pm 0.2$\\
        \hline
        \rule[-.5ex]{0pt}{3.ex}
  148184 ($\chi$~Oph)  & $<1.9$ & (3) & $4.7^{+2.9}_{-1.5}$ & (4) & $14.2^{+3.1}_{-2.3}$ & (4) & $23.6^{+5.9}_{-3.9}$ & $18.6 \pm 0.6$\\
        \hline
        \rule[-.5ex]{0pt}{3.ex}
  149404   & $<14.1^{\rm b}$ & (3) & $6.4^{+1.2}_{-0.9}$ & (9) & $26.4^{+11.6}_{-6.8}$ & (9) & $39.1^{+11.5}_{-7.1}$ & $36.6\pm 0.6$\\
        \hline
        \rule[-.5ex]{0pt}{3.ex}
  149757 ($\zeta$~Oph) & $<2.1$ & (3) & $4.6^{+1.0}_{-0.7}$ & (4) & $5.2^{+0.2}_{-0.2}$ & (5) & $14.4^{+2.0}_{-1.5}$ & $12.6\pm 0.4$\\
        \hline
  \end{tabular}
\tablerefs{
(1)~\cite{Indriolo2012},
(2)~\cite{Albertsson2014},
(3)~Reevaluated upper limits from \cite{Indriolo2012},
(4)~\cite{Savage1977},
(5)~\cite{Bohlin1978},
(6)~\cite{Rachford2002},
(7)~\cite{Rachford2009},
(8)~\cite{Jenkins2019},
(9)~\cite{VanDePutte2023}.
}
$^{\rm a}\,$Measured values of $N({\rm H}_3^+)$ with $1\sigma$ errors for detection cases, and reevaluated upper limits for non-detection (see Section~\ref{LoS}). 

$^{\rm b}\,$The reported upper limit is a sum of the upper limits, 4.2+5.5+4.4 (for three distinct velocity components, see Sections~\ref{revisited} and \ref{149404}), corresponding to the three density peaks along the line of sight (see Figure~\ref{f_app1_11}). 

$^{\rm c}\,$Measured values of $N({\rm H}_2)$ and $N({\rm H})$. Their distributions for each sight line are assumed to be log-normal, the respective mean $\langle\log N\rangle$ and standard deviation $\sigma_{\log}$ are reported in the corresponding references. The shown errors represent 68\% confidence limits around the mean $\langle N\rangle = 10^{\langle\log N\rangle}\exp[\frac12(\ln 10\,\sigma_{\log})^2]$.

$^{\rm d}\,$The total column density computed from the measurements, $N({\rm H}_{\rm tot})=2N({\rm H}_2)+N({\rm H})$, the errors represent 68\% confidence limits around the mean value.

$^{\rm e}\,$The total column density $N({\rm H}_{\rm tot}^{\rm map})$ with $1\sigma$ errors derived from 12 posterior samples of the extinction map \citep{Edenhofer2024}.
\label{table1}
\end{table*}

\begin{figure*}
\begin{center}
	\includegraphics[width=.8\textwidth]{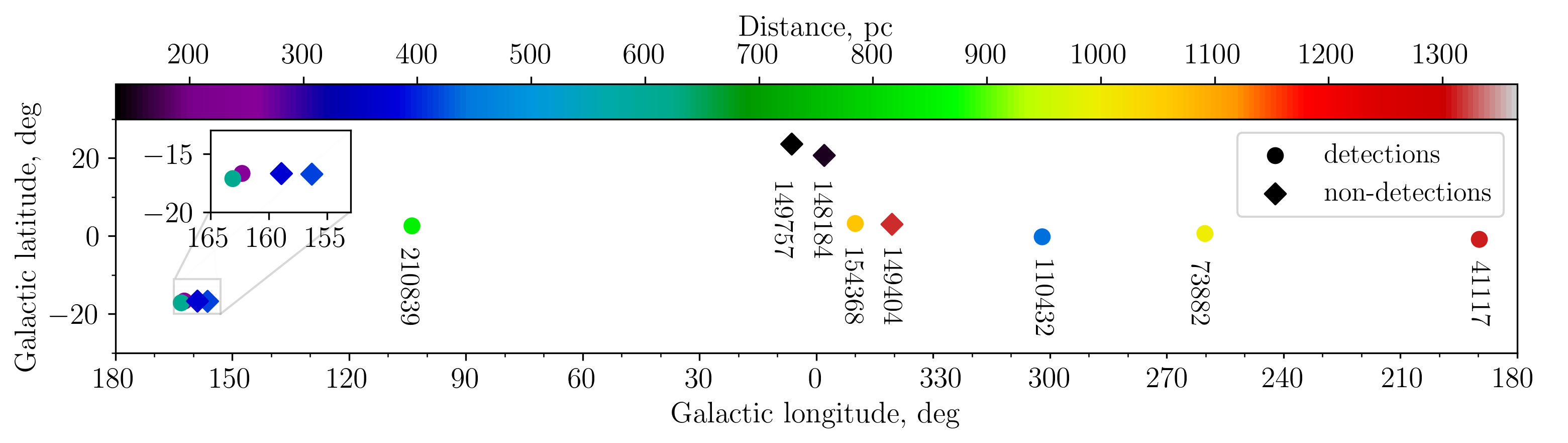}
    \caption{Selected sight lines (HD numbers of the target stars, see Section~\ref{LoS}) in Galactic coordinates. The bullets represent observations with H$_3^+$ detection, the diamonds represent non-detections, the color coding shows the distance to the target star. The zoom-in region contains a cluster of two detections (HD~24398 and HD~24534) and two non-detections (HD~21856 and HD~22951). The gas density distribution along each sight line is presented in Appendix~\ref{app1}.}
    \label{f2}
\end{center}
\end{figure*}

The most reliable of the available methods to infer the column densities of atomic and molecular hydrogen is to measure Ly$\alpha$ and H$_2$ Lyman absorption \citep{Savage1977, Bohlin1978, Diplas1994, Snow2000, Rachford2001, Rachford2002, Rachford2009, Shull2021,VanDePutte2023}, which give the total hydrogen column $N({\rm H}_{\rm tot})=2N({\rm H}_2)+N({\rm H})$. From all H$_3^+$ measurements that have been reported so far in the literature toward local, diffuse molecular clouds \citep{Indriolo2012, Albertsson2014} we therefore selected only those ``reliable'' targets where both H and H$_2$ columns were measured using the above methods. This yields a total of seven sight lines with H$_3^+$ detection, and 15 non-detection sight lines (out of which we selected five representative cases, corresponding to the highest peak densities derived from the extinction map along these sight lines). Measured values of $N({\rm H}_3^+)$ (or estimated upper limits for non-detection cases), $N({\rm H}_2)$, and $N({\rm H})$ for these targets are summarized in Table~\ref{table1}. The target locations in the Galaxy are illustrated in Figure~\ref{f2}, the gas distributions for each sight line are plotted in Appendix~\ref{app1}. 

Out of the 12 selected sight lines, eight contain a single dominant density peak (see Appendix~\ref{app1}). For the remaining four, we used available information on distinct velocity components derived from molecular and atomic absorption features to disentangle the respective contributions of the multiple peaks. We were able to identify multiple velocity components for two of such sight lines: two distinct components observed toward HD~41117 (detection) in neutral potassium, which is used as a proxy for H$_{\rm tot}$ \citep{Welty2001}, and three components observed toward HD~149404 (non-detection) in CH, tracing H$_2$ \citep{Sheffer2008}. In Section~\ref{CRIR} we demonstrate that these results allow us to reasonably estimate the individual contributions to $N({\rm H}_{\rm tot})$ from the clumps seen in Figure~\ref{f_app1_3} for HD~41117, and to $N({\rm H}_2)$ from the clumps in Figure~\ref{f_app1_11} for HD~149404. Furthermore, this analysis puts independent constraints on the column density values for the two targets (see discussion in Sections~\ref{41117} and \ref{149404}). The two remaining targets, HD~21856 and 22951, represent non-detections; each has two closely separated clumps along the sight lines (see Figures~\ref{f_app1_8} and \ref{f_app1_9}) where no separate velocity components were identified.

One should keep in mind that the dust reconstruction in the currently available extinction maps has a spatially varying line-of-sight resolution that degrades with distance \citep[from 0.6~pc at a distance of 100~pc to about 6~pc at 1~kpc,][]{Edenhofer2024}. Therefore, the sight lines containing gas clumps at $\sim1$~kpc distances (HD~41117, 73882, and 149404) may suffer from insufficient radial resolution, i.e., robust analysis of such clumps may require development of higher-resolution maps at larger distances.

We stress that available data on H and H$_2$ column measurements along all sight lines considered here are given with {\it symmetric} log errors \citep[][is the only exception]{Bohlin1978}. This may be related to the $\propto\log N$ scaling for the equivalent width of saturated lines used for the analysis \citep{Shull2024}. Hence, we assume log-normal distributions for $N({\rm H})$ and $N({\rm H}_2)$, and treat the given errors as corresponding standard deviations. This approach is apparently followed in \citet{Rachford2001, Rachford2002, Rachford2009}: the authors explicitly assume a normal distribution of $\log N({\rm H}_2)$ for each sight line \citep[see Section~3.2 in][]{Rachford2002}. While the resulting distribution of $N({\rm H}_{\rm tot})$ is, strictly speaking, not log-normal, it can be fairly accurately approximated with it, i.e., 68\% confidence limits around mean $N({\rm H}_{\rm tot})$ can be approximately treated as $1\sigma$ bounds.

\subsection{Revisited upper limits on $N({\rm H}_3^+)$}
\label{revisited}

Several sight lines where H$_3^+$ was not detected in the observations presented by \citet{Indriolo2012} have been re-observed at higher S/N and spectral resolution. In many cases, the subsequent observations reveal H$_3^+$ detections at levels above the previously quoted upper limits \citep[e.g., HD~27778 and HD~41117 in][]{Albertsson2014}. For this reason, we decided to revisit the upper limits on $N({\rm H}_3^+)$ reported by \citet{Indriolo2012}. In that work, uncertainties on equivalent widths were computed as $\sigma(W_{\lambda})=\sigma\lambda_{\rm pix}\sqrt{\mathcal{N}_{\rm pix}}$ where $\sigma$ is the standard deviation on the continuum at the expected location of the H$_3^+$ lines, $\lambda_{\rm pix}$ is the wavelength per pixel, and $\mathcal{N}_{\rm pix}$ is the number of pixels over which an absorption line is expected. These values were used to separately compute uncertainties in the column densities of the $(J,K)=(1,0)$ and (1,1) states of H$_3^+$, which were then added in quadrature to determine the uncertainties on the total H$_3^+$ column density. As upper limits are by definition one-sided probability distributions, the uncertainties on state specific column densities should not be added in quadrature. Additionally, the method for obtaining $\sigma(W_{\lambda})$ described above does not fully capture the signal that may reside in weak but coherent features (discussed in the following paragraph) that occur at wavelengths near the H$_3^+$ transitions. Together, these two effects produce systematically low upper limits. 

In order to re-calculate upper limits on $N({\rm H}_3^+)$ for the selected non-detection targets, we adopt the following method. First, we predict the expected line-center velocity and width of the H$_3^+$ absorption lines using the 4300~\AA\ line of CH as a proxy. As H$_3^+$ is formed from H$_2$, and as $N({\rm CH})$ and $N({\rm H}_2)$ have a linear relationship \citep{Sheffer2008}, we expect CH and H$_3^+$ to have similar absorption line profiles because line broadening is dominated by radial velocity gradients along the line of sight, rather than thermal motion.  Fully processed VLT/UVES spectra of four target stars (HD 22951, HD 148184, HD 149404, and HD 149757) were downloaded from the ESO Science Archive Facility,\footnote{\url{https://doi.org/10.18727/archive/50}} and degraded to the resolution of the corresponding H$_3^+$ spectra in \citet{Indriolo2012}. In the case of HD 21856, no CH data are available, so the spectrum from the nearby star HD 22951 was used.\footnote{The density peaks seen in Figures~\ref{f_app1_8} and \ref{f_app1_9} along the the lines of sight to HD 21856 and 22951 are about 14~pc apart in 3D (both for peaks at around 150 and 300~pc).} The CH absorption line profile was then fit with the sum of $n$ Gaussian functions to determine the line-center velocities and widths ($n=3$ for HD 149404 and $n=1$ for all other sight lines). Taking these parameters to be fixed, but allowing the line depths to vary as a free parameter, we then perform Gaussian fits to the H$_3^+$ spectra from \citet{Indriolo2012}. Using the optimized fit parameters and covariance matrix we compute the equivalent width and its uncertainty for each velocity component of both the $R(1,0)$ and $R(1,1)^u$ transitions of H$_3^+$. We note that these ``measured'' equivalent widths do not correspond to detections of H$_3^+$ absorption, but are simply being used to quantify the potential signal from absorption in coherent features that appear at the expected locations of H$_3^+$ absorption lines. These values are then used to compute column densities in the respective states, and the upper limit on the total H$_3^+$ column density is taken to be $N(1,0)+\sigma(N(1,0))+N(1,1)+\sigma(N(1,1))$, as reported in Table~\ref{table1}. 

\subsection{Total column density: Extinction maps versus measurements}
\label{maps_vs_measurements}

For individual targets, we still expect some deviations between the total column densities obtained from measurements, $N({\rm H}_{\rm tot})$, and the values of $N({\rm H}_{\rm tot}^{\rm map})$ derived from the extinction map. \citet{Rachford2002, Rachford2009} showed that values of $N({\rm H}_{\rm tot})$ measured for different sight lines and plotted versus the respective measured reddening exhibit noticeable deviations from the adopted average dependence ($\sim30\%$, up to a factor of 2 in extreme cases). 

There are several possible reasons why the ratio of $N({\rm H_{ tot}})$ to reddening may vary from one sight line to another, or exhibit a systematic offset from the expected value. First, the gas-to-dust mass ratio may deviate from the canonical value of 100 in turbulent diffuse regions.\footnote{The large-scale systematic increase of the gas-to-dust mass ratio with the galactocentric distance \citep{Giannetti2017} is negligible for the selected targets, as this distance varies by less than 1~kpc between the probed gas clumps (see the right panel of Figure~\ref{f6} below).} Second, the value of $R_V$ may vary between different sight lines \citep[see, e.g., Table 2 in][]{Rachford2009}. Third, $R_V$ may not be the only parameter that controls the ratio of dust mass to extinction. Finally, sub-parsec density spikes/structures that are not resolved in the map may affect the measured column densities.
    
The last two columns in Table~\ref{table1} give the values of $N({\rm H}_{\rm tot})$ and $N({\rm H}_{\rm tot}^{\rm map})$, indeed revealing a moderate (typically around 10--25\%) systematic underestimate of the total column seen in the map. Out of all selected targets HD~41117 is the only outlier here, with the map column density being a factor of $2.2$ smaller than the measure value (see Section~\ref{41117}). To compensate for this systematic effect, in our simulations we multiply the gas density deduced from the extinction map by the scaling factor $N({\rm H}_{\rm tot})/N({\rm H}_{\rm tot}^{\rm map})$, as discussed in the following section.

\section{Simulations with {\sc 3d-pdr}}
\label{3D_PDR}

In this work, we derive the CR ionization rate for the selected lines of sight by evaluating H$_2$ and H$_3^+$ column densities for the expected range of $\zeta_{{\rm H}_{2}}$. Along each sight line, we identify dense gas clumps where substantial amounts of H$_2$ are expected to form. Apart from HD~41117, all targets with H$_3^+$ detection contain only a single such clump (see Appendix~\ref{app1}), which allows us to localize the value of $\zeta_{{\rm H}_{2}}$ in 3D space.

We note that what is described as the ionization rate of H$_2$ in this paper \citep[and in previous works deriving it from H$_3^+$ measurements, e.g.,][]{Indriolo2012, Neufeld2017} is the rate of the reaction ${\rm H}_2 + {\rm CR} \to {\rm H}_2^+ + e^-$, which does not include dissociative ionization (${\rm H}_2 + {\rm CR} \to {\rm H}^+ +{\rm H} + e^-$). The latter channel comprises only 2--3\% of the total H$_2$ ionization \citep[see, e.g.,][]{Cravens1978}, and therefore is usually neglected. Hence, $\zeta_{{\rm H}_{2}}$ denotes the CR ionization rate for the above main reaction.  

Atomic-to-molecular hydrogen transition occurs inside the predominantly neutral photodissociation regions (PDR), where the heating and the chemical processes are driven by far-UV photons with energies of 6--13.6~eV \citep[see, e.g.,][]{DraineBook2011, TielensBook2005}. We use the {\sc 3d-pdr}\footnote{\url{https://uclchem.github.io/3dpdr/}} code \citep{Bisbas2012} to solve the PDR chemistry and estimate molecular abundances and temperatures for each region of interest. Three-dimensional modeling offers an advantage in accurately considering H$_2$ self-shielding within clouds with complex filamentary structures. The temperature distribution is self-consistently computed taking into account both radiative and chemical processes, until the system achieves chemical equilibrium and thermal balance.

To reduce the amount of computational cost in the three-dimensional models, we use the standard subset of the UMIST 2012 chemical network described in \citet{McElroy13}. It is restricted to four elements with the abundances (listed in Table~\ref{table_abundances}) typical to the solar neighborhood \citep{Sofia2004, Gerin2015, Cartledge2004}, and contains 33 species (including electrons) with 330 reactions. We model H$_2$ formation using the treatment of \citet{Cazaux_Tielens_2002,Cazaux_Tielens_2004} and assuming the initial fraction of 0.3 for H$_2$. In addition, we assume the dust-to-gas mass ratio of $10^{-2}$ and a microturbulence velocity of 1~km~s$^{-1}$. The temperature of dust grains heated by the interstellar FUV field is self-consistently computed from Equations~(5) and (6) of \citet{Dust_temperature}. All our models are evolved for a chemical time of $10~\,{\rm Myr}$, which ensures that the cloud has reached thermal equilibrium.

\begin{table}[h!]
  \caption{Gas-phase elemental abundances (relative to H) assumed in the simulations}
  \centerline{
  \begin{tabular}{ |l | c |}
        \hline
        \rule[-.5ex]{0pt}{3.ex}
Element & Abundance\\
        \hline\hline
        \rule[-.5ex]{0pt}{3.ex}
He & $1.0\times10^{-1}$\\
        \hline
        \rule[-.5ex]{0pt}{3.ex}
C & $1.6 \times 10^{-4}$\\
        \hline
        \rule[-.5ex]{0pt}{3.ex}
O & $3.9 \times 10^{-4}$\\
        \hline
  \end{tabular}
  }    
\label{table_abundances}
\end{table}

We use the 3D dust map by \citet{Edenhofer2024}, which provides a parsec-scale spatial resolution, to derive the spatial distribution of the total gas density $n_{\rm tot}$ (see Section~\ref{3D_maps}). For each individual gas clump, we define an ellipsoidal simulation domain centered at the density peak. The ellipsoid typically has semi-major axes of 70--80~pc oriented along the sight line, and a semi-minor axis of about 30~pc. If the filament is not aligned with the sight line, we take an ellipsoid with a somewhat larger semi-minor axis to include more material in other directions. The extinction map within the simulation domain is interpolated on a uniform Cartesian grid with a 1~pc resolution, resulting in a total of $(1-5)\times 10^{5}$ elements.

We apply the scaling factor of $N({\rm H}_{\rm tot})/N({\rm H}_{\rm tot}^{\rm map})$ to the map density inside a simulation domain, to compensate for systematic discrepancy between the measured column densities and the values derived from the map (see Section~\ref{maps_vs_measurements}). Three simulations with different scaling factors are run for each sight line, corresponding to the mean value and the upper and lower limits of the confidence interval for the measured total column density $N({\rm H}_{\rm tot})$ displayed in Table~\ref{table1}. From this point $n_{\rm tot}$ always denotes the rescaled total density.

\subsection{Catalogue of FUV-active stars}
\label{catalogue}

OB stars serve as the primary source of the FUV field that influences the PDR chemistry \citep{TielensBook2005}. Due to their highly heterogeneous spatial distribution, the presence of a single star near the simulation domain can significantly impact the local FUV field. The Gaia DR3 release is the most complete catalog of stars with {\it G}-band magnitude $< 19$~mag \citep{GaiaEDR3_Validation}, but it misses most of the brightest stars with {\it G} $< 3$~mag. We expand the Gaia DR3 source list with stars from the Tycho-2 catalog, which is 99\% complete up to {\it V} = 11.0~mag \citep{Tycho2Catalog}. The new version of Tycho-2,\footnote{\url{https://gaia.aip.de/metadata/gaiadr3/tycho2tdsc_merge}} merged with the Tycho Double Star Catalogue (TDSC) \citep{TDSC_Catalog}, is used to account for additional components in double and multiple star systems.

We perform a cross-match with other catalogs to find information about the parallax and effective temperature of the stars. Parallax measurements from the Gaia DR3 data are utilized if available, otherwise we rely on the Hipparcos~2 catalog \citep{Hipparcos2}. Values of the effective temperature $T_{\rm eff}$ are taken from the PASTEL catalog of stellar atmospheric parameters \citep{PastelCatalog} or from the the Tycho-2 Spectral Type Catalog \citep{Tycho2_SpectralTypeCatalog}. The spectra are selected from the ATLAS9 grid of stellar atmosphere models \citep{ATLAS9_Castelli2003} corresponding to a given $T_{\rm eff}$; since the surface gravity has no significant effect on the total FUV flux, we adopt the highest available values.\footnote{\url{https://wwwuser.oats.inaf.it/fiorella.castelli/grids/gridp00k0odfnew/fp00k0tab.html}} 

For a star with a given synthetic spectrum $I(\lambda)$, the observed flux density is proportional to the geometrical dilution factor $R_*^2/d^2$, where $R_*$ is the stellar radius and $d$ is the distance to the star. We estimate $R_*$ by utilizing the definition of the Hipparcos $V_T$ band apparent magnitude \citep{Hipparcso_synthetic_photometry}:
\begin{equation}
    -2.5\log{\frac{\int f(\lambda) S(\lambda)\lambda d \lambda}{\int cS(\lambda) d\lambda / \lambda}} - 48.60 - ZP,
    \label{V_T}
\end{equation}
where 
\begin{equation}
        f(\lambda) = \frac{R^{2}_{*}}{d^{2}}\pi I(\lambda) e^{-A_{\lambda}},
        \label{f_lambda}
\end{equation}
is the observed flux density (in erg~cm$^{-2}$~s$^{-1}$~\AA$^{-1}$), $A_{\lambda}$ is the extinction at wavelength $\lambda$ (in \AA), $S(\lambda)$ is the passband response function, and $ZP=-0.037$ is zero-point magnitude for the $V_T$ band. The flux is corrected for dust extinction by using the publicly available interstellar extinction curve for diffuse Milky Way\footnote{\url{https://www.stsci.edu/hst/instrumentation/reference-data-for-calibration-and-tools/astronomical-catalogs/interstellar-extinction-curves}}. The total FUV flux of the star at a given point is then obtained by integrating Equation~(\ref{f_lambda}) over the energy range of 6--13.6~eV, with $d$ being the distance to the point. The local FUV field produced by all stars, $u(6-13.6~{\rm eV})$, follows the \citet{DraineBook2011} normalization, i.e.,
\begin{equation}
G_{\rm D}=\frac{u(6-13.6~{\rm eV})}{8.94\times10^{-14}~{\rm erg}~{\rm cm}^{-3}}\,.    
\end{equation}

For stars whose effective temperatures are missing in PASTEL and Tycho-2 catalogs, we use stellar parameters from the StarHorse catalog of photo-astrometric distances and astrophysical parameters for Gaia EDR3 stars \citep{StarHorse2021}. The catalog provides effective temperatures, masses, and surface gravity, so that stellar radii can be straightforwardly calculated. For self-consistency, we also replace measured parallaxes for these stars with the modified StarHorse distances. One should keep in mind the potential influence of the extinction prior in \citet{StarHorse2021} using the Bayestar19 extinction map \citep{Green2019}, which leads to data being incorporated multiple times into the model. However, we believe that the indirect biases on the intrinsic stellar parameters and the parallax are not significant for our purposes.

To facilitate the calculations of FUV field, we exclude stars whose contribution is found statistically negligible. For this, we first calculate the FUV spectrum in the solar neighborhood to verify that the above procedure reproduces the spectrum by \citet{Mathis1983} as well as the available measurements \citep[see Figure~12.2 in][]{DraineBook2011}. Then, from a cumulative plot of the FUV field versus $T_{\rm eff}$ we conclude that stars with $T_{\rm eff}\lesssim 10,000$~K have only minor contribution. Thus, for our calculations we only include stars with higher effective temperatures.

\subsection{Modifications to {\sc 3d-pdr}}

We have made several corrections and updates to the publicly available {\sc 3d-pdr} code, which are listed in Appendix~\ref{app2}.

In addition, we have modified the routine that computes the FUV field. The {\sc 3d-pdr} code has two possible implementations of the field for 3D simulations -- plane-parallel or isotropic radiation; neither of those is suitable for our purposes. The Galactic FUV radiation is extremely anisotropic, as stars are concentrated near the disk midplane. Furthermore, it is necessary to incorporate point-source contributions of nearby stars.

The anisotropy of the radiation field is implemented by taking advantage of the {\sc HEALpix} routine (\citealt{HEALPIX_1,HEALPIX_2}) that performs ray-tracing in {\sc 3d-pdr}. In our 3D models, the $\ell=0$ level of {\sc HEALpix} refinement is considered, corresponding to a 12-ray structure emanated uniformly on the celestial sphere. We compute the stellar radiation impinging from each of these {\sc HEALpix} pixels separately, and the attenuated field is then obtained by summing up over the 12 contributions. To give an example: for the FUV field in the center of the clump depicted in Figure~\ref{f1}, partial contributions of HEALPix pixels in nested order are 5.6, 20.5, 0.6, 8.1, 14.2, 9.7, 16.7, 10.9, 0.8, 0.9, 9.7, and 1.8\% (without accounting for extinction inside the simulation domain). 

\begin{figure*}
\begin{center}
	\includegraphics[width=\textwidth]{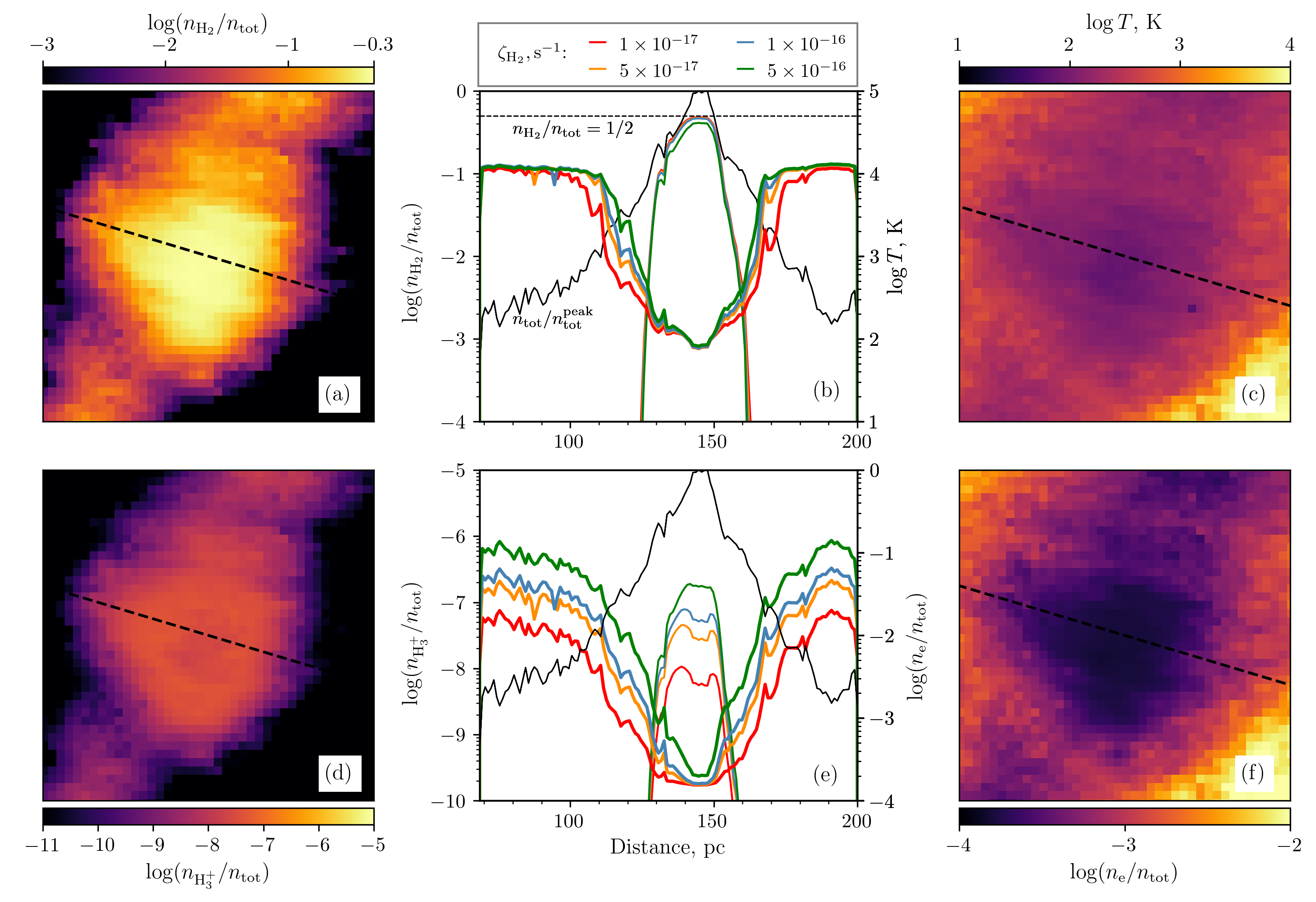}
    \caption{Results of the simulations for the line of sight to HD~24398. The figure illustrates the spatial distributions of relevant physical parameters in the gas clump depicted in Figure~\ref{f1}. In the two central panels (b,e), four parameters are plotted versus the distance in the range indicated by the shading in Figure~\ref{f1}(d): abundance of molecular hydrogen $n_{{\rm H}_2}/n_{\rm tot}$ (thin colored curves) and gas temperature $T$ (thick colored curves) in panel (b), as well as H$_3^+$ abundance $n_{{\rm H}_3^+}/n_{\rm tot}$ (thin colored curves) and ionization fraction $n_e/n_{\rm tot}$ (thick colored curves) in panel (e). The parameters are computed for different values of $\zeta_{\rm H_2}$ (listed in the legend), the horizontal dashed line in panel (b) indicates the saturation level $n_{{\rm H}_2}/n_{\rm tot}=1/2$, the thin black curve in both panels shows the gas density $n_{\rm tot}$ (normalized by the peak value $n_{\rm tot}^{\rm peak}$) for the selected sample of the extinction map. The four side panels (a,c,d,f) illustrate two-dimensional distributions of these parameters (for $\zeta_{\rm H_2}=5\times10^{-17}$~s$^{-1}$) within the zoom-in cross section in Figure~\ref{f1}(c), the dashed lines show the line of sight.}
    \label{f3}
\end{center}
\end{figure*}

FUV-active stars are distributed across the extinction map, but high-resolution simulations at large scales are not feasible due to computational limitations. However, the absolute majority of such stars are located at distances much larger than the size of the simulation domain, i.e., their contributions to the local field practically do not vary at the domain scales. This fact allows us to divide the external FUV field into two components: point-source contributions of individual ``nearby'' stars and a continuous ``far-field'' component.

Stars contributing to the far-field component are selected such that their combined field does not vary by more than 20\% across the simulation domain (neglecting extinction within the domain). Individual low-temperature stars located close to the simulation domain may also be included into this group if their contributions are sufficiently small. The group of nearby point sources typically includes B stars with effective temperatures around 20,000--30,000~K located at distances of $\lesssim60$~pc from the density peak, and O stars with $T_{\rm eff} \gtrsim 30,000$~K at $\lesssim100$~pc. Associations of stars with lower $T_{\rm eff}$ may have a noticeable effect on the FUV field if they are located sufficiently close to the simulation domain, in which case they are included into the group of nearby stars.

Extinction outside the simulation domain is computed using a 10~pc resolution map and employing the procedure discussed in Section~\ref{catalogue}. Within the domain, we combine incoming radiation from nearby point sources with the corresponding far-field component to obtain the unattenuated field for a given solid-angle segment, and compute extinction for each simulation cell using the method implemented in the code \citep{Bisbas2012}.

\subsection{Simulated physical structure}

Individual simulations were run for each gas clump located on the selected lines of sight. Figure~\ref{f3} illustrates the results of such a simulation for a clump on the sight line to HD~24398, chosen as a representative example of key features that we found in common for most of the studied cases.

The temperature profile shown in Figure~\ref{f3}(b) reflects the typical transition between the warm neutral medium (WNM, $3\times10^3~{\rm K}\lesssim T \lesssim 10^4$~K) and the cold neutral medium \citep[CNM, $T \lesssim 300$~K; see, e.g.,][]{Wolfire2003, DraineBook2011}. The FUV radiation of the surrounding OB stars, the main heating source in PDR regions, is efficiently absorbed by dust. This leads to a rapid temperature decrease toward the dense region, from $T\sim10^{4}$~K in the WNM to $\approx80$~K in the density peak for HD~24398. The molecular hydrogen abundance rapidly increases toward the center, saturating at $n_{{\rm H}_{2}}/n_{\rm tot}=1/2$ near the peak for $\zeta_{\rm H_2}\lesssim 10^{-16}$~s$^{-1}$. 

For most of the detections (HD~24398, 24534, 73882, 110432, 154368) and for one non-detection (HD~148184), a fully molecular region develops in each dense clump. The temperature of molecular gas varies between $T\approx50$~K and $\approx80$~K, depending on the peak density: the lower temperature value corresponds the peak density of $n_{\rm tot}\approx100$~cm$^{-3}$ in HD~154368, and the higher one to $\approx35$~cm$^{-3}$ in HD~24398. The temperature can increase in the presence of OB stars nearby: for example, $T\approx90$~K in the gas clump of HD~148184, despite the peak density of $\approx90$~cm$^{-3}$. On the other hand, there is an exceptional case of HD~73882, with a fully developed molecular region at $T\approx110$~K and the peak density of 22~cm$^{-3}$. However, we note uncertainty in the position of the target star HD~73882, which leads to noticeable variations in the scaling correction applied to the map density in this case (see Section~\ref{73882}).

For detection HD~210839 and most of the non-detections (HD~21856, 22951, 149404) the peak density is lower, varying between $\approx10$~cm$^{-3}$ and 17~cm$^{-3}$. Hydrogen in such gas is only partially molecular and temperatures are higher, changing from $n_{{\rm H}_{2}}/n_{\rm tot}\approx0.4$ and $\approx110$~K for HD~210839 to $\approx0.1$ and $\approx140$~K for HD~21856. The lowest peak density among the selected sight lines is achieved in the remaining detection HD~41117 (which is the outlier in terms of the deviation between $N({\rm H}_{\rm tot})$ and $N({\rm H}_{\rm tot}^{\rm map})$, see Table~\ref{table1}). Even after the density rescaling, the peak value (in the densest  clump, see Figure~\ref{f_app1_3}) is below $9$~cm$^{-3}$, and the simulations give $n_{{\rm H}_{2}}/n_{\rm tot}\approx0.3$ and $\approx120$~K. The remaining non-detection HD~149757 has the peak density of 36~cm$^{-3}$, which is comparable to the clumps with fully developed molecular hydrogen. However, the gas temperature in this case is about 110~K and the H$_{2}$ abundance does not exceed 0.4, because of the proximity to OB stars (see Section~\ref{149757}).

In Figure~\ref{f3}(e) we see the ionization fraction $n_{e}/n_{\rm tot}$ reducing toward the center of the HD~24398 clump. For $\zeta_{\rm H_2}\lesssim 10^{-16}$~s$^{-1}$, it saturates near the density peak at the level of $n_{e}/n_{\rm tot}=1.6 \times 10^{-4}$, the carbon initial abundance. This regime is realized for most of the clumps, except for a slightly lower ionization fraction of $1.3 \times 10^{-4}$ at the highest density peak of HD~154368, where the extinction starts affecting ionization of carbon. Thus, no significant reduction (below the carbon abundance level) is observed in $n_{e}/n_{\rm tot}$  even for the highest gas densities. This conclusion is particularly important, as a strong reduction in $n_{e}/n_{\rm tot}$ would be accompanied by the proportional increase in $n_{{\rm H}_{3}^{+}}/n_{\rm tot}$ \citep[see Figure~1 in][]{Neufeld2017}, which could then lead to a situation where $N({\rm H}_3^+)$ is dominated by this phenomenon. 

We also see that most of H$_{3}^{+}$ is formed where the H$_{2}$ abundance exceeds $\approx0.2$. The H$_{3}^{+}$ abundance increases with $n_{{\rm H}_{2}}/n_{\rm tot}$, and exhibits a local minimum at the density peak. Given that $n_{{\rm H}_{3}^{+}}\propto n_{{\rm H}_{2}}/n_e$ (see Section~\ref{discussion}), this minimum in $n_{{\rm H}_{3}^{+}}/n_{\rm tot}$ is because both $n_{{\rm H}_{2}}/n_{\rm tot}$ and $n_{e}/n_{\rm tot}$ show a broad plateau near the peak while $n_{\rm tot}$ continues to increase.

Finally, the four right panels (a,c,d,f) in Figure~\ref{f3} illustrate two-dimensional distributions of the parameters discussed above, showing that their shapes essentially follow the gas density distribution in Figure~\ref{f1}(c).

\section{CR ionization rate in individual clouds}
\label{CRIR}

For each selected sight line, we perform a set of simulations for different characteristic values of $N({\rm H}_{\rm tot})$ and $\zeta_{\rm H_2}$. For $N({\rm H}_{\rm tot})$ we use three values, corresponding to the measured mean and 68\% confidence limits listed in Table~\ref{table1}. To probe the possible expected range of CR ionization rate, we adopt four values of $\zeta_{\rm H_2}=1\times$, $5\times$, $10\times$, and $50\times10^{-17}$~s$^{-1}$. Values of $N({\rm H}_2)$ and $N({\rm H}_3^+)$ computed for each combination are summarized in Appendix~\ref{app3} (Tables~\ref{table3} and \ref{table4}). In order to explore potential impact of different map realizations, simulations for each sight line are run for the density distribution from one of the extinction map samples, and for the density obtained by averaging over all samples.

As explained in Section~\ref{catalogue}, the distances to all target stars as well as to individual nearby stars, i.e., those contributing significantly to the local FUV field in the gas clumps, are taken from Gaia DR3 or StarHorse catalogue (if available), or from Hipparcos~2 data. The only two exceptions are the distances to the target star HD~73882 (see Section~\ref{73882}), and to one UV-active star toward the sight line HD~154368 (see Section~\ref{154368}). In these cases, we use the recently updated distance values and reevaluate the stellar radii as well as the observed flux densities accordingly (see Section~\ref{catalogue}). 

\begin{figure}[h!]
\begin{center}
	\includegraphics[width=\columnwidth]{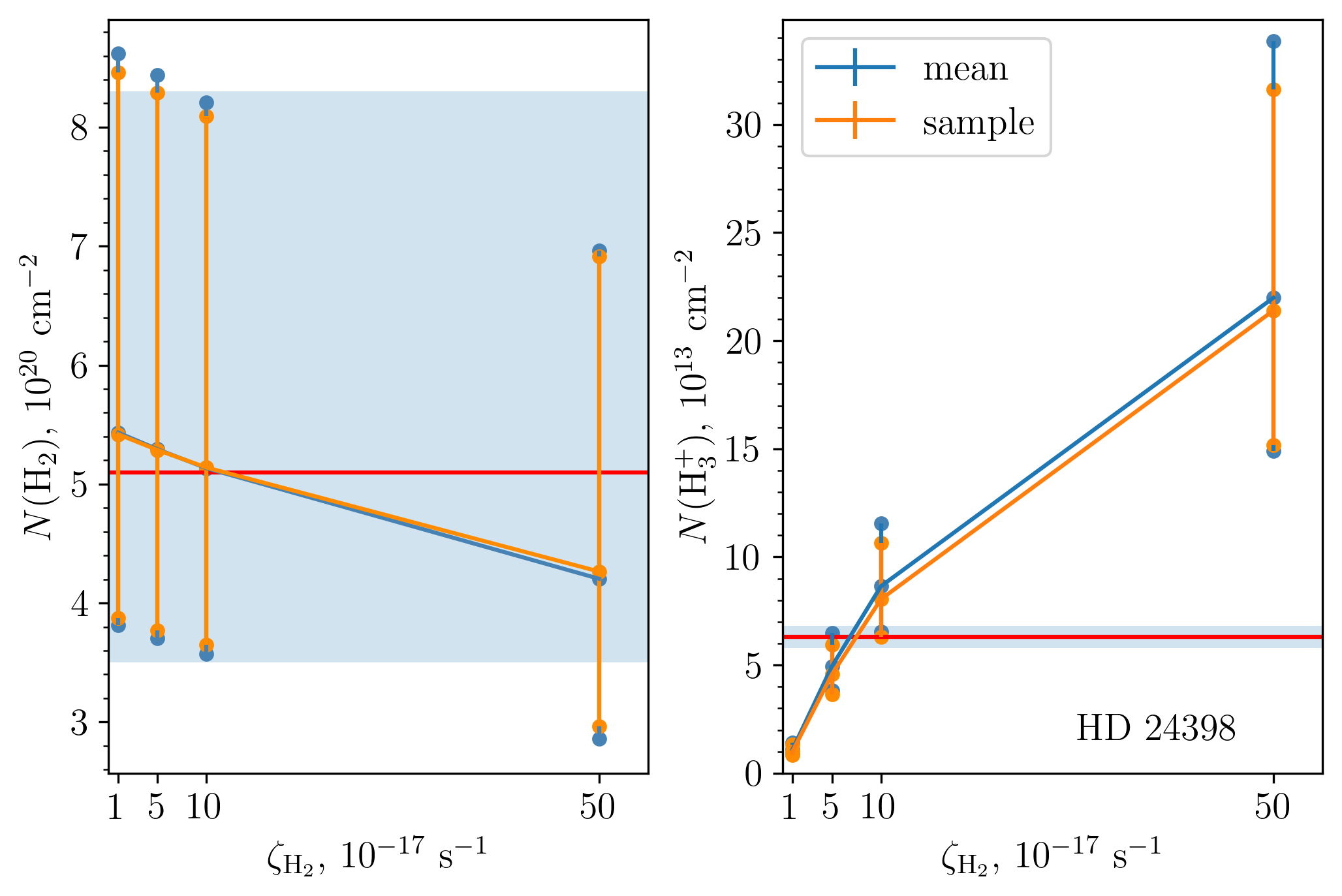}
    \caption{Line of sight to HD~24398 with H$_3^+$ detection. The bullets show computed column densities $N({\rm H}_2)$ (left panel) and $N({\rm H}_3^+)$ (right panel), the horizontal lines with shaded stripes indicate the respective measured mean values with 68\% confidence limits. For each of the four values of $\zeta_{\rm H_2}$, the central bullets (connected by the straight lines) show the results computed for the measured mean $N({\rm H}_{\rm tot})$ while the upper and lower bullets represent the respective 68\% confidence limits of $N({\rm H}_{\rm tot})$ (see Table~\ref{table1}). The orange symbols correspond to the density distribution from one of the extinction map samples, the blue symbols are for the density obtained by averaging over all samples.} 
    \label{f4_1}
\end{center}
\end{figure}

\begin{figure}[h!]
\begin{center}
	\includegraphics[width=\columnwidth]{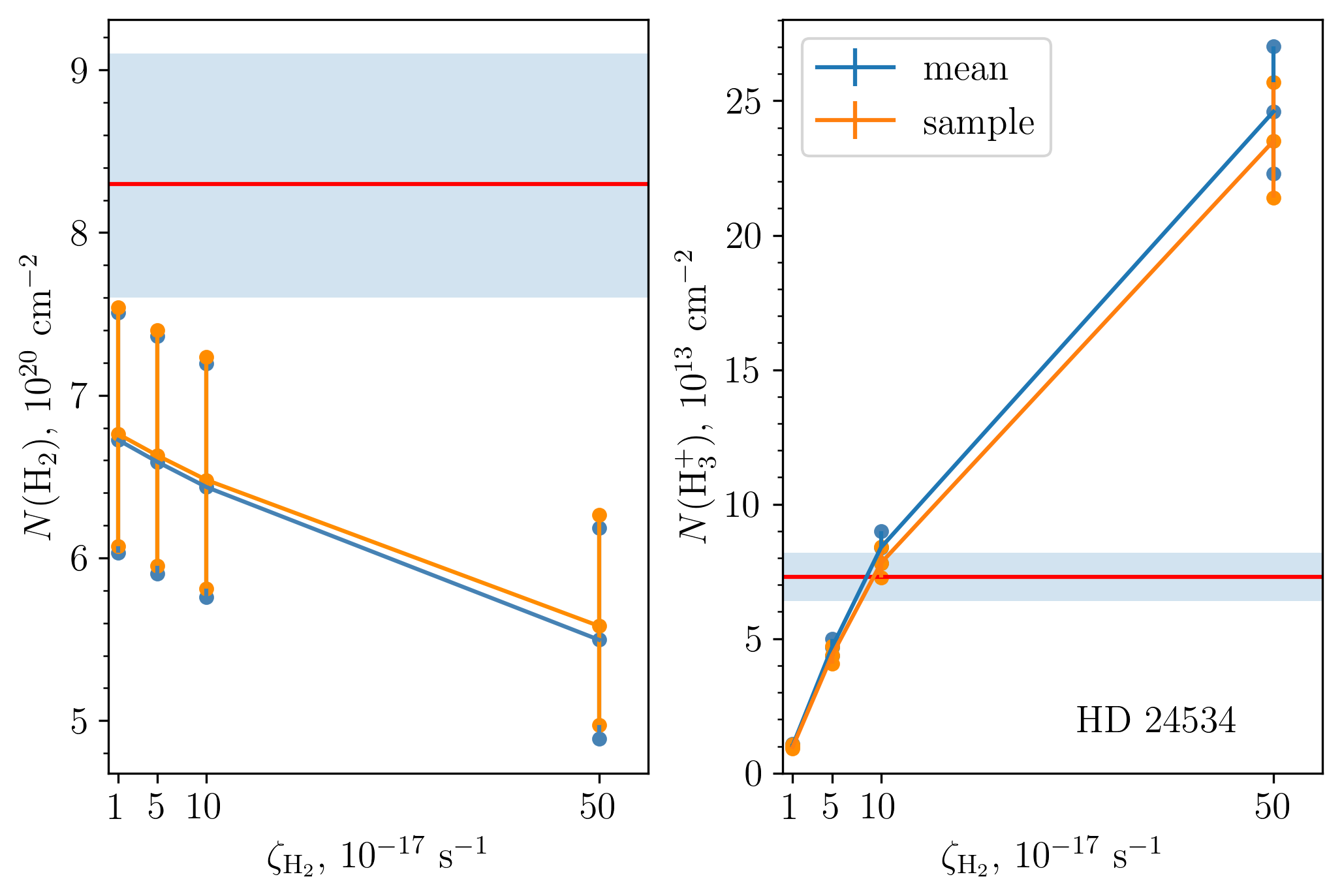}
    \caption{Same as in Figure~\ref{f4_1} for HD~24534.} 
    \label{f4_2}
\end{center}
\end{figure}

Below we present the results of our simulations for each selected sight line (Figures~\ref{f4_1}--\ref{f4_12}), plotting the computed values of $N({\rm H}_2)$ and $N({\rm H}_3^+)$ versus $\zeta_{\rm H_2}$ together with the observed values from Table~\ref{table1}. Their detailed comparison and analysis is given in Section~\ref{discussion}. We point out that the deviations between the results obtained for the selected map sample and for the average is practically negligible for all sight lines, with the biggest difference of about 10\% seen for HD~41117. 

The far-field component of the FUV field is found to be between $G_{\rm D}=$~0.4--0.8 in the proximity of all studied clumps. However, the contribution of individual nearby stars or their associations varies dramatically between the clumps for many sight lines, where the FUV field within the corresponding simulation domains is completely dominated by the nearby stars. Their specific effect for each of such sight lines is then summarized in short subsections below and, where appropriate, the effect of distance uncertainty on the results is also discussed. To compare the contributions of the two FUV components, we quote the field magnitude without accounting for extinction {\it inside} the simulation domain; for nearby stars, we report their total field at the closest point of the domain boundary and at the domain center. No separate subsections are included for the three detection sight lines, HD~24398 (Figure~\ref{f4_1}), 24534 (Figure~\ref{f4_2}), and 210839 (Figure~\ref{f4_7}), where the far-field component dominates and no additional discussion is needed.

For the four targets with multiple gas clumps along the line of sight (detection HD~41117 and non-detections HD~21856, 22951, and 149404, see their discussion in Section~\ref{LoS}), we report contributions of the individual clumps into the computed results. For two of these targets, HD~41117 and 149404, we describe the additional analysis performed using distinct velocity components (derived from available observations) that allow us to constrain amount of gas in the individual clumps. Given that no separate components have been detected in H$_3^+$ measurements, we cannot disentangle CR ionization produced in the individual clumps, i.e., the same $\zeta_{\rm H_2}$ is assumed for multiple clumps along a given sight line. 

\subsection{H$_3^+$ detections}
\label{detections}


\subsubsection{HD~41117 ($\chi^{2}$ Ori)}
\label{41117}

\begin{figure}[h!]
\begin{center}
	\includegraphics[width=\columnwidth]{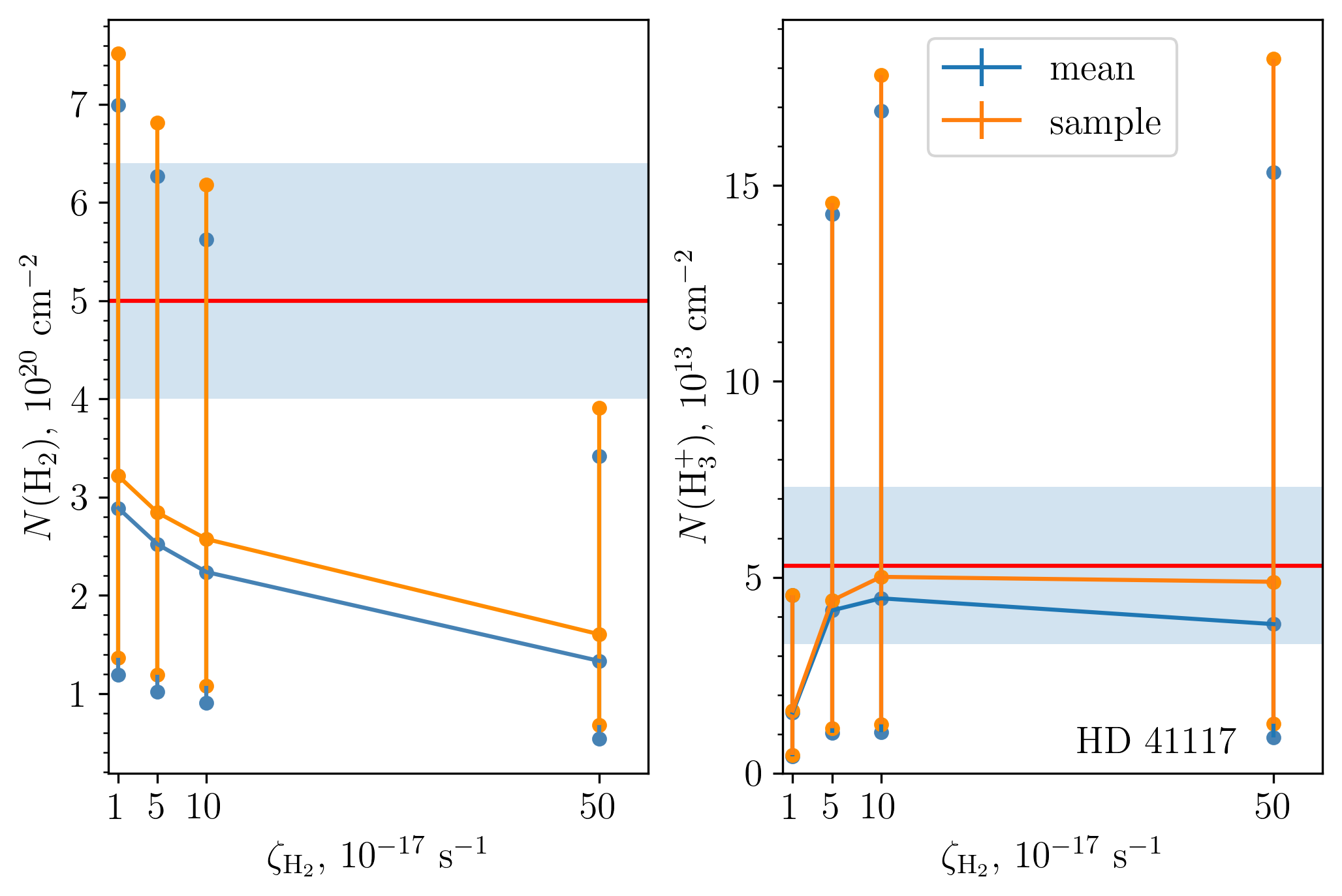}
    \caption{Same as in Figure~\ref{f4_1} for HD~41117.} 
    \label{f4_3}
\end{center}
\end{figure}

The sight line to HD~41117 is the only outlier in terms of the deviation between $N({\rm H}_{\rm tot})$ and $N({\rm H}_{\rm tot}^{\rm map})$ (see Table~\ref{table1}). Nevertheless, after the density rescaling we see that the computed $N({\rm H}_{2})$ shows reasonably good agreement with observations: the results plotted in the left panel of Figure~\ref{f4_3} agree well within the $N({\rm H}_{2})$ errors set by the confidence interval for $N({\rm H}_{\rm tot})$. The three clumps along the sight line in Figure~\ref{f_app1_3} have comparable computed contributions, e.g., $N({\rm H}_{2}) \approx 0.7\times$, $1.3\times$, and $1.0\times10^{20}$~cm$^{-2}$, respectively, for the mean $N({\rm H}_{\rm tot})$ and $\zeta_{\rm H_2}=5\times10^{-17}$~s$^{-1}$. On the other hand, one should bear in mind that the peak densities for all three clumps -- even after the rescaling -- are below 10~cm$^{-3}$, which makes gas along this line of sight the most diffuse out of all considered targets. As discussed below in Section~\ref{discussion}, this is the reason why the computed $N({\rm H}_3^+)$ saturates and starts decreasing with $\zeta_{\rm H_2}$, whereas in all other cases it monotonically increases. By comparing the middle panel in Figure~\ref{f_app1_3} with that for the other sight lines, we also point out that the morphology of gas distribution in the clumps of HD~41117 appears much less compact. 

{\it Analysis of multiple velocity components.} The absorption profile of CH at 4300~\AA\ toward this star is asymmetric, hinting at multiple components, but they are not particularly well-separated in the available spectrum \citep{Crane1995}. Absorption from the 7698~\AA\ line of K~\textsc{i}, however, clearly shows two minima at about $v_{\rm LSR}=-3$~km/s and $+3$~km/s \citep{Welty2001}. While the authors separate the absorption into eight total components (see their Table 2), we take the sum of their components 2 and 3, $N({\rm K~\textsc{i}})=4.6\times10^{11}$~cm$^{-2}$, and of components 5 and 6, $N({\rm K~\textsc{i}})=9.2\times10^{11}$~cm$^{-2}$, as the K~\textsc{i} column densities in the $-3$ and $+3$ km~s$^{-1}$ features, respectively. \citet{Welty2001} also derive an empirical relationship between $N({\rm K~\textsc{i}})$ and $N({\rm H}_{\rm tot})$ that is expressed as $\log{N({\rm K~\textsc{i}})} = -27.21+1.84\,\log{N({\rm H}_{\rm tot})}$, and that we use to estimate total column densities of $N({\rm H}_{\rm tot})=1.3\times10^{21}$~cm$^{-2}$ and $2.0\times10^{21}$~cm$^{-2}$ in the resulting two components. The sum of these values is in excellent agreement with the total column density along the line of sight derived from direct $N({\rm H}_2)$ and $N({\rm H})$ measurements (see Table~\ref{table1}). Furthermore, the former value of $N({\rm H}_{\rm tot})$ agrees within the errors with the rescaled $N({\rm H}_{\rm tot}^{\rm map})$ derived from the map for the first clump in Figure~\ref{f_app1_3}, while the latter value corresponds to the sum of those for the second and third clumps.

Note that while the K~\textsc{i} absorption and density maps show two separate components for this sight line, the analysis of H$_3^+$ in \citet{Albertsson2014} uses only one component.

\subsubsection{HD~73882}
\label{73882}

\begin{figure}
\begin{center}
	\includegraphics[width=\columnwidth]{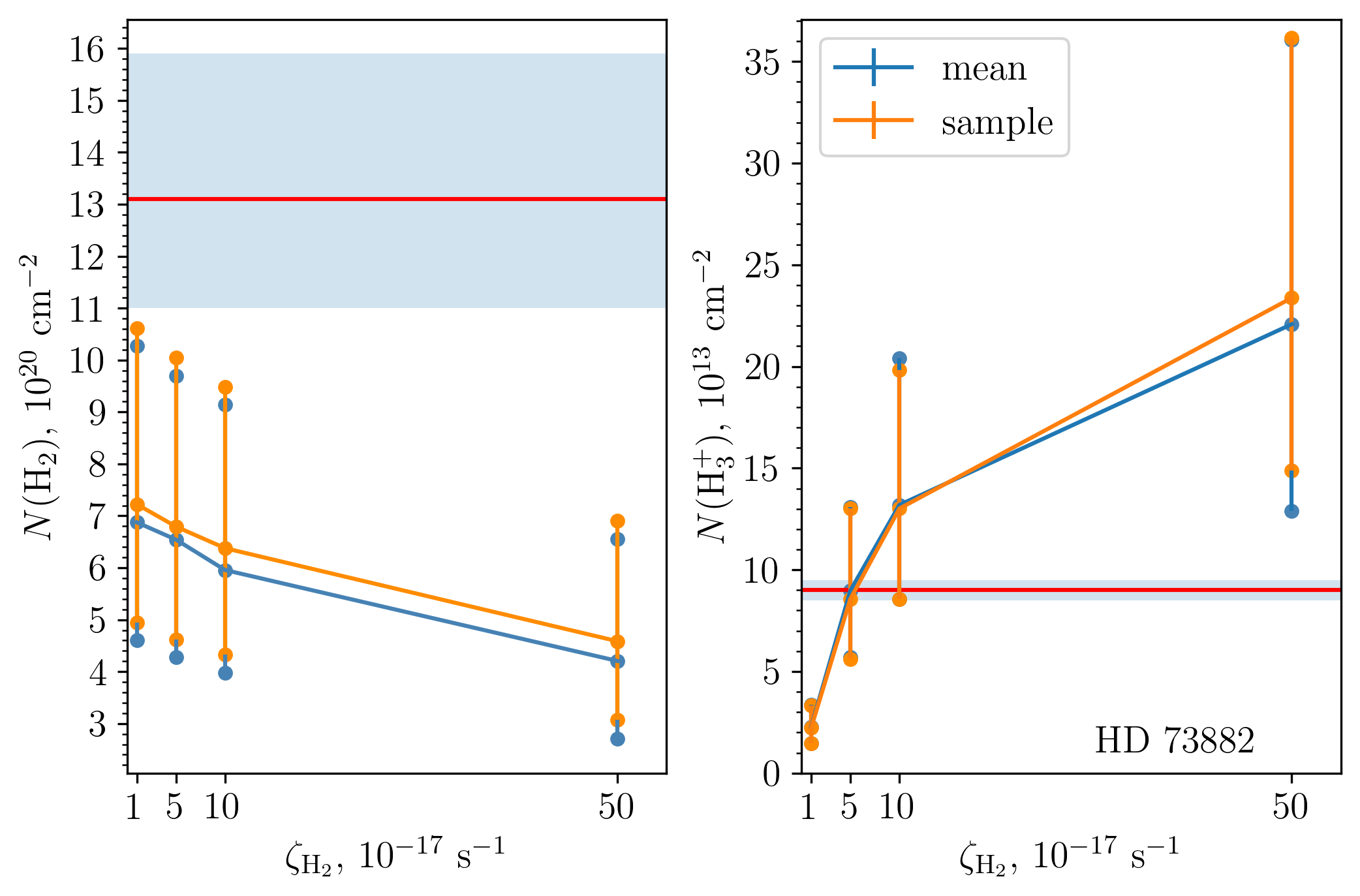}
    \caption{Same as in Figure~\ref{f4_1} for HD~73882.} 
    \label{f4_4}
\end{center}
\end{figure}

\begin{figure}
\begin{center}
	\includegraphics[width=\columnwidth]{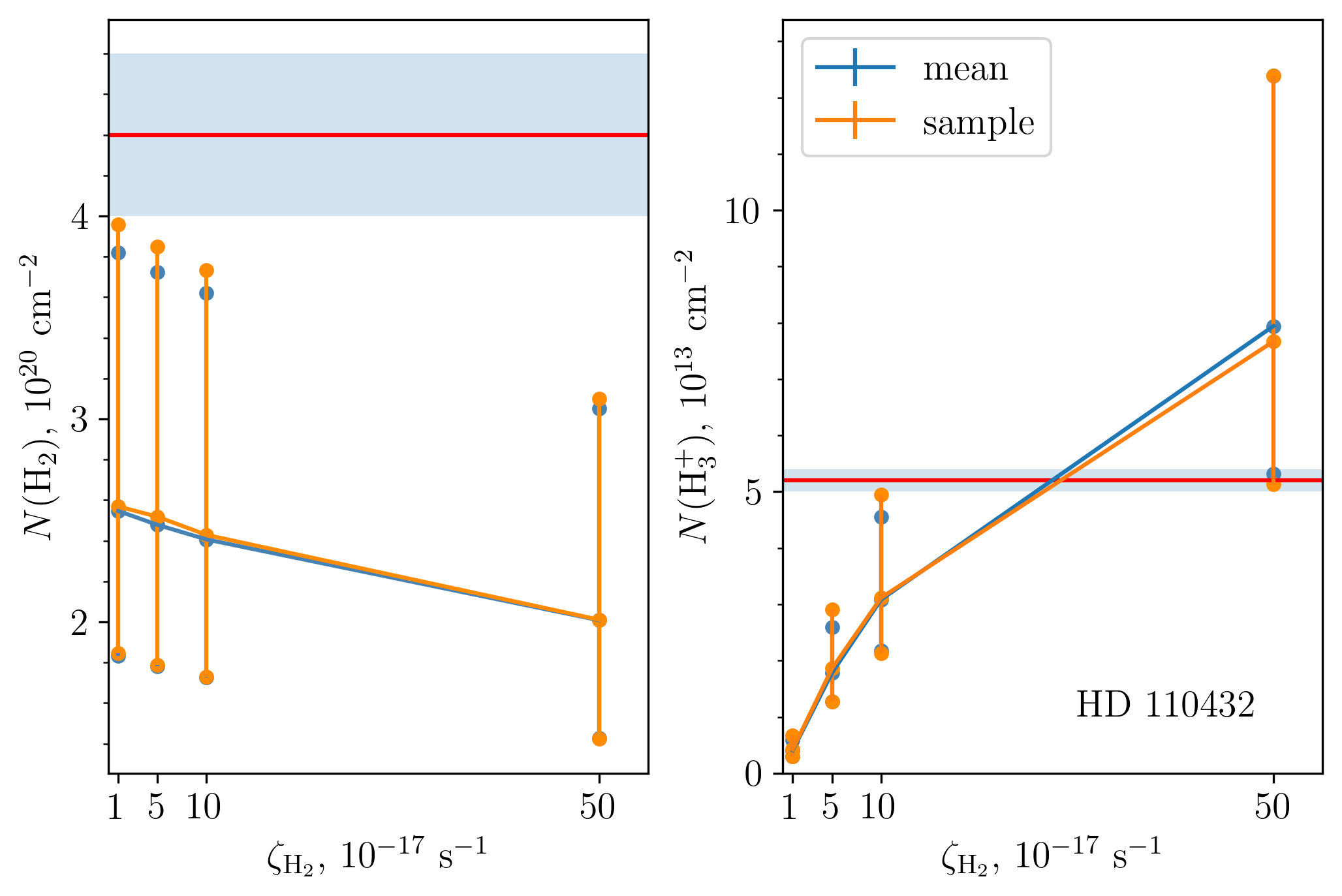}
    \caption{Same as in Figure~\ref{f4_1} for HD~110432.} 
    \label{f4_5}
\end{center}
\end{figure}

The distance to HD~73882 is highly uncertain. Parallax measurements disagree at a level of many $\sigma$: 346.5 $\pm$ 64 pc (Gaia DR2), 744 $\pm$ 32 pc (Gaia DR3), 461 $\pm$ 333 pc (Hipparcos). At the same time, recent photometric measurements based on the UV spectroscopy \citep{Shull2019} recommend a distance of 1000~pc. 

Placing the target star at the distance recommended in the latest Gaia release leads to an unrealistically low value of $N({\rm H}_{\rm tot}^{\rm map})\approx 3.4\times10^{20}$~cm$^{-2}$, whereas the distance recommended by \citet{Shull2019} yields $N({\rm H}_{\rm tot}^{\rm map})$ which is only 30\% lower than the measured value $N({\rm H}_{\rm tot})$ (see Table~\ref{table1}). Then, changing the distance from 1000 pc to 1050 pc has practically no impact on $N({\rm H}_{\rm tot}^{\rm map})$, and the resulting computed $N({\rm H}_{2})$ decreases by about 6\%. On the other hand, for a similar reduction of the distance, from 1000 pc to to 940 pc (where the local minimum between two density peaks is located, see Figure~\ref{f_app1_4}) the resulting computed $N({\rm H}_{2})$ increases by more than 30\%. This occurs due to a sharp increase of the ratio $N({\rm H}_{\rm tot})/N({\rm H}_{\rm tot}^{\rm map})$, from $\approx1.3$ to $\approx2.0$, and therefore of the rescaled gas density in the simulations (see Section~\ref{3D_PDR}). Given that this ratio typically varies between 1.1 and 1.3 for the other sight lines, we choose the recommended distance of 1000~pc for the representative simulations in Figure~\ref{f4_4}. In this case the star (with $T_{\rm eff}\approx 32,000$~K) is placed fairly close to the second density peak, which somewhat reduces contribution of the peak to the computed $N({\rm H}_{2})$.

\subsubsection{HD~110432}

An association of about 40 OB stars at a 90~pc distance from the center of the simulation domain for Figure~\ref{f4_5} increases the total FUV field to $G_{\rm D}\approx5.0$ at the closest point of the domain boundary, and to $G_{\rm D}\approx1.0$ at the center; the far-field component is about 0.6. 

\subsubsection{HD~154368}
\label{154368}

\begin{figure}
\begin{center}
	\includegraphics[width=\columnwidth]{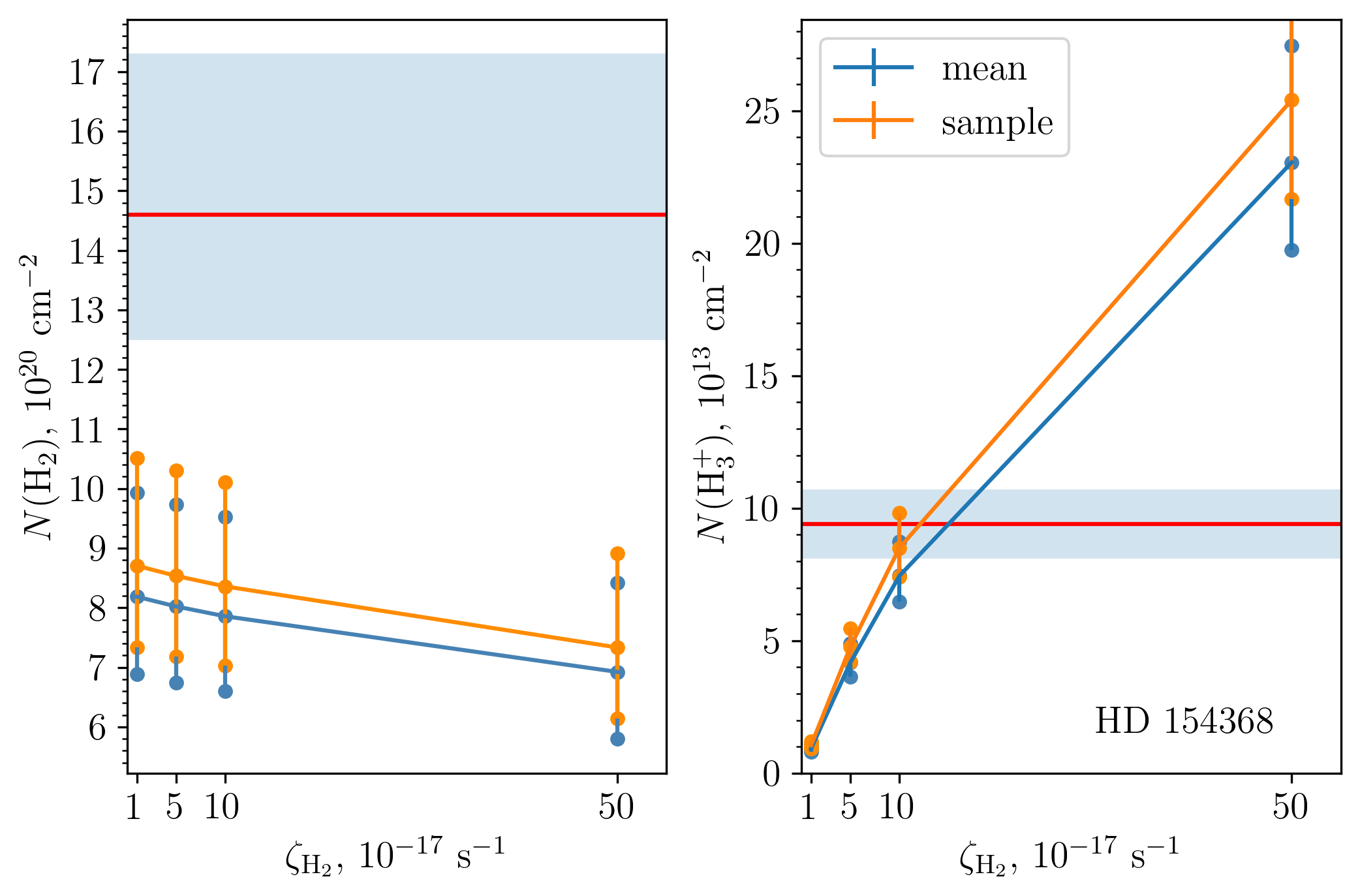}
    \caption{Same as in Figure~\ref{f4_1} for HD~154368.} 
    \label{f4_6}
\end{center}
\end{figure}

An association of about 50 OB stars at a 100~pc distance from the center of the simulation domain for Figure~\ref{f4_6} increases the total FUV field to $G_{\rm D}\approx1.9$ at the closest point of the domain boundary, and to $G_{\rm D}\approx1.5$ at the center; the far-field component is about 0.6. For one of the stars of the association, HD~158926, we adopted a distance of 112~pc based on the newer interferometric measurements \citep{Tango2006}, instead of the Hipparcos value of 175~pc.


\begin{figure}
\begin{center}
	\includegraphics[width=\columnwidth]{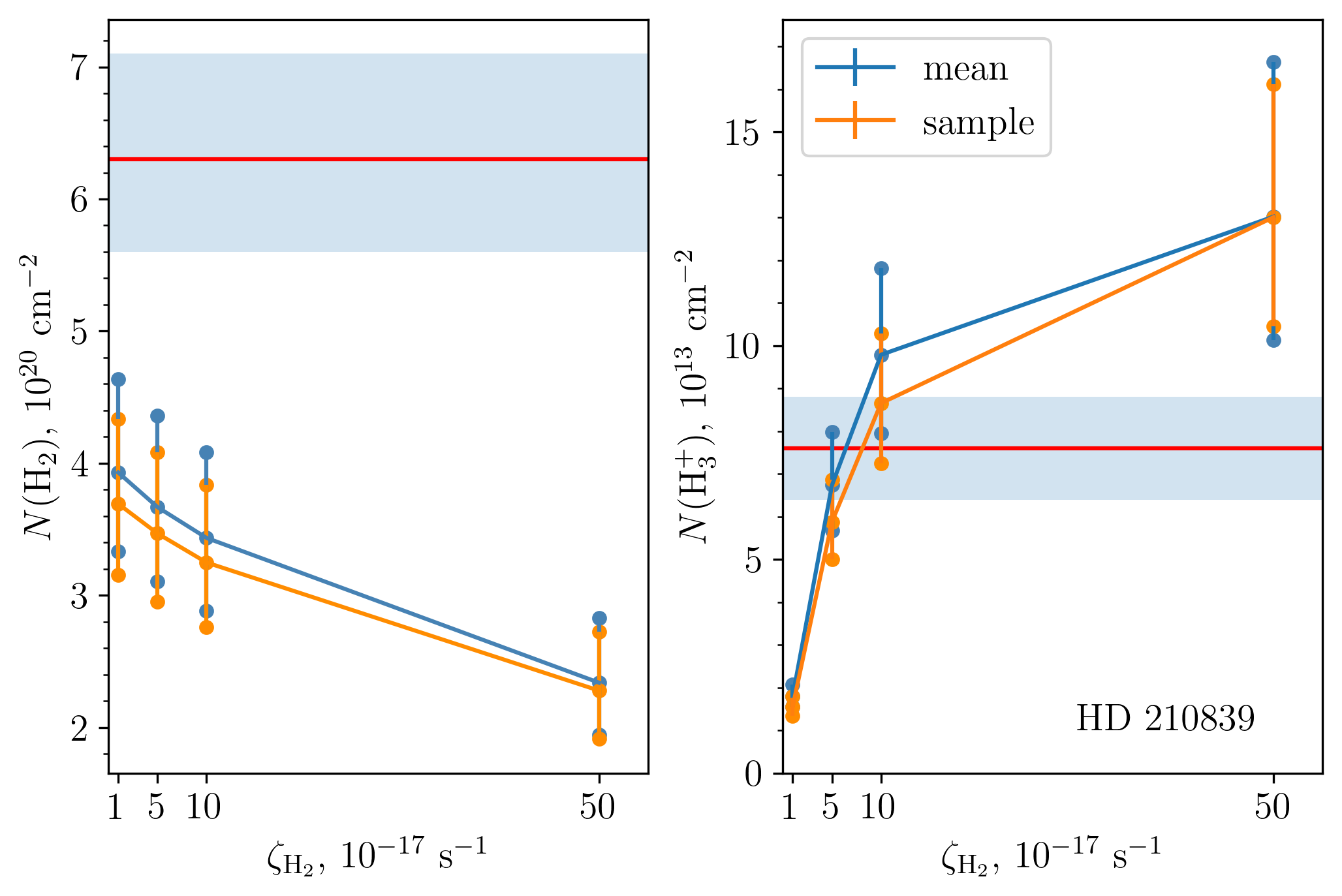}
    \caption{Same as in Figure~\ref{f4_1} for HD~210839.} 
    \label{f4_7}
\end{center}
\end{figure}

\subsection{H$_3^+$ non-detections}
\label{non-detections}

\subsubsection{HD~21856}
\label{21856}

\begin{figure}
\begin{center}
	\includegraphics[width=\columnwidth]{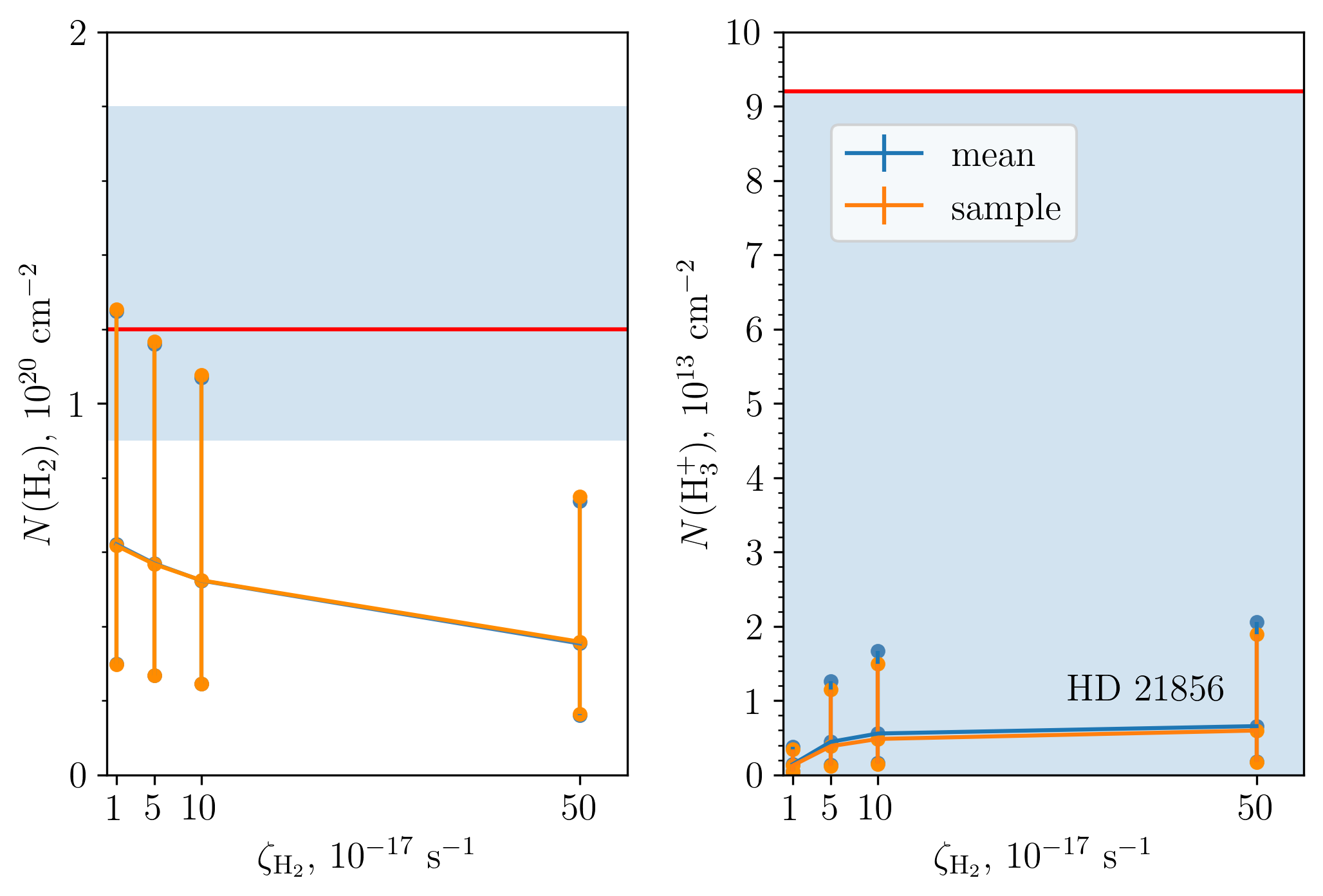}
    \caption{Line of sight to HD~21856 without H$_3^+$ detection. In the right panel, the upper edge of the shaded region shows the limit on $N({\rm H}_3^+)$ detection derived from observations (see Section~\ref{LoS}), otherwise the legend is the same as in Figure~\ref{f4_1}.} 
    \label{f4_8}
\end{center}
\end{figure}

Table~\ref{table1} shows that HD~21856 has the lowest measured value of $N({\rm H}_{2})$ among all selected sight lines. At the same time, the noise level in the measured H$_3^+$ signal is 3--8 times higher than that for the non-detection sight lines HD~22951, 148184, and 149757 \citep{Indriolo2012}. This leads to a relatively high upper limit on $N({\rm H}_3^+)$, and therefore no useful constraints can be provided for the upper $\zeta_{\rm H_2}$ limit, as evident from Figure~\ref{f4_8}. On the other hand, we note that the computed value of $N({\rm H}_2)$ is about 50\% of the measured value, which is within the typical range for all selected sight lines. The first clump along the sight line in Figure~\ref{f_app1_8} has a minor contribution compared to the second clump, e.g., $N({\rm H}_{2}) \approx 0.1\times$ and $0.5\times10^{20}$~cm$^{-2}$, respectively, for the mean $N({\rm H}_{\rm tot})$ and $\zeta_{\rm H_2}=5\times10^{-17}$~s$^{-1}$. 

It is noteworthy that, unlike the first clump (where the radiation is mostly determined by the far-field component with $G_{\rm D}\approx0.5$), the FUV field in the second clump is completely dominated by an extended association of OB stars located near the boundary of the simulation domain. This leads to $G_{\rm D}\approx7.5$ at the boundary and $G_{\rm D}\approx2.7$ at the domain center, while the far-field contribution is only $\approx0.4$. 

The dominant FUV source in this association is HD~23180 ($o$~Per). Fortunately, both $N({\rm H}_{2})$ and $N({\rm H})$ have been measured toward this star \citep{Savage1977, Bohlin1978}, which allows us to estimate the total gas column, $N({\rm H}_{\rm tot})=16.9^{+4.6}_{-3.1}$~cm$^{-2}$. This is much smaller than the value of $N({\rm H}_{\rm tot}^{\rm map})\approx42.6$~cm$^{-2}$ derived from the map for the distance of 331~pc, taken according to the Gaia DR3 release. However, decreasing the distance within the parallax uncertainty to 301~pc reduces the computed total column down to $N({\rm H}_{\rm tot}^{\rm map})\approx14.8$~cm$^{-2}$. This yields the ratio of $N({\rm H}_{\rm tot})/N({\rm H}_{\rm tot}^{\rm map})\approx1.14$, which is quite close to the values obtained for the other three targets in the proximity to HD~21856 (see Figure~\ref{f2}). Hence, we adopt the distance of 301~pc for our simulations. The resulting distance from HD~23180 to the center of the simulation domain for the second clump is $\approx23$~pc.

\subsubsection{HD~22951 (40~Per)}

\begin{figure}
\begin{center}
\includegraphics[width=\columnwidth]{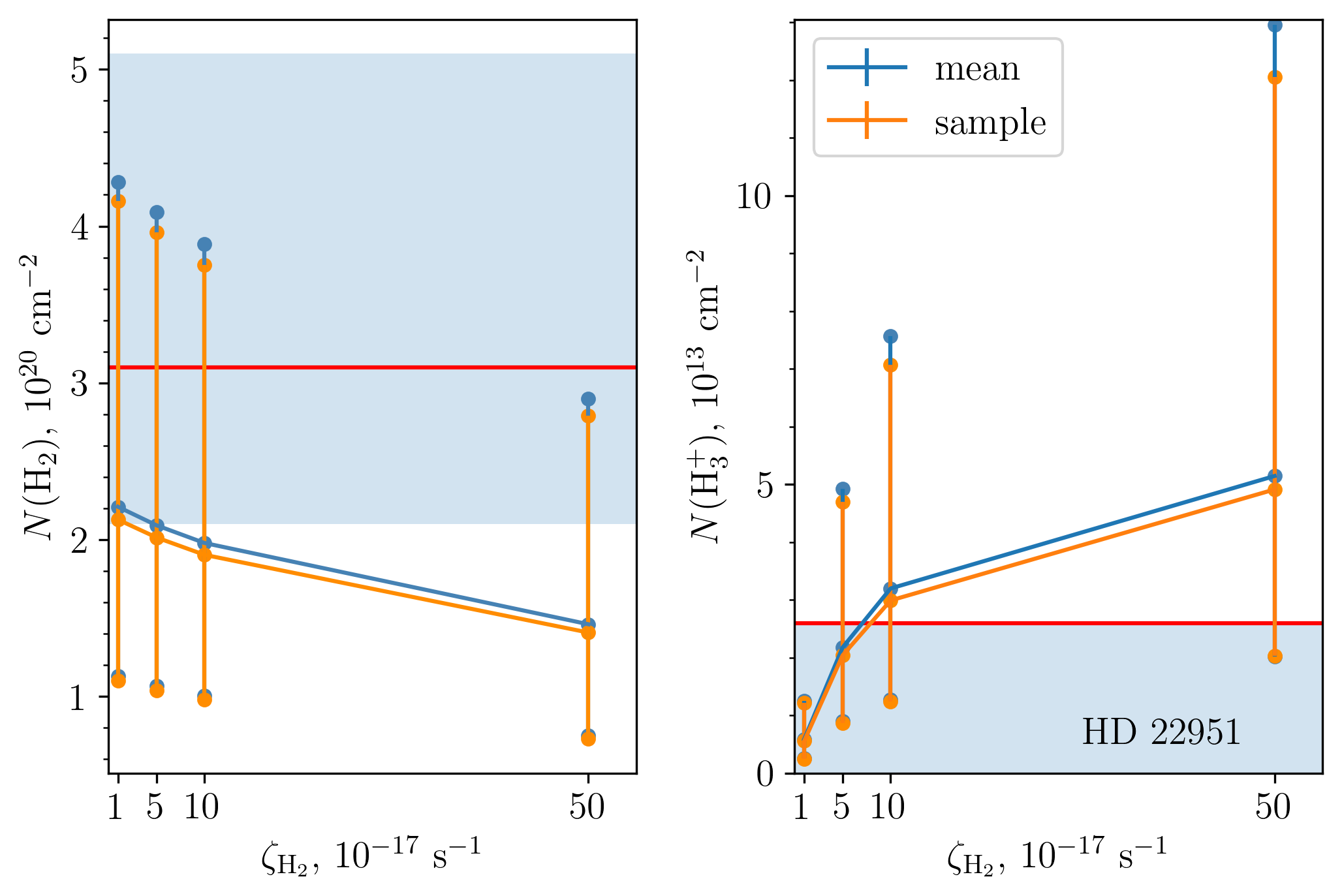}
    \caption{Same as in Figure~\ref{f4_8} for HD~22951.} 
    \label{f4_9}
\end{center}
\end{figure}

Measurements toward HD~22951 show a substantial amount of H$_2$ and a reasonably low noise level in H$_3^+$ signal. These data are compared with the simulation results in Figure~\ref{f4_9}, showing that HD~22951 is well suited for constraining $\zeta_{\rm H_2}$. The first of two clumps in Figure~\ref{f_app1_9} dominates the computed H$_2$ column, e.g., $N({\rm H}_{2}) \approx 1.5\times$ and $0.5\times10^{20}$~cm$^{-2}$, respectively, for the mean $N({\rm H}_{\rm tot})$ and $\zeta_{\rm H_2}=5\times 10^{-17}$~s$^{-1}$. 

The sight line to HD~22951 is very close to that toward HD~21856, as one can see from Figure~\ref{f2}. These sight lines probe the same two large-scale gas structures (that can be recognized by comparing Figures~\ref{f_app1_8} and \ref{f_app1_9}) with the clumps located at distances around 150 pc and 300 pc. The far-field FUV components in the first and second clumps of HD~22951 are therefore similar to those of HD~21856, and the field in the second clump is also dominated by the nearby OB stars discussed in Section~\ref{21856}. The FUV field in the center of the second-clump domain, $G_{\rm D}\approx4.3$, is substantially higher than that in HD~21856 because of a closer proximity ($\approx 13$~pc) to the dominant star HD~23180.

\subsubsection{HD~148184 ($\chi$~Oph)}

\begin{figure}
\begin{center}
	\includegraphics[width=\columnwidth]{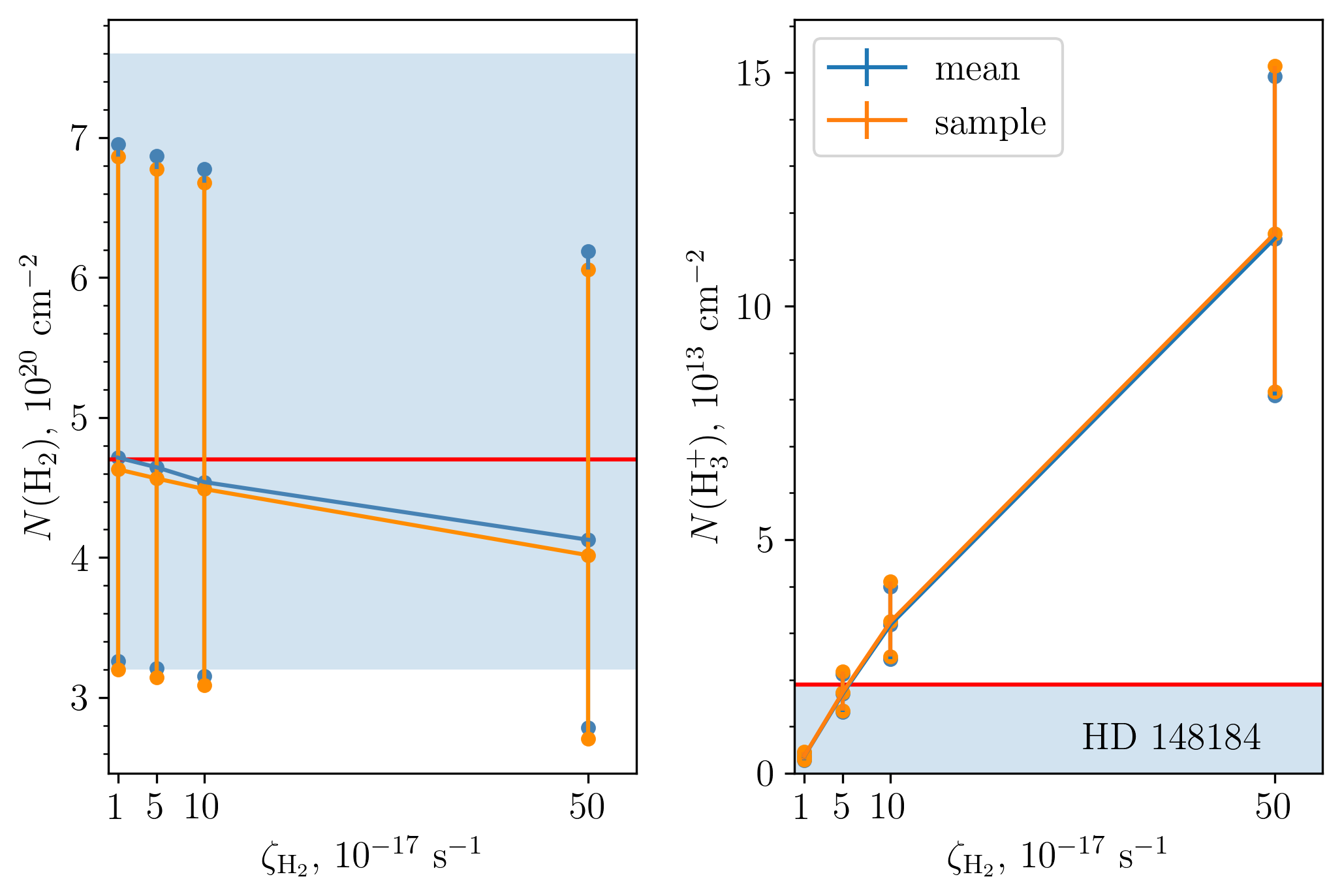}
    \caption{Same as in Figure~\ref{f4_8} for HD~148184.} 
    \label{f4_10}
\end{center}
\end{figure}

A compact association of OB stars, including four stars with $T_{\rm eff} > 27,000$~K, is located at a 20~pc distance from the center of the simulation domain for Figure~\ref{f4_10}. This increases the total FUV field to $G_{\rm D}\approx5.4$ at the domain boundary, and to $\approx4.1$ at the center. The far-field component in this clump is somewhat higher than on average, $G_{\rm D} \approx0.8$. 

\subsubsection{HD~149404}
\label{149404}

\begin{figure}
\begin{center}
	\includegraphics[width=\columnwidth]{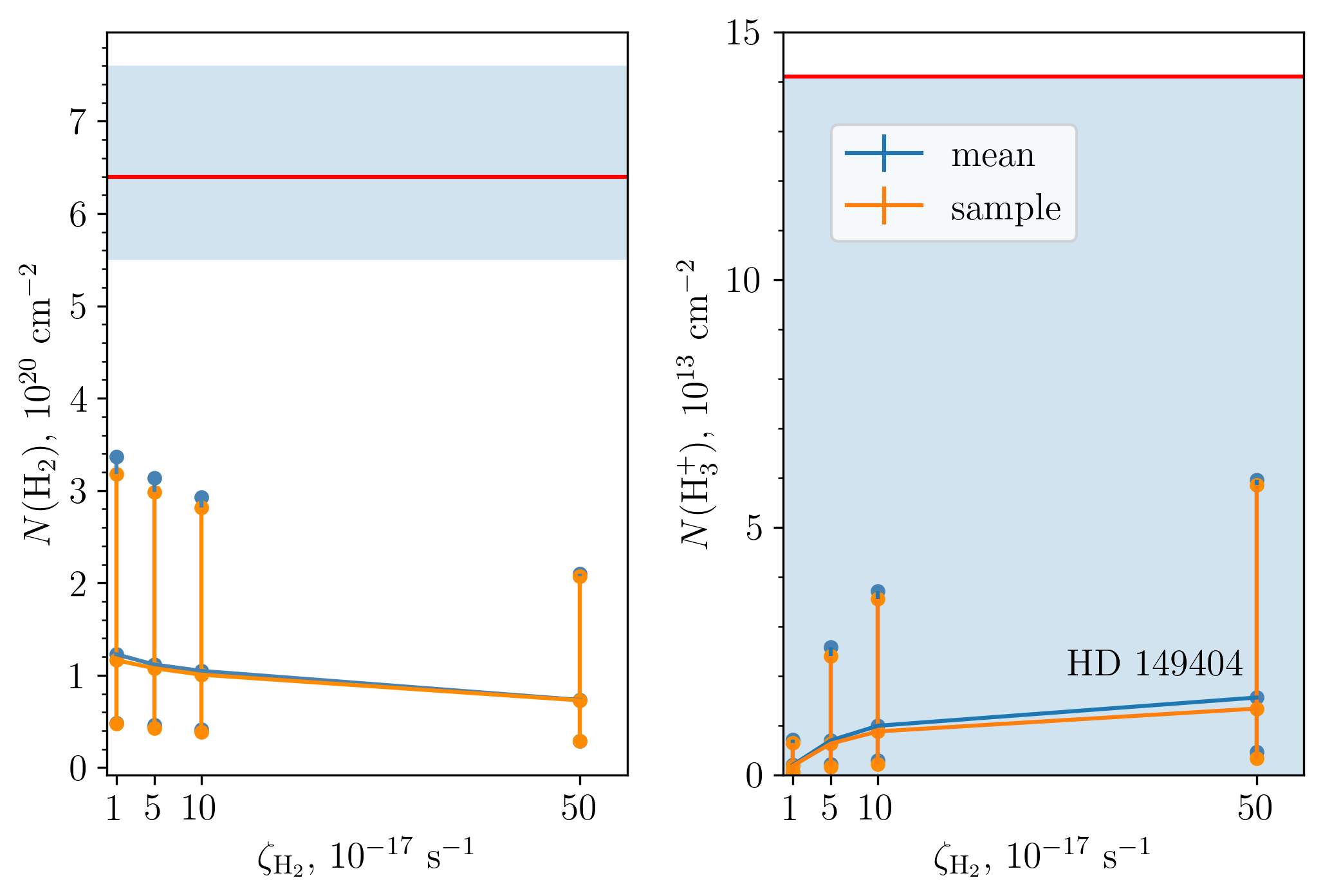}
    \caption{Same as in Figure~\ref{f4_8} for HD~149404.} 
    \label{f4_11}
\end{center}
\end{figure}

The $N({\rm H}_3^+)$ upper limit in the right panel of Figure~\ref{f4_11} is substantially higher than for the other selected non-detections, because the noise level in H$_3^+$ signal measured toward HD~149404 is the highest \citep{Indriolo2012}. Therefore, no useful constraints can be put for the upper $\zeta_{\rm H_2}$ limit in this case. Furthermore, the left panel shows that the computed $N({\rm H}_2)$ is particularly strongly underestimated: for the mean and the upper-limit values of $N({\rm H}_{\rm tot})$, it is smaller than the mean measured value by factors of $\approx6$ and $\approx2$, respectively. Such a sharp dependence on $N({\rm H}_{\rm tot})$ is associated with a low computed fraction of molecular hydrogen in all three clumps along the sight line in Figure~\ref{f_app1_11}, and therefore the computed $N({\rm H}_2)$ is highly sensitive to the gas density (see discussion in the beginning of Section~\ref{discussion}). This, in turn, indicates that the physical parameters assumed for HD~149404 (such as, e.g., gas distribution and/or FUV field in the clumps) may be inaccurate. Further evidence of inaccuracy is the fact that the computed H$_2$ column is completely dominated by the first two clumps, for all combinations of $N({\rm H}_{\rm tot})$ and $\zeta_{\rm H_2}$: e.g., $N({\rm H}_{2}) \approx 0.6\times$, $0.5\times$, and $0.03\times10^{20}$~cm$^{-2}$, respectively, for the mean $N({\rm H}_{\rm tot})$ and $\zeta_{\rm H_2}=5\times10^{-17}$~s$^{-1}$. This contradicts results of the analysis of multiple velocity components (presented below), where all three contributions are comparable with each other. 

In all three clumps, the FUV field is completely dominated by nearby OB stars. For the first clump, with the far-field component of $G_{\rm D}\approx0.4$, an extended association of OB stars located near the boundary of the simulation domain increases the total field to $G_{\rm D}\approx1.8$ at the domain boundary, and to $\approx1.0$ at the center. For the second clump, the far-field component is $G_{\rm D}\approx0.7$, and 13 nearby OB stars increase the total FUV field to $\approx1.2$ at the domain boundary, and to $\approx1.0$ at the center. Similarly, in the third clump with the far-field component of $G_{\rm D}\approx0.6$, a group of nine nearby OB stars increase the total FUV field to $\approx2.0$ at the domain boundary, and to $\approx1.0$ at the center. 

{\it Analysis of multiple velocity components.} Observations of the 4300~\AA\ line of CH toward this sight line were obtained with VLT/UVES, and the standard pipeline processed spectrum was retrieved from the ESO Science Archive Facility. These observations reveal three distinct absorption components at about $v_{\rm LSR}=-13$~km/s, $-2$~km/s, and $+6$~km/s. By fitting the absorption profile with the sum of three Gaussian functions we determine equivalent widths for each component, and compute CH column densities in the optically thin limit using the oscillator strength from \citet{Larsson1983}. The resulting values are $N({\rm CH})=6.5\times10^{12}$~cm$^{-2}$, $9.3\times10^{12}$~cm$^{-2}$, and $1.3\times10^{13}$~cm$^{-2}$, respectively, for the above three components. Using the empirically derived linear relation $N({\rm CH}) = 3.5^{+2.1}_{-1.4}\times10^{-8}\,N({\rm H}_2)$ from \citet{Sheffer2008}, we derive the respective contributions $N({\rm H}_2)=1.9^{+1.2}_{-0.7}\times10^{20}$~cm$^{-2}$, $2.7^{+1.7}_{-1.0}\times10^{20}$~cm$^{-2}$, and $3.6^{+2.4}_{-1.3}\times10^{20}$~cm$^{-2}$. Their sum agrees well within the errors with the value of $N({\rm H}_2)$ from direct measurements (see Table~\ref{table1}). 

\begin{figure}[h!]
\begin{center}
	\includegraphics[width=\columnwidth]{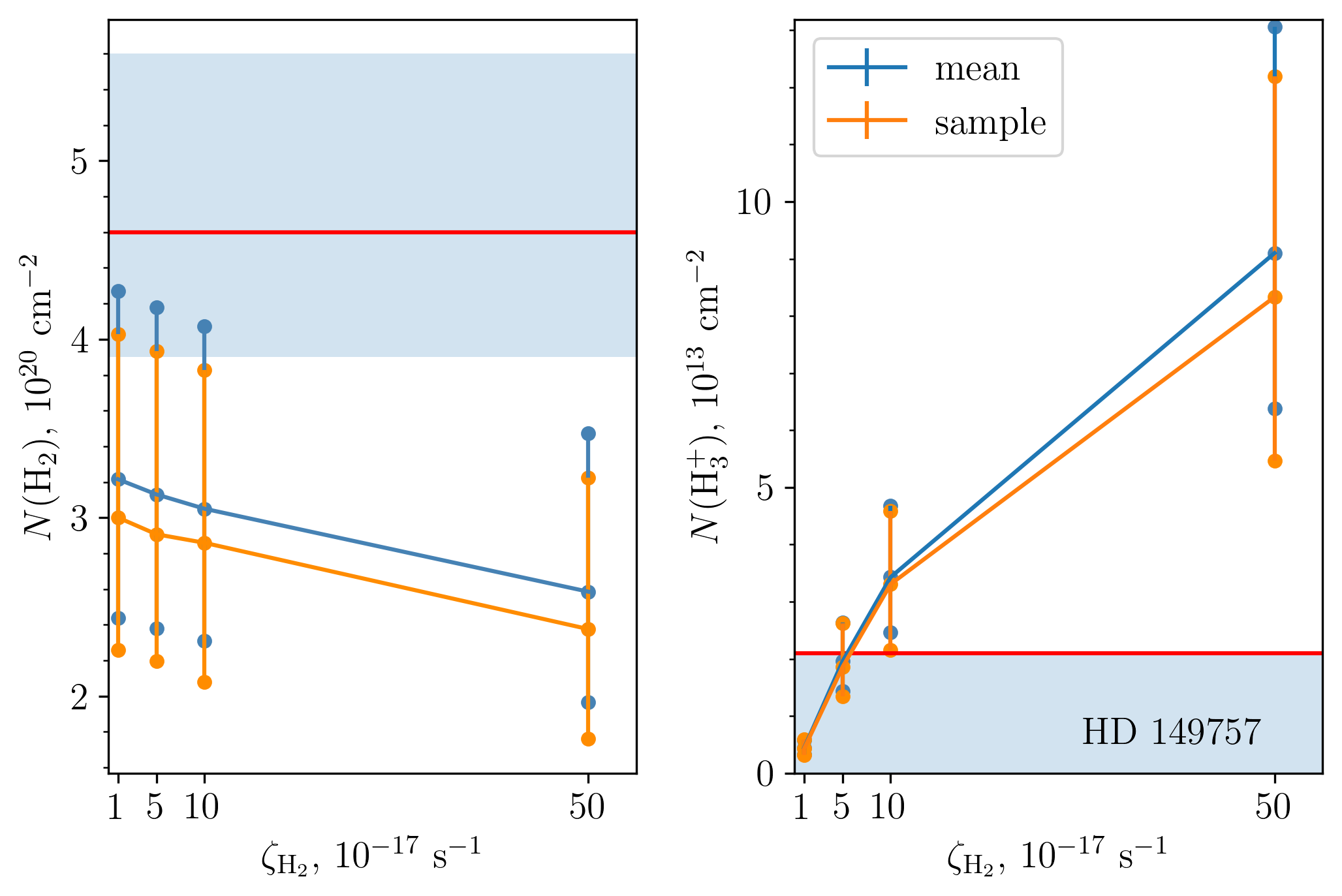}
    \caption{Same as in Figure~\ref{f4_8} for HD~149757.} 
    \label{f4_12}
\end{center}
\end{figure}

\subsubsection{HD~149757 ($\zeta$~Oph)}
\label{149757}

The sight line to HD~149757 probes the same large-scale gas structure as that toward HD~148184 (which can be seen from Figures~\ref{f_app1_10} and \ref{f_app1_12}), with the same far-field component of $G_{\rm D}\approx0.8$. The nearby field is similarly affected by the same OB association, located somewhat further away from the domain center for Figure~\ref{f4_12}, at $\approx45$~pc. The resulting FUV field varies between $G_{\rm D}\approx5.4$ at the domain boundary to $\approx1.5$ at the center.

\section{Discussion}
\label{discussion}

As expected, in the left panels of Figures~\ref{f4_1}--\ref{f4_12} we see a rather insignificant reduction of $N({\rm H}_2)$ at $\zeta_{\rm H_2}\lesssim 10^{-16}$~s$^{-1}$ \citep{Neufeld2017}. The right panels typically show an almost linear increase of $N({\rm H}_3^+)$ in this range of $\zeta_{\rm H_2}$. This behavior follows from the H$_3^+$ balance equation, 
\begin{equation}
\epsilon({\rm H}_3^+)\zeta_{\rm H_2} n_{\rm H_2} =kn_{{\rm H}_3^+}n_e,
\label{H3+_balance}
\end{equation}
where $\epsilon({\rm H}_3^+)$ is the fraction of CR ionization events followed by H$_3^+$ production (which tends to unity in the predominantly molecular gas), and $k$ is the total rate of dissociative H$_3^+$ recombination \citep{Neufeld2017}. For sufficiently small $\zeta_{\rm H_2}/n_{\rm tot}$, where both $n_{\rm H_2}$ and $n_e$ remain approximately constant (see Figure~\ref{f3}), we obtain $N({\rm H}_3^+)\propto \zeta_{\rm H_2}$. In lower-density clumps (such as HD~41117), $N({\rm H}_3^+)$ starts saturating or even decreasing at $\zeta_{\rm H_2}\gtrsim 10^{-16}$~s$^{-1}$, since $n_{\rm H_2}$ decreases and $n_e$ increases with $\zeta_{\rm H_2}$ in this regime.

Analysis of the results presented in Section~\ref{CRIR} shows that the computed $N({\rm H}_2)$ exhibits a fairly sharp increase with $N({\rm H}_{\rm tot})$ for lower-density clumps. Comparing the range of $N({\rm H}_{\rm tot})$ used for each sight line (Table~\ref{table1}) with the respective computed range of $N({\rm H}_2)$ (Figures~\ref{f4_1}--\ref{f4_12}), we find this trend to be particularly pronounced for HD~41117 (Figure~\ref{f4_3}), HD~21856 (Figure~\ref{f4_8}), and also HD~22951 (Figure~\ref{f4_9}), where the slope $d\log N({\rm H}_2)/d\log N({\rm H}_{\rm tot})$ is about or exceeds 3. Such behavior can be explained with the H$_2$ balance equation \citep{DraineBook2011}, 
\begin{equation}
G_{\rm D}\zeta_{\rm diss}^0 f_{\rm dust}f_{\rm sh}n_{{\rm H}_2}= Rn_{\rm tot}n_{\rm H}, 
\label{H2_balance}
\end{equation}
where $R$ is the H$_2$ formation rate, $f_{\rm dust}[N({\rm H}_{\rm tot})]$ describes dust extinction, and $f_{\rm sh}[N({\rm H}_2)]$ is the H$_2$ self-shielding function. The latter scales approximately as $f_{\rm sh}\propto 1/\sqrt{N({\rm H}_2)}$ for $N({\rm H}_2)$ between $\sim10^{16}$~cm$^{-2}$ and $\sim3\times 10^{20}$~cm$^{-2}$ \citep{Draine_self_shielding}. Assuming predominantly atomic gas (which is a reasonable approximation for the low-density clumps) and neglecting dust extinction, we obtain the scaling dependence $N({\rm H}_2)\propto (Rn_{\rm tot}^2/G_{\rm D})^2$, showing that the slope may be as high as 4.  

At the same time we note that in sufficiently dense clumps, such as HD~24534, 73882, 110432, 154368 (detections), or HD~148184 (non-detection), H$_2$ fraction is very close to 1/2 near the peaks. Thus, unlike the lower-density clumps, $N({\rm H}_2)$ in the dense clumps is approximately proportional to $N({\rm H}_{\rm tot})$ -- and therefore it is practically insensitive to (a factor of two) variations in the values of $R$ and/or $G_{\rm D}$. 

\subsection{Underestimate of $N({\rm H}_2)$}
\label{underestimate}

Figures~\ref{f4_1}--\ref{f4_12} generally show a moderate systematic underestimate of H$_2$ amount in our simulations, overall by about 30--50\% for the mean values of $N({\rm H}_2)$. For two targets, HD~24398 and HD~148184, the computed and measured mean values of $N({\rm H}_2)$ nearly coincide. On the other hand, for HD~149404 the mean computed $N({\rm H}_2)$ is a factor of $\approx6$ below the mean measured value. 

We stress that the overall underestimate in $N({\rm H}_2)$ is observed after re-scaling the map density distribution (to match the measured $N({\rm H}_{\rm tot})$, as discussed in Section~\ref{3D_PDR}), and has no noticeable correlation with the peak clump density or the computed H$_2$ fraction. Bearing in mind the above discussion, it is therefore unlikely that this systematic deviation is related to uncertainties in the model parameters (such as, e.g., possibly underestimated $R$ and/or overestimated $G_{\rm D}$). 

In statistical terms, the overall H$_2$ underestimate is moderate as well. Consider our results plotted in Figures~\ref{f4_1}--\ref{f4_12} for the upper limits of $N({\rm H}_{\rm tot})$ confidence intervals: we see that the corresponding computed values of $N({\rm H}_2)$ occur to be within (or very close to the lower limit of) the respective confidence intervals for measurements. The three exceptions are HD~154368 and HD~210839, where the remaining underestimate of $N({\rm H}_2)$ is in a range of 2--3 $\sigma$, and HD~149404 with 3--4 $\sigma$.

A possible reason of the moderate H$_2$ underprediction in our simulations may be related to the fact that we compute {\it equilibrium} H$_2$ abundance in {\it quiescent} clumps. However, H$_2$ formation is a very slow process whose timescale is set by the local photo-dissociation rate. Furthermore, \citet{Bialy2019} showed that the presence of strong compressive turbulence ($\mathcal{M}_s=4.5$) in the diffuse ISM may additionally affect H$_2$ abundance, leading to systematically higher values compared to equilibrium (up to a factor of two for the median values, see the upper panels in their Figure~4). The origin of this effect is the non-linearity of H$_2$ self-shielding in local density perturbations produced by the turbulence: while the total column exhibits minor variations, the non-linearity results in enhanced H$_2$ production near the local density peaks (assuming predominantly atomic gas). Hence, one can expect that this effect operating within sharp unresolved (sub-pc) peaks which may be present in the clumps due strong turbulence would lead to a certain increase in $N({\rm H}_2)$ compared to the computed values.
 
The potential impact of systematically underestimated $N({\rm H}_2)$ on the optimum $\zeta_{\rm H_2}$ is discussed in Section~\ref{correction}.

\subsection{Optimum values of $N({\rm H}_2)$, $N({\rm H}_{\rm tot})$, and $\zeta_{\rm H_2}$}
\label{optimum}

\begin{figure*}
\begin{center}
	\includegraphics[width=0.268\textwidth]{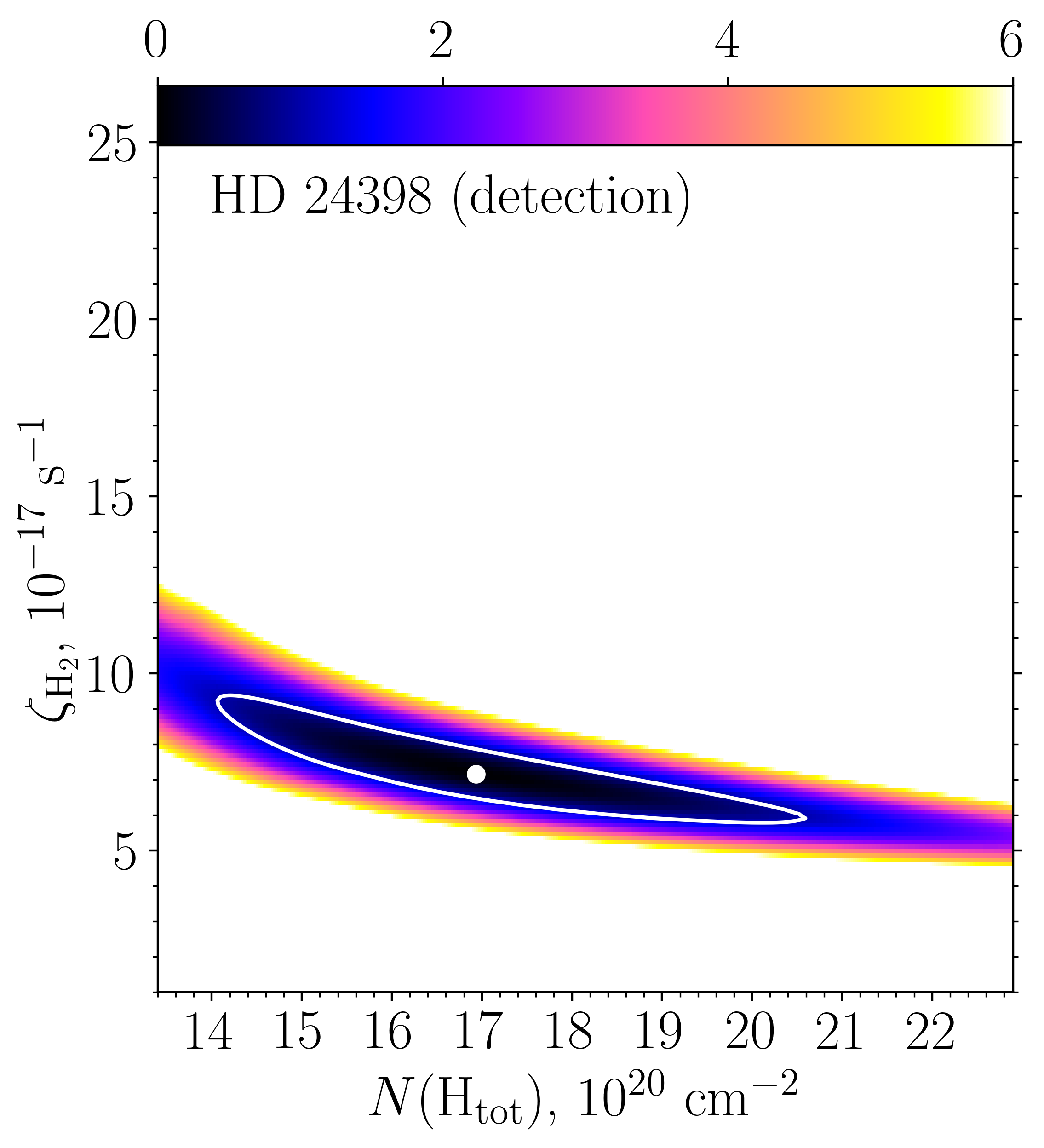}
        \includegraphics[width=0.235\textwidth]{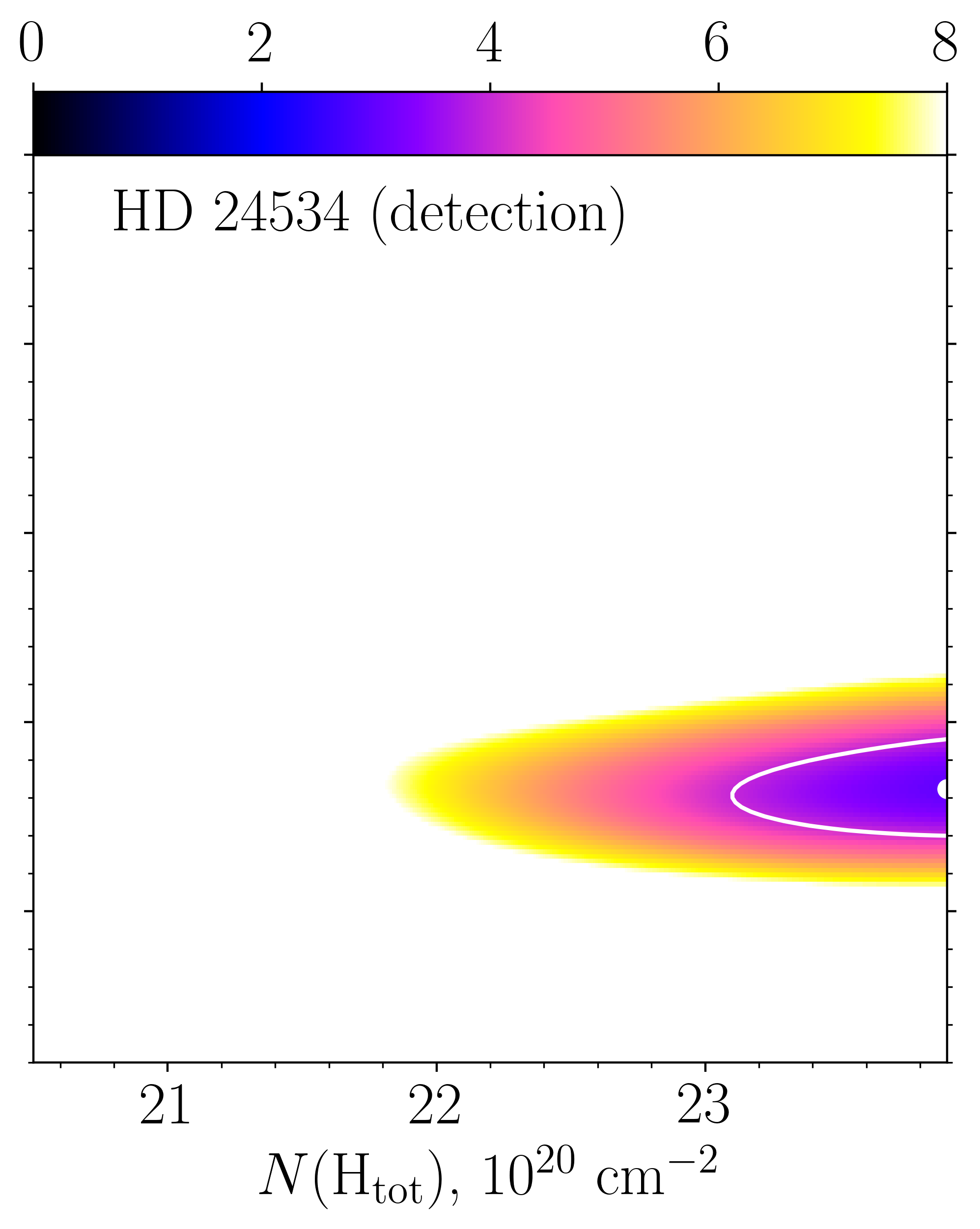}
        \includegraphics[width=0.241\textwidth]{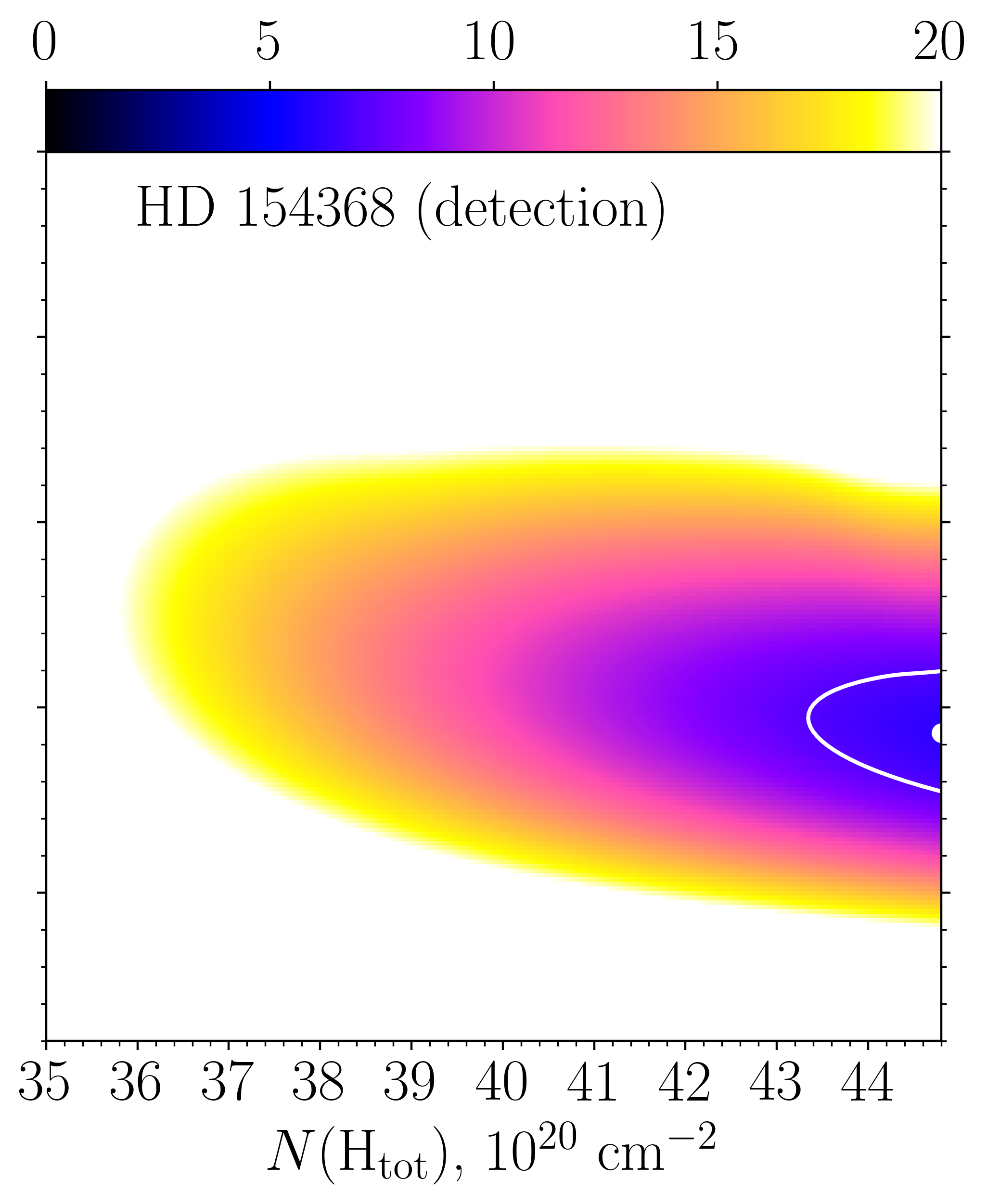}
        \includegraphics[width=0.235\textwidth]{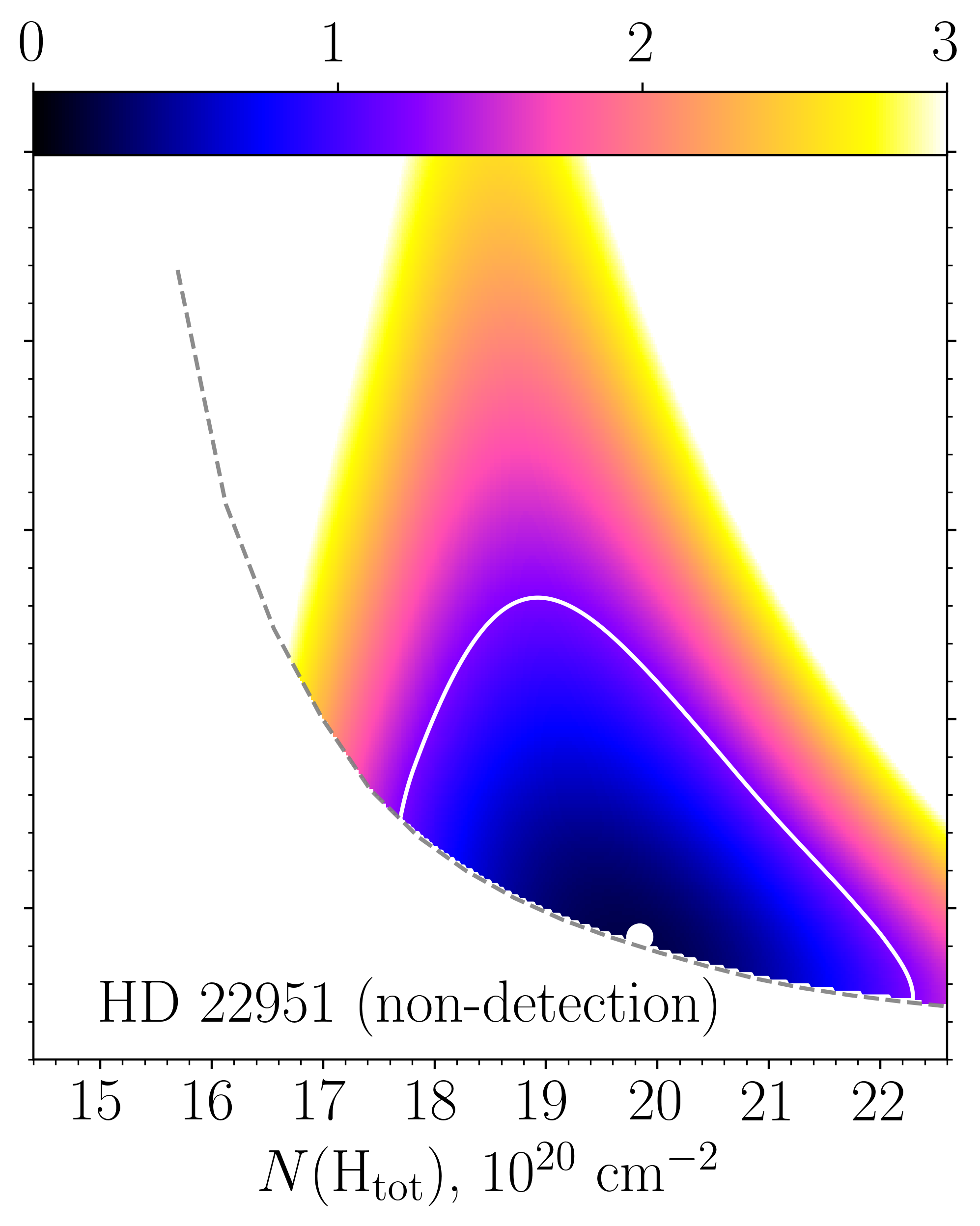}
    \caption{Reduced $\chi^2$ error (indicated in the color bars) in the plane of $\zeta_{\rm H_2}$ and $N({\rm H}_{\rm tot})$, plotted for three detection cases HD~24398, HD~24534, and HD~154368, as well as for one non-detection HD~22951. All plots are computed with Equation~(\ref{chi2}) by using the bicubic interpolation of the simulation results in Tables~\ref{table3} and \ref{table4}. In each panel, the white bullet indicates the minimum, and the white contour line traces where $\chi^2$ exceeds the minimum by unity. For the non-detection, the dashed curve is the locus of points corresponding to the $N({\rm H}_3^+)$ upper limit in the right panel of Figure~\ref{f4_9}. The range of $N({\rm H}_{\rm tot})$ for each sight line corresponds to the respective 68\% confidence interval (see Table~\ref{table1}).}
    \label{f5}
\end{center}
\end{figure*}

The optimum values for each sight line with H$_3^+$ detection can be estimated by minimizing a function analogous to the reduced $\chi^2$ error,
\begin{eqnarray}
\chi^2=\sum_{k={\rm H}_{\rm tot}\,,\,{\rm H}_2}\frac{[\log N(k)-\langle\log N(k)\rangle]^2}{\sigma_{\log,k}^2} \hspace{2cm}\label{chi2}\\
+ \frac{[N({\rm H}_3^+)-\langle N({\rm H}_3^+)\rangle]^2}{\sigma_{{\rm H}_3^+}^2}\,.
\nonumber
\end{eqnarray}
Here we take into account that $N({\rm H}_{\rm tot})$ and $N({\rm H}_2)$ are assumed to have log-normal distributions, with the mean log values $\langle\log N(k)\rangle$ and the standard deviations $\sigma_{\log,k}$ derived from observations [see Note (c) under Table~\ref{table1}]. The reasoning for Equation~(\ref{chi2}) is as follows. We aim to minimize errors for three parameters, and have two variables to tune, $\zeta_{\rm H_2}$ and $N({\rm H}_{\rm tot})$. Hence, the number of degrees of freedom is $3-2$ and therefore we use a pre-factor of one for the sum. A contour in the plane of $\zeta_{\rm H_2}$ and $N({\rm H}_{\rm tot})$ where $\chi^2$ exceeds the minimum by unity can be considered as ``$1\sigma$'' limits. 

For non-detections, the upper limit for $\zeta_{\rm H_2}$ as well as the corresponding optimum $N({\rm H}_2)$ and $N({\rm H}_{\rm tot})$ are estimated by setting to zero the last term in Equation~(\ref{chi2}) if $N({\rm H}_3^+)<\langle N({\rm H}_3^+)\rangle$, where $\langle N({\rm H}_3^+)\rangle$ denotes the upper limit for $N({\rm H}_3^+)$ (see  Table~\ref{table1}).

The optimum values of $N({\rm H}_2)$, $N({\rm H}_{\rm tot})$, and $\zeta_{\rm H_2}$ (or the upper $\zeta_{\rm H_2}$ limits for non-detections) are listed in Table~\ref{table2} for each sight line. To compute these results, we apply the bicubic interpolation of $N({\rm H}_3^+)$ and $N({\rm H}_2)$ between the simulated values given in Tables~\ref{table3} and \ref{table4} in Appendix~\ref{app3}.

\begin{table}
  \caption{Optimum $N({\rm H}_2)$ and $N({\rm H}_{\rm tot})$ with the optimum/upper-limit value of $\zeta_{\rm H_2}$, derived for each sight line from simulations (Section~\ref{optimum}). The numbers in parentheses are the optimum $\zeta_{\rm H_2}$ corrected for $N({\rm H}_2)$ underestimate (Section~\ref{correction})$^{\rm a}$}
  \vspace{-.6cm}
\begin{center}
  \begin{tabular}{ |l | c | c | c |}
        \hline
        \rule[-1.5ex]{0pt}{4.5ex}
  HD~number & $N({\rm H}_2)$ & $N({\rm H}_{\rm tot})$ & $\zeta_{\rm H_2}$ \\
     & ($10^{20}$~cm$^{-2}$) & ($10^{20}$~cm$^{-2}$) & ($10^{-17}$~s$^{-1}$) \\
        \hline\hline
        \rule[-.5ex]{0pt}{3.ex}
  24398  & 5.4  & 16.9 &  $7.1^{+0.7}_{-0.7}$ (7.6)\\
        \hline
        \rule[-.5ex]{0pt}{3.ex}
  24534  & 7.3 & 23.9 &  $8.2^{+1.3}_{-1.2}$ (7.0)\\
        \hline
        \rule[-.5ex]{0pt}{3.ex}
  41117  & 4.9 & 42.6 & $2.7^{+1.2}_{-1.2}$ (3.8) \\
        \hline
        \rule[-.5ex]{0pt}{3.ex}
  73882  & 10.3 & 47.9 & $3.1^{+0.3}_{-0.1}$ (2.7)\\
        \hline
        \rule[-.5ex]{0pt}{3.ex}
  110432 & 3.7 & 19.9 &  $10.6^{+0.6}_{-0.6}$ (9.0)\\
        \hline
        \rule[-.5ex]{0pt}{3.ex}
  154368 & 10.1 & 44.8 &  $9.3^{+1.7}_{-1.6}$ (6.0)\\
        \hline
        \rule[-.5ex]{0pt}{3.ex}
  210839 & 4.1 & 32.9 & $5.1^{+1.3}_{-1.1}$ (3.8)\\
        \hline\hline
        \rule[-.5ex]{0pt}{3.ex}
  21856  & -- & -- & --\\
        \hline
        \rule[-.5ex]{0pt}{3.ex}
  22951  & 2.8 & 19.8 & $< 11.4$\\
        \hline
        \rule[-.5ex]{0pt}{3.ex}
  148184 & 4.7 & 24.0 & $< 11.6$\\
        \hline
        \rule[-.5ex]{0pt}{3.ex}
  149404 &  -- &  -- & --  \\
        \hline
        \rule[-.5ex]{0pt}{3.ex}
  149757 & 4.0 & 16.4 & $< 8.2$\\
        \hline
  \end{tabular}
\end{center}
\vspace{0cm}
$^{\rm a}\,$The noise level in $N({\rm H}_3^+)$ measurements was too high for HD~21856 and HD~149404, and therefore no useful constraints can be put in these cases (see Sections~\ref{21856} and \ref{149404}).
\label{table2}
\end{table}

In Figure~\ref{f5} we present examples of the reduced $\chi^2$ error plotted in the plane of $\zeta_{\rm H_2}$ and $N({\rm H}_{\rm tot})$. These results, illustrating typical contours of constant $\chi^2$, allow us to estimate the resulting uncertainty in $\zeta_{\rm H_2}$. 

Let us first analyze the three detection cases plotted in Figure~\ref{f5}. By comparing Tables~\ref{table1} and \ref{table2} we see that the optimum $N({\rm H}_2)$ is quite close to the measured mean value for HD~24398, while for HD~24534 and HD~154368 it is slightly below the lower limit of the measured confidence interval. This is a simple reflection of the fact that $N({\rm H}_2)$ computed for the mean $N({\rm H}_{\rm tot})$ is about the measured value for HD~24398 (Figure~\ref{f4_1}), while for HD~24534 (Figure~\ref{f4_2}) and HD~154368 (Figure~\ref{f4_6}) the computed values are smaller. For this reason, the $\chi^2$ minimum for HD~24398 is located well within the $N({\rm H}_{\rm tot})$ confidence interval, whereas for the other two detection targets it is pushed to the upper $N({\rm H}_{\rm tot})$ limit. These latter two cases illustrate the overall trend of overestimating the optimum $N({\rm H}_{\rm tot})$, resulting from the systematic moderate underestimate of the computed $N({\rm H}_2)$. In Section~\ref{correction} we present an approach to compensate for this effect when estimating the optimum $\zeta_{\rm H_2}$.

We note that $\zeta_{\rm H_2}$ and $N({\rm H}_{\rm tot})$ show anticorrelation for the detection cases HD~24398 and HD~154368, because the $N({\rm H}_3^+)$ versus $\zeta_{\rm H_2}$ dependence for these targets is noticeably sensitive to variations in $N({\rm H}_{\rm tot})$ (see the right panels in Figure~\ref{f4_1} and \ref{f4_6}). The sensitivity is much lower for HD~24534 (see the right panel in Figure~\ref{f4_2}), for which also the $N({\rm H}_{\rm tot})$ confidence interval is relatively narrow. The upper and lower uncertainty bounds in $\zeta_{\rm H_2}$ for each sight line, defined by the respective variations between the computed minimum and the ``$+1\sigma$'' contour of $\chi^2$ (and listed in Table~\ref{table2} next to the optimum), are largely set by uncertainty in the measured $N({\rm H}_3^+)$.  

Non-detections are illustrated in the last panel of Figure~\ref{f5} with HD~22951, in which case the computed optimum $N({\rm H}_2)$ and $N({\rm H}_{\rm tot})$ are both within the respective confidence intervals (compare Tables~\ref{table1} and \ref{table2}). The dashed curve shows the locus of points corresponding to the $N({\rm H}_3^+)$ upper limit in the right panel of Figure~\ref{f4_9}. As the region below cannot be used to explicitly constrain ionization, the $\chi^2$ minimum lies on that curve. In analogy to detections, now we define the upper limit for $\zeta_{\rm H_2}$ as the value at the ``$+1\sigma$'' contour above the $\chi^2$ minimum. The upper limits are listed in Table~\ref{table2} for all non-detections except for HD~21856 and HD~149404: as discussed in Sections~\ref{21856} and \ref{149404}, the noise level in $N({\rm H}_3^+)$ measurements was too high for these sight lines (see also Figures~\ref{f4_8} and \ref{f4_11}), and therefore no useful constraints can be put on $\zeta_{\rm H_2}$.

\subsection{Correction for optimum $\zeta_{\rm H_2}$}
\label{correction}

\begin{figure*}[t!]
\begin{center}
	\includegraphics[width=0.97\columnwidth]{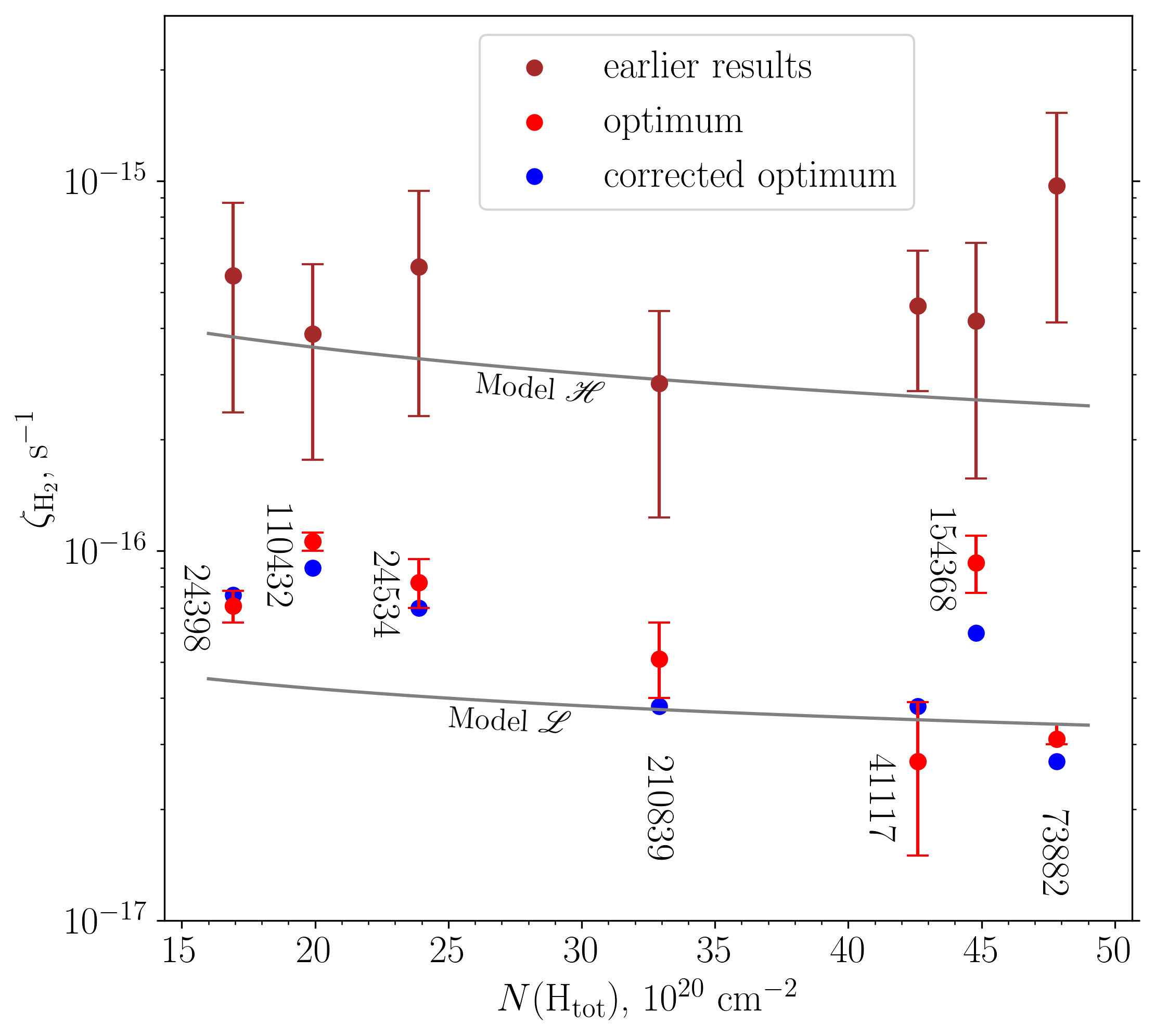}
    \includegraphics[width=1.110\columnwidth]{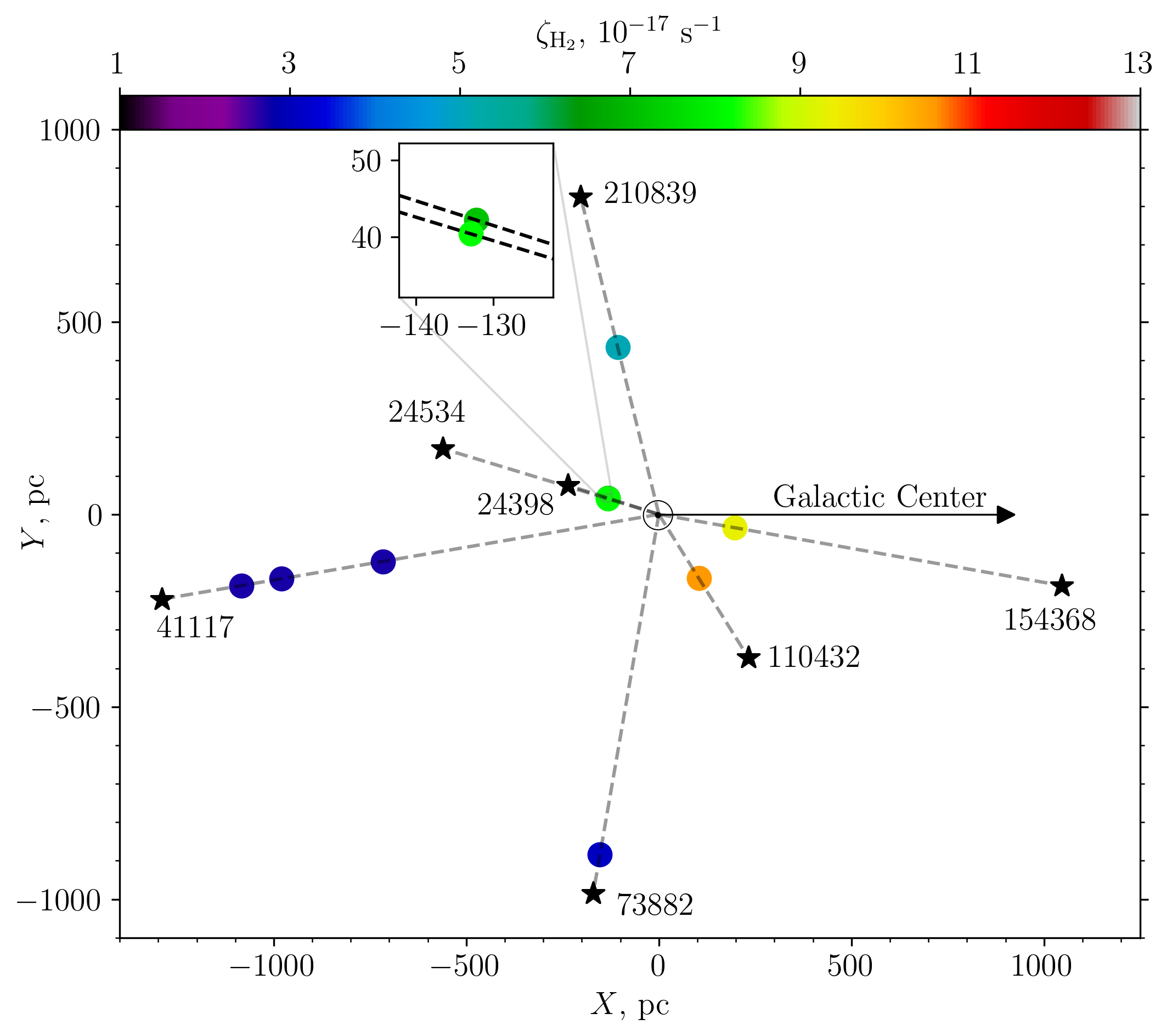}
    \caption{CR ionization rate obtained from the analysis of H$_3^+$ measurements toward the target stars (HD numbers, see Table~\ref{table1}). The left panel presents a comparison of $\zeta_{\rm H_2}$ values derived from our analysis (for the detection sight lines, see Table~\ref{table2}) with earlier estimates from \citet{Indriolo2012} and \citet{Neufeld2017}, the error bars show $1\sigma$ uncertainties. The red bullets depict the derived optimum values (see Section~\ref{optimum}), the blue bullets are the values corrected for $N({\rm H}_2)$ underestimate (see Section~\ref{correction}; errors cannot be computed self-consistently in this case). The two gray curves are the dependencies corresponding to models $\mathscr{H}$ and $\mathscr{L}$ of the CR spectrum discussed in \citet{Padovani2018}. The bullets in the right panel show locations of the gas clumps that were probed along the detection sight lines (projected onto the Galactic plane), the color coding indicates the value of the optimum $\zeta_{\rm H_2}$.}
    \label{f6}
\end{center}
\end{figure*}

In Section~\ref{underestimate} we discussed that our simulations systematically underestimate $N({\rm H}_2)$, typically by 30--50\%. For the detection sight lines, this is expected to have a relatively moderate but noticeable effect on the values of optimum $\zeta_{\rm H_2}$ derived from Equation~(\ref{chi2}). 

To qualitatively understand the effect on $\zeta_{\rm H_2}$, we note that the relative uncertainty in $N({\rm H}_3^+)$ is substantially smaller than those in $N({\rm H}_2)$ and $N({\rm H}_{\rm tot})$ (see Table~\ref{table1}). Accordingly, Equation~(\ref{chi2}) tends to push the optimum $N({\rm H}_2)$ up, toward the mean measured value, by increasing $N({\rm H}_{\rm tot})$ to the upper limit of the confidence interval -- whereas the optimum $N({\rm H}_3^+)$ remains close to the measured mean. As evident from comparing Tables~\ref{table1} and \ref{table2}, the resulting optimum $N({\rm H}_{\rm tot})$ exceeds the measured mean, while the corresponding $N({\rm H}_2)$ still remains somewhat below the measured mean for most of the sight lines. Both trends cause $\zeta_{\rm H_2}$ to increase, as it follows from Equation~(\ref{H3+_balance}) (keeping in mind that $n_e\propto n_{\rm tot}$ for most of the detection sight lines). Thus, we conclude that the optimum values of $\zeta_{\rm H_2}$ listed in Table~\ref{table2} are systematically overestimated for most of the sight lines, i.e., the corrected values are expected to be even lower.

The corrected $\zeta_{\rm H_2}$ for a given detection sight line can be reasonably evaluated as follows: we require the ratio $N({\rm H}_3^+)/N({\rm H}_2)$ that is derived from the simulation for the mean $N({\rm H}_{\rm tot})$ to be equal to the ratio of the measured mean values. Such an approach allows us to simultaneously compensate for the bias in $N({\rm H}_{\rm tot})$, as $n_e$ in this case corresponds to the mean $N({\rm H}_{\rm tot})$, and ensure that CRs generate $N({\rm H}_3^+)$ in measured proportions to $N({\rm H}_2)$. To obtain the corrected $\zeta_{\rm H_2}$, we use quadratic interpolation for the computed $N({\rm H}_3^+)$ and $N({\rm H}_2)$ between the first three points of $\zeta_{\rm H_2}=1\times$, $5\times$, and $10\times10^{-17}$~s$^{-1}$. The resulting values for each detection sight line are given in Table~\ref{table2} in parentheses next to the optimum values; we do not report errors for the corrected $\zeta_{\rm H_2}$, as they cannot be computed self-consistently. We see that, apart from the outlier case HD~41117, all detection sight lines with underestimated $N({\rm H}_2)$ show a moderate correction of optimum $\zeta_{\rm H_2}$ down by 20--40\%, i.e., somewhat less than the magnitude of $N({\rm H}_2)$ underestimate.

\subsection{Comparison with previous results}
\label{comparison}

Comparison between the ionization rates derived in this paper and the earlier results for sight lines with H$_3^+$ detection \citep{Indriolo2012, Neufeld2017} shows a dramatic reduction. The optimum values of $\zeta_{\rm H_2}$ listed in Table~\ref{table2} are approximately 4 to 30 times lower than the previous estimates. In Figure~\ref{f6} we illustrate the reduction for all detection sight lines, where the earlier estimate for HD~41117 is from \citet{Neufeld2017} \citep[see their Figure~10 for $\zeta_p\equiv \zeta_{\rm H_2}/2.3$, based on H$_3^+$ measurements reported in][]{Albertsson2014}, and the rest is from \citet{Indriolo2012}.

We found that the main reason behind the reduction seen in the left panel of Figure~\ref{f6} is that the peak values of gas density derived from the map are significantly lower than the values assumed previously (see Section~\ref{3D_maps}). The reduction factor for $\zeta_{\rm H_2}$ closely follows the ratio of the peak density in our simulations (with taking into account the map density re-scaling we made to match $N({\rm H}_{\rm tot}^{\rm map})$ and the measured $N({\rm H}_{\rm tot})$, see Section~\ref{3D_PDR}) to the gas density listed in Table~4 of \citet{Indriolo2012}. The only exception here is the sight line HD~41117, where the computed $\zeta_{\rm H_2}$ is a factor of $\approx13$ smaller than the value reported in \citet{Neufeld2017}, whereas the density reduction is about 30. It is worth to remind that HD~41117 is also the only outlier in terms of the deviation between $N({\rm H}_{\rm tot})$ and $N({\rm H}_{\rm tot}^{\rm map})$ (see Table~\ref{table1}).

We also point out that uncertainty of our results for $\zeta_{\rm H_2}$ is significantly smaller than that of \citet{Indriolo2012}. This is because the electron density in \citet{Indriolo2012} was treated as a free parameter with the assumed uncertainty of $\pm50\%$, whereas in our case it is computed self-consistently. This allows us to claim that $\zeta_{\rm H_2}$ for the analyzed sight lines shows a factor of $\sim3$ variability with respect to one another. The spatial distribution of the ionization rate in the Galactic plane is illustrated in the right panel of Figure~\ref{f6}.

\section{Conclusions and outlook}

We re-evaluated the rate of H$_2$ ionization by CRs for 12 sight lines, inferred from available observations of H$_3^+$, H$_2$ and H column densities. In contrast to previous analyses, we used the 3D gas density derived from dust extinction maps \citep{Edenhofer2024}, and employed the {\sc 3d-pdr} code \citep{Bisbas2012} to solve for the equilibrium chemistry. Compared with earlier values of $\zeta_{\rm H_2}$ reported for the same sight lines \citep{Indriolo2012, Neufeld2017}, we found an average reduction by a factor of $\sim10$. 

Our main findings and conclusions can be summarized as follows:
\begin{itemize}
    \item The gas density determined from extinction mapping is systematically lower than the previously-used estimates for the gas along these sight lines, deduced from absorption-line observations of different molecular tracers.  Our densities are broadly consistent with new density estimates based on fitting of C$_2$ rotational lines, which have also recently been revised downwards by an order of magnitude \citep{Neufeld2024}.
    \item The high-resolution extinction map \citep{Edenhofer2024} does not only allow us to evaluate the magnitude of the gas density, but also yields the spatial gas distribution and thus makes it possible to localize the gas clumps where CR ionization is probed in individual measurements.
    \item The physical structure of the localized clumps is reconstructed in simulations with the {\sc 3d-pdr} code \citep{Bisbas2012}, assuming $\zeta_{\rm H_2}$ to be the only unconstrained parameter. The simulations show that $\zeta_{\rm H_2}$ is systematically reduced compared to the earlier estimates, and the reduction factor closely follows the ratio of the peak density derived from the map to the density assumed earlier.
    \item Our new ionization rates (see Figure~\ref{f6}) show a factor of $\sim3$ variability with respect to one another.  While our error bars do not reflect all possible uncertainties in the modeling, this is at least strongly suggestive of a true spatial variability to the CR ionization rate, which may reflect the presence of CR sources internal to some clouds, proximity of clouds to recent supernovae, or CR transport effects.
    \item The high ionization rates in \citet{Indriolo2012} prompted speculation that the spectrum of low-energy protons may be much higher than would be suggested by the measurements from the Voyager probes \citep{Cummings2016}.  Our new results bring the average ionization rates into reasonable agreement with the Voyager measurements, represented by the ``model $\mathscr{L}$'' curve in Figure~\ref{f6}, and strongly disfavor the ``model $\mathscr{H}$'' spectrum discussed in \citet{Ivlev2015} and \citet{Padovani2018} -- which was developed in response to the high ionization rates reported in \citet{Indriolo2012}.
    \item In general, the parameter that is derived from absorption observations of different CR-generated ions is a ratio of the CR ionization rate to the density of species that lead to the ion destruction. This parameter is $\zeta_{\rm H_2}/n_{\rm tot}$ for H$_3^+$ observations, where the relation $n_e\propto n_{\rm tot}$ reasonably holds, and the same relation is usually employed for OH$^+$ observations of atomic gas \citep[used to estimate the rate of atomic hydrogen ionization, see][]{Indriolo2015}. Also in denser molecular clouds, where ortho-H$_2$D$^+$ ions are used to trace H$_3^+$ \citep{Sabatini2020, Sabatini2023}, the destruction occurs in reactions with CO molecules whose density is assumed to follow H$_2$ density. Given the above discussion of uncertainty in earlier density estimates, a careful re-evaluation for all previously published estimates of $\zeta_{\rm H_2}$ in molecular clouds may be useful. 
\end{itemize}

The last point above is particularly relevant for the available OH$^+$ measurements \citep{Indriolo2015, Bacalla2019}. The current estimates of the ionization rate in atomic gas are unreliable, as they are based on rather arbitrary assumed gas density values of 35~cm$^{-3}$ \citep[][obtained by extrapolating from the fiducial value used for molecular gas]{Indriolo2015} or 100~cm$^{-3}$ \citep{Bacalla2019}. If extinction mapping can be applied also to the lower-density gas where OH$^+$ ions primarily form, it will be possible to greatly expand the sample of clouds with well-constrained ionization rates. Moreover, sight lines with detections of both OH$^+$ and H$_3^+$ will provide unbiased insight into CR attenuation, as the ionization rate can be constrained at different depths into the same cloud.

Our findings about the low gas densities are in harmony with parallel work on the Local Bubble expansion \citep{Zucker2022} and the Per-Tau shell formation \citep{Bialy2021, Soler2023}. Furthermore, our results support the recent studies by \citet{Redaelli2022}, who suggest the presence of low volume density (tens of cm$^{-3}$ at most) molecular material surrounding the L1544 prestellar core in the nearby Taurus Molecular Cloud Complex. We stress that the reduced gas density and the accompanying reduction in the CR ionization rate may have profound implications on the chemical composition of diffuse and translucent clouds. 

\section*{Acknowledgments}
The authors gratefully acknowledge the support of the Max Planck Society. We thank the anonymous referee for their constructive comments, Jorma Harju for valuable suggestions and advice, and Anna Ivleva for assistance with compiling the figures and tables. TGB acknowledges support from the Leading Innovation and Entrepreneurship Team of Zhejiang Province of China (Grant No.~2023R01008). This research has used tools developed as part of the EXPLORE project that has received funding from the European Union's Horizon 2020 research and innovation programme under grant agreement No 101004214.  

\bibliographystyle{apj}
\bibliography{paper_arXiv}

\begin{thebibliography}{}
\expandafter\ifx\csname natexlab\endcsname\relax\def\natexlab#1{#1}\fi

\bibitem[{{Albertsson} {et~al.}(2014){Albertsson}, {Indriolo}, {Kreckel},
  {Semenov}, {Crabtree}, \& {Henning}}]{Albertsson2014}
{Albertsson}, T., {Indriolo}, N., {Kreckel}, H., {et~al.} 2014, \apj, 787, 44

\bibitem[{{Anders} {et~al.}(2019){Anders}, {Khalatyan}, {Chiappini}, {Queiroz},
  {Santiago}, {Jordi}, {Girardi}, {Brown}, {Matijevi{\v{c}}}, {Monari},
  {Cantat-Gaudin}, {Weiler}, {Khan}, {Miglio}, {Carrillo}, {Romero-G{\'o}mez},
  {Minchev}, {de Jong}, {Antoja}, {Ramos}, {Steinmetz}, \& {Enke}}]{Anders2019}
{Anders}, F., {Khalatyan}, A., {Chiappini}, C., {et~al.} 2019, \aap, 628, A94

\bibitem[{{Anders} {et~al.}(2022){Anders}, {Khalatyan}, {Queiroz}, {Chiappini},
  {Ard{\`e}vol}, {Casamiquela}, {Figueras}, {Jim{\'e}nez-Arranz}, {Jordi},
  {Mongui{\'o}}, {Romero-G{\'o}mez}, {Altamirano}, {Antoja}, {Assaad},
  {Cantat-Gaudin}, {Castro-Ginard}, {Enke}, {Girardi}, {Guiglion}, {Khan},
  {Luri}, {Miglio}, {Minchev}, {Ramos}, {Santiago}, \&
  {Steinmetz}}]{StarHorse2021}
{Anders}, F., {Khalatyan}, A., {Queiroz}, A.~B.~A., {et~al.} 2022, \aap, 658,
  A91

\bibitem[{{Bacalla} {et~al.}(2019){Bacalla}, {Linnartz}, {Cox}, {Cami},
  {Roueff}, {Smoker}, {Farhang}, {Bouwman}, \& {Zhao}}]{Bacalla2019}
{Bacalla}, X.~L., {Linnartz}, H., {Cox}, N. L.~J., {et~al.} 2019, \aap, 622,
  A31

\bibitem[{{Bergin} {et~al.}(1995){Bergin}, {Langer}, \&
  {Goldsmith}}]{Bergin1995}
{Bergin}, E.~A., {Langer}, W.~D., \& {Goldsmith}, P.~F. 1995, \apj, 441, 222

\bibitem[{{Bessell} \& {Murphy}(2012)}]{Hipparcso_synthetic_photometry}
{Bessell}, M., \& {Murphy}, S. 2012, \pasp, 124, 140

\bibitem[{{Bialy} {et~al.}(2022){Bialy}, {Belli}, \& {Padovani}}]{Bialy2022}
{Bialy}, S., {Belli}, S., \& {Padovani}, M. 2022, \aap, 658, L13

\bibitem[{{Bialy} {et~al.}(2019){Bialy}, {Neufeld}, {Wolfire}, {Sternberg}, \&
  {Burkhart}}]{Bialy2019}
{Bialy}, S., {Neufeld}, D., {Wolfire}, M., {Sternberg}, A., \& {Burkhart}, B.
  2019, \apj, 885, 109

\bibitem[{{Bialy} {et~al.}(2021){Bialy}, {Zucker}, {Goodman}, {Foley}, {Alves},
  {Semenov}, {Benjamin}, {Leike}, \& {En{\ss}lin}}]{Bialy2021}
{Bialy}, S., {Zucker}, C., {Goodman}, A., {et~al.} 2021, \apjl, 919, L5

\bibitem[{{Bisbas} {et~al.}(2012){Bisbas}, {Bell}, {Viti}, {Yates}, \&
  {Barlow}}]{Bisbas2012}
{Bisbas}, T.~G., {Bell}, T.~A., {Viti}, S., {Yates}, J., \& {Barlow}, M.~J.
  2012, \mnras, 427, 2100

\bibitem[{{Black} \& {Dalgarno}(1977)}]{Black1977}
{Black}, J.~H., \& {Dalgarno}, A. 1977, \apjs, 34, 405

\bibitem[{{Bohlin} {et~al.}(1978){Bohlin}, {Savage}, \& {Drake}}]{Bohlin1978}
{Bohlin}, R.~C., {Savage}, B.~D., \& {Drake}, J.~F. 1978, \apj, 224, 132

\bibitem[{{Bovino} {et~al.}(2020){Bovino}, {Ferrada-Chamorro}, {Lupi},
  {Schleicher}, \& {Caselli}}]{Bovino2020}
{Bovino}, S., {Ferrada-Chamorro}, S., {Lupi}, A., {Schleicher}, D.~R.~G., \&
  {Caselli}, P. 2020, \mnras, 495, L7

\bibitem[{{Cartledge} {et~al.}(2004){Cartledge}, {Lauroesch}, {Meyer}, \&
  {Sofia}}]{Cartledge2004}
{Cartledge}, S. I.~B., {Lauroesch}, J.~T., {Meyer}, D.~M., \& {Sofia}, U.~J.
  2004, \apj, 613, 1037

\bibitem[{{Caselli} {et~al.}(1999){Caselli}, {Walmsley}, {Tafalla}, {Dore}, \&
  {Myers}}]{Caselli1999}
{Caselli}, P., {Walmsley}, C.~M., {Tafalla}, M., {Dore}, L., \& {Myers}, P.~C.
  1999, \apjl, 523, L165

\bibitem[{{Caselli} {et~al.}(1998){Caselli}, {Walmsley}, {Terzieva}, \&
  {Herbst}}]{Caselli1998}
{Caselli}, P., {Walmsley}, C.~M., {Terzieva}, R., \& {Herbst}, E. 1998, \apj,
  499, 234

\bibitem[{{Castelli} \& {Kurucz}(2003)}]{ATLAS9_Castelli2003}
{Castelli}, F., \& {Kurucz}, R.~L. 2003, in Modelling of Stellar Atmospheres,
  ed. N.~{Piskunov}, W.~W. {Weiss}, \& D.~F. {Gray}, Vol. 210, A20

\bibitem[{{Cazaux} \& {Tielens}(2002)}]{Cazaux_Tielens_2002}
{Cazaux}, S., \& {Tielens}, A.~G.~G.~M. 2002, \apjl, 575, L29

\bibitem[{{Cazaux} \& {Tielens}(2004)}]{Cazaux_Tielens_2004}
---. 2004, \apj, 604, 222

\bibitem[{{Ceccarelli} {et~al.}(2014){Ceccarelli}, {Dominik},
  {L{\'o}pez-Sepulcre}, {Kama}, {Padovani}, {Caux}, \&
  {Caselli}}]{Ceccarelli2014}
{Ceccarelli}, C., {Dominik}, C., {L{\'o}pez-Sepulcre}, A., {et~al.} 2014,
  \apjl, 790, L1

\bibitem[{{Crane} {et~al.}(1995){Crane}, {Lambert}, \& {Sheffer}}]{Crane1995}
{Crane}, P., {Lambert}, D.~L., \& {Sheffer}, Y. 1995, \apjs, 99, 107

\bibitem[{{Cravens} \& {Dalgarno}(1978)}]{Cravens1978}
{Cravens}, T.~E., \& {Dalgarno}, A. 1978, \apj, 219, 750

\bibitem[{{Cummings} {et~al.}(2016){Cummings}, {Stone}, {Heikkila}, {Lal},
  {Webber}, {J{\'o}hannesson}, {Moskalenko}, {Orlando}, \&
  {Porter}}]{Cummings2016}
{Cummings}, A.~C., {Stone}, E.~C., {Heikkila}, B.~C., {et~al.} 2016, \apj, 831,
  18

\bibitem[{{Dalgarno}(2006)}]{Dalgarno2006}
{Dalgarno}, A. 2006, Proceedings of the National Academy of Science, 103, 12269

\bibitem[{{Diplas} \& {Savage}(1994)}]{Diplas1994}
{Diplas}, A., \& {Savage}, B.~D. 1994, \apjs, 93, 211

\bibitem[{{Draine}(2011)}]{DraineBook2011}
{Draine}, B.~T. 2011, {Physics of the Interstellar and Intergalactic Medium}
  (Princeton, NJ: Princeton Univ. Press)

\bibitem[{{Draine} \& {Bertoldi}(1996)}]{Draine_self_shielding}
{Draine}, B.~T., \& {Bertoldi}, F. 1996, \apj, 468, 269

\bibitem[{{Edenhofer} {et~al.}(2022){Edenhofer}, {Leike}, {Frank}, \&
  {En{\ss}lin}}]{Edenhofer2022}
{Edenhofer}, G., {Leike}, R.~H., {Frank}, P., \& {En{\ss}lin}, T.~A. 2022,
  arXiv e-prints, arXiv:2206.10634

\bibitem[{{Edenhofer} {et~al.}(2024{\natexlab{a}}){Edenhofer}, {Zucker},
  {Frank}, {Saydjari}, {Speagle}, {Finkbeiner}, \&
  {En{\ss}lin}}]{Edenhofer2024}
{Edenhofer}, G., {Zucker}, C., {Frank}, P., {et~al.} 2024{\natexlab{a}}, \aap,
  685, A82

\bibitem[{{Edenhofer} {et~al.}(2024{\natexlab{b}}){Edenhofer}, {Frank}, {Roth},
  {Leike}, {Guerdi}, {Scheel-Platz}, {Guardiani}, {Eberle}, {Westerkamp}, \&
  {En{\ss}lin}}]{Edenhofer2024NIFTyRE}
{Edenhofer}, G., {Frank}, P., {Roth}, J., {et~al.} 2024{\natexlab{b}}, arXiv
  e-prints, arXiv:2402.16683

\bibitem[{{Fabricius} {et~al.}(2002){Fabricius}, {H{\o}g}, {Makarov}, {Mason},
  {Wycoff}, \& {Urban}}]{TDSC_Catalog}
{Fabricius}, C., {H{\o}g}, E., {Makarov}, V.~V., {et~al.} 2002, \aap, 384, 180

\bibitem[{{Fabricius} {et~al.}(2021){Fabricius}, {Luri}, {Arenou}, {Babusiaux},
  {Helmi}, {Muraveva}, {Reyl{\'e}}, {Spoto}, {Vallenari}, {Antoja}, {Balbinot},
  {Barache}, {Bauchet}, {Bragaglia}, {Busonero}, {Cantat-Gaudin}, {Carrasco},
  {Diakit{\'e}}, {Fabrizio}, {Figueras}, {Garcia-Gutierrez}, {Garofalo},
  {Jordi}, {Kervella}, {Khanna}, {Leclerc}, {Licata}, {Lambert}, {Marrese},
  {Masip}, {Ramos}, {Robichon}, {Robin}, {Romero-G{\'o}mez}, {Rubele}, \&
  {Weiler}}]{GaiaEDR3_Validation}
{Fabricius}, C., {Luri}, X., {Arenou}, F., {et~al.} 2021, \aap, 649, A5

\bibitem[{{Favre} {et~al.}(2018){Favre}, {Ceccarelli}, {L{\'o}pez-Sepulcre},
  {Fontani}, {Neri}, {Manigand}, {Kama}, {Caselli}, {Jaber Al-Edhari},
  {Kahane}, {Alves}, {Balucani}, {Bianchi}, {Caux}, {Codella}, {Dulieu},
  {Pineda}, {Sims}, \& {Theul{\'e}}}]{Favre2018}
{Favre}, C., {Ceccarelli}, C., {L{\'o}pez-Sepulcre}, A., {et~al.} 2018, \apj,
  859, 136

\bibitem[{{Fontani} {et~al.}(2017){Fontani}, {Ceccarelli}, {Favre}, {Caselli},
  {Neri}, {Sims}, {Kahane}, {Alves}, {Balucani}, {Bianchi}, {Caux}, {Jaber
  Al-Edhari}, {Lopez-Sepulcre}, {Pineda}, {Bachiller}, {Bizzocchi},
  {Bottinelli}, {Chacon-Tanarro}, {Choudhury}, {Codella}, {Coutens}, {Dulieu},
  {Feng}, {Rimola}, {Hily-Blant}, {Holdship}, {Jimenez-Serra}, {Laas},
  {Lefloch}, {Oya}, {Podio}, {Pon}, {Punanova}, {Quenard}, {Sakai}, {Spezzano},
  {Taquet}, {Testi}, {Theul{\'e}}, {Ugliengo}, {Vastel}, {Vasyunin}, {Viti},
  {Yamamoto}, \& {Wiesenfeld}}]{Fontani2017}
{Fontani}, F., {Ceccarelli}, C., {Favre}, C., {et~al.} 2017, \aap, 605, A57

\bibitem[{{Gaches} {et~al.}(2019){Gaches}, {Offner}, \& {Bisbas}}]{Gaches2019}
{Gaches}, B. A.~L., {Offner}, S. S.~R., \& {Bisbas}, T.~G. 2019, \apj, 878, 105

\bibitem[{{Geballe} \& {Oka}(1996)}]{Geballe1996}
{Geballe}, T.~R., \& {Oka}, T. 1996, \nat, 384, 334

\bibitem[{{Gerin} {et~al.}(2015){Gerin}, {Ruaud}, {Goicoechea}, {Gusdorf},
  {Godard}, {de Luca}, {Falgarone}, {Goldsmith}, {Lis}, {Menten}, {Neufeld},
  {Phillips}, \& {Liszt}}]{Gerin2015}
{Gerin}, M., {Ruaud}, M., {Goicoechea}, J.~R., {et~al.} 2015, \aap, 573, A30

\bibitem[{{Giannetti} {et~al.}(2017){Giannetti}, {Leurini}, {K{\"o}nig},
  {Urquhart}, {Pillai}, {Brand}, {Kauffmann}, {Wyrowski}, \&
  {Menten}}]{Giannetti2017}
{Giannetti}, A., {Leurini}, S., {K{\"o}nig}, C., {et~al.} 2017, \aap, 606, L12

\bibitem[{{Gibb} {et~al.}(2004){Gibb}, {Whittet}, {Boogert}, \&
  {Tielens}}]{Gibb2004}
{Gibb}, E.~L., {Whittet}, D.~C.~B., {Boogert}, A.~C.~A., \& {Tielens},
  A.~G.~G.~M. 2004, \apjs, 151, 35

\bibitem[{{Glassgold} {et~al.}(2012){Glassgold}, {Galli}, \&
  {Padovani}}]{Glassgold2012}
{Glassgold}, A.~E., {Galli}, D., \& {Padovani}, M. 2012, \apj, 756, 157

\bibitem[{{Glassgold} \& {Langer}(1973)}]{Glassgold73}
{Glassgold}, A.~E., \& {Langer}, W.~D. 1973, \apj, 186, 859

\bibitem[{{Gloeckler} \& {Fisk}(2015)}]{Gloeckler2015}
{Gloeckler}, G., \& {Fisk}, L.~A. 2015, \apjl, 806, L27

\bibitem[{{Goldsmith}(2013)}]{Goldsmith2013}
{Goldsmith}, P.~F. 2013, \apj, 774, 134

\bibitem[{{G{\'o}rski} {et~al.}(2005){G{\'o}rski}, {Hivon}, {Banday},
  {Wandelt}, {Hansen}, {Reinecke}, \& {Bartelmann}}]{HEALPIX_2}
{G{\'o}rski}, K.~M., {Hivon}, E., {Banday}, A.~J., {et~al.} 2005, \apj, 622,
  759

\bibitem[{{Green} {et~al.}(2019){Green}, {Schlafly}, {Zucker}, {Speagle}, \&
  {Finkbeiner}}]{Green2019}
{Green}, G.~M., {Schlafly}, E., {Zucker}, C., {Speagle}, J.~S., \&
  {Finkbeiner}, D. 2019, \apj, 887, 93

\bibitem[{{Hasegawa} \& {Herbst}(1993)}]{Hasegawa1993}
{Hasegawa}, T.~I., \& {Herbst}, E. 1993, \mnras, 261, 83

\bibitem[{{Hayakawa} {et~al.}(1961){Hayakawa}, {Nishimura}, \&
  {Takayanagi}}]{Hayakawa1961}
{Hayakawa}, S., {Nishimura}, S., \& {Takayanagi}, T. 1961, \pasj, 13, 184

\bibitem[{{Heays} {et~al.}(2017){Heays}, {Bosman}, \& {van
  Dishoeck}}]{Heays_2017}
{Heays}, A.~N., {Bosman}, A.~D., \& {van Dishoeck}, E.~F. 2017, \aap, 602, A105

\bibitem[{{Herbst} \& {Klemperer}(1973)}]{Herbst1973}
{Herbst}, E., \& {Klemperer}, W. 1973, \apj, 185, 505

\bibitem[{{H{\o}g} {et~al.}(2000){H{\o}g}, {Fabricius}, {Makarov}, {Urban},
  {Corbin}, {Wycoff}, {Bastian}, {Schwekendiek}, \& {Wicenec}}]{Tycho2Catalog}
{H{\o}g}, E., {Fabricius}, C., {Makarov}, V.~V., {et~al.} 2000, \aap, 355, L27

\bibitem[{{Hollenbach} {et~al.}(2012){Hollenbach}, {Kaufman}, {Neufeld},
  {Wolfire}, \& {Goicoechea}}]{Hollenbach2012}
{Hollenbach}, D., {Kaufman}, M.~J., {Neufeld}, D., {Wolfire}, M., \&
  {Goicoechea}, J.~R. 2012, \apj, 754, 105

\bibitem[{{Hollenbach} {et~al.}(1991){Hollenbach}, {Takahashi}, \&
  {Tielens}}]{Dust_temperature}
{Hollenbach}, D.~J., {Takahashi}, T., \& {Tielens}, A.~G.~G.~M. 1991, \apj,
  377, 192

\bibitem[{{Indriolo} {et~al.}(2007){Indriolo}, {Geballe}, {Oka}, \&
  {McCall}}]{Indriolo2007}
{Indriolo}, N., {Geballe}, T.~R., {Oka}, T., \& {McCall}, B.~J. 2007, \apj,
  671, 1736

\bibitem[{{Indriolo} \& {McCall}(2012)}]{Indriolo2012}
{Indriolo}, N., \& {McCall}, B.~J. 2012, \apj, 745, 91

\bibitem[{{Indriolo} {et~al.}(2015){Indriolo}, {Neufeld}, {Gerin}, {Schilke},
  {Benz}, {Winkel}, {Menten}, {Chambers}, {Black}, {Bruderer}, {Falgarone},
  {Godard}, {Goicoechea}, {Gupta}, {Lis}, {Ossenkopf}, {Persson},
  {Sonnentrucker}, {van der Tak}, {van Dishoeck}, {Wolfire}, \&
  {Wyrowski}}]{Indriolo2015}
{Indriolo}, N., {Neufeld}, D.~A., {Gerin}, M., {et~al.} 2015, \apj, 800, 40

\bibitem[{{Ivlev} {et~al.}(2015){Ivlev}, {Padovani}, {Galli}, \&
  {Caselli}}]{Ivlev2015}
{Ivlev}, A.~V., {Padovani}, M., {Galli}, D., \& {Caselli}, P. 2015, \apj, 812,
  135

\bibitem[{{Jenkins}(2019)}]{Jenkins2019}
{Jenkins}, E.~B. 2019, \apj, 872, 55

\bibitem[{{K{\'a}losi} {et~al.}(2023){K{\'a}losi}, {Gamer}, {Grieser}, {von
  Hahn}, {Isberner}, {J{\"a}ger}, {Kreckel}, {Neufeld}, {Paul}, {Savin},
  {Schippers}, {Schmidt}, {Wolf}, {Wolfire}, \& {Novotn{\'y}}}]{Kalosi2023}
{K{\'a}losi}, {\'A}., {Gamer}, L., {Grieser}, M., {et~al.} 2023, \apjl, 955,
  L26

\bibitem[{{Keto} \& {Caselli}(2008)}]{Keto2008}
{Keto}, E., \& {Caselli}, P. 2008, \apj, 683, 238

\bibitem[{{Kovalenko} {et~al.}(2018){Kovalenko}, {Dung Tran}, {Rednyk},
  {Rou{\v{c}}ka}, {Dohnal}, {Pla{\v{s}}il}, {Gerlich}, \&
  {Glos{\'\i}k}}]{Kovalenko2018}
{Kovalenko}, A., {Dung Tran}, T., {Rednyk}, S., {et~al.} 2018, \apj, 856, 100

\bibitem[{{Lallement} {et~al.}(2022){Lallement}, {Vergely}, {Babusiaux}, \&
  {Cox}}]{Lallement2022}
{Lallement}, R., {Vergely}, J.~L., {Babusiaux}, C., \& {Cox}, N.~L.~J. 2022,
  \aap, 661, A147

\bibitem[{{Larsson} \& {Siegbahn}(1983)}]{Larsson1983}
{Larsson}, M., \& {Siegbahn}, P. E.~M. 1983, \jcp, 79, 2270

\bibitem[{{Leger} {et~al.}(1985){Leger}, {Jura}, \& {Omont}}]{Leger1985}
{Leger}, A., {Jura}, M., \& {Omont}, A. 1985, \aap, 144, 147

\bibitem[{{Leike} {et~al.}(2020){Leike}, {Glatzle}, \&
  {En{\ss}lin}}]{Leike2020}
{Leike}, R.~H., {Glatzle}, M., \& {En{\ss}lin}, T.~A. 2020, \aap, 639, A138

\bibitem[{{Li} {et~al.}(2014){Li}, {Goodman}, {Sridharan}, {Houde}, {Li},
  {Novak}, \& {Tang}}]{Li2014}
{Li}, H.~B., {Goodman}, A., {Sridharan}, T.~K., {et~al.} 2014, in Protostars
  and Planets VI, ed. H.~{Beuther}, R.~S. {Klessen}, C.~P. {Dullemond}, \&
  T.~{Henning}, 101

\bibitem[{{Luo} {et~al.}(2023{\natexlab{a}}){Luo}, {Zhang}, {Bisbas}, {Li},
  {Zhou}, {Tang}, {Wang}, {Zuo}, \& {Yue}}]{Luo2023b}
{Luo}, G., {Zhang}, Z.-Y., {Bisbas}, T.~G., {et~al.} 2023{\natexlab{a}}, \apj,
  946, 91

\bibitem[{{Luo} {et~al.}(2023{\natexlab{b}}){Luo}, {Zhang}, {Bisbas}, {Li},
  {Tang}, {Wang}, {Zhou}, {Zuo}, {Yue}, {Zhou}, \& {Lin}}]{Luo2023a}
---. 2023{\natexlab{b}}, \apj, 942, 101

\bibitem[{{Mathis} {et~al.}(1983){Mathis}, {Mezger}, \& {Panagia}}]{Mathis1983}
{Mathis}, J.~S., {Mezger}, P.~G., \& {Panagia}, N. 1983, \aap, 128, 212

\bibitem[{{McCall} {et~al.}(2003){McCall}, {Huneycutt}, {Saykally}, {Geballe},
  {Djuric}, {Dunn}, {Semaniak}, {Novotny}, {Al-Khalili}, {Ehlerding},
  {Hellberg}, {Kalhori}, {Neau}, {Thomas}, {{\"O}sterdahl}, \&
  {Larsson}}]{McCall2003}
{McCall}, B.~J., {Huneycutt}, A.~J., {Saykally}, R.~J., {et~al.} 2003, \nat,
  422, 500

\bibitem[{{McCall} {et~al.}(2004){McCall}, {Huneycutt}, {Saykally}, {Djuric},
  {Dunn}, {Semaniak}, {Novotny}, {Al-Khalili}, {Ehlerding}, {Hellberg},
  {Kalhori}, {Neau}, {Thomas}, {Paal}, {{\"O}sterdahl}, \&
  {Larsson}}]{McCall2004}
---. 2004, \pra, 70, 052716

\bibitem[{{McElroy} {et~al.}(2013){McElroy}, {Walsh}, {Markwick}, {Cordiner},
  {Smith}, \& {Millar}}]{McElroy13}
{McElroy}, D., {Walsh}, C., {Markwick}, A.~J., {et~al.} 2013, \aap, 550, A36

\bibitem[{{McKee}(1989)}]{McKee1989}
{McKee}, C.~F. 1989, \apj, 345, 782

\bibitem[{{Najar} \& {Kalugina}(2020)}]{Najar2020}
{Najar}, F., \& {Kalugina}, Y. 2020, RSC Advances, 10, 8580

\bibitem[{{Neufeld} \& {Wolfire}(2017)}]{Neufeld2017}
{Neufeld}, D.~A., \& {Wolfire}, M.~G. 2017, \apj, 845, 163

\bibitem[{{Neufeld} {et~al.}(2024){Neufeld}, {Welty}, {Ivlev}, {Caselli},
  {Edenhofer}, {Indriolo}, {Obolentseva}, {Silsbee}, {Sonnentrucker}, \&
  {Wolfire}}]{Neufeld2024}
{Neufeld}, D.~A., {Welty}, D., {Ivlev}, A.~V., {et~al.} 2024, submitted

\bibitem[{{O'Donnell} \& {Watson}(1974)}]{Odonnell1974}
{O'Donnell}, E.~J., \& {Watson}, W.~D. 1974, \apj, 191, 89

\bibitem[{{O'Neill} {et~al.}(2024){O'Neill}, {Zucker}, {Goodman}, \&
  {Edenhofer}}]{ONeill2024}
{O'Neill}, T.~J., {Zucker}, C., {Goodman}, A.~A., \& {Edenhofer}, G. 2024,
  arXiv e-prints, arXiv:2403.04961

\bibitem[{{Ormel} {et~al.}(2009){Ormel}, {Paszun}, {Dominik}, \&
  {Tielens}}]{Ormel2009}
{Ormel}, C.~W., {Paszun}, D., {Dominik}, C., \& {Tielens}, A.~G.~G.~M. 2009,
  \aap, 502, 845

\bibitem[{{Ossenkopf} \& {Henning}(1994)}]{Ossenkopf1994}
{Ossenkopf}, V., \& {Henning}, T. 1994, \aap, 291, 943

\bibitem[{{Padovani} {et~al.}(2018){Padovani}, {Ivlev}, {Galli}, \&
  {Caselli}}]{Padovani2018}
{Padovani}, M., {Ivlev}, A.~V., {Galli}, D., \& {Caselli}, P. 2018, \aap, 614,
  A111

\bibitem[{{Padovani} {et~al.}(2020){Padovani}, {Ivlev}, {Galli}, {Offner},
  {Indriolo}, {Rodgers-Lee}, {Marcowith}, {Girichidis}, {Bykov}, \&
  {Kruijssen}}]{Padovani2020}
{Padovani}, M., {Ivlev}, A.~V., {Galli}, D., {et~al.} 2020, \ssr, 216, 29

\bibitem[{{Phan} {et~al.}(2023){Phan}, {Recchia}, {Mertsch}, \&
  {Gabici}}]{Phan2023}
{Phan}, V. H.~M., {Recchia}, S., {Mertsch}, P., \& {Gabici}, S. 2023, \prd,
  107, 123006

\bibitem[{{Rachford} {et~al.}(2001){Rachford}, {Snow}, {Tumlinson}, {Shull},
  {Roueff}, {Andre}, {Desert}, {Ferlet}, {Vidal-Madjar}, \&
  {York}}]{Rachford2001}
{Rachford}, B.~L., {Snow}, T.~P., {Tumlinson}, J., {et~al.} 2001, \apj, 555,
  839

\bibitem[{{Rachford} {et~al.}(2002){Rachford}, {Snow}, {Tumlinson}, {Shull},
  {Blair}, {Ferlet}, {Friedman}, {Gry}, {Jenkins}, {Morton}, {Savage},
  {Sonnentrucker}, {Vidal-Madjar}, {Welty}, \& {York}}]{Rachford2002}
---. 2002, \apj, 577, 221

\bibitem[{{Rachford} {et~al.}(2009){Rachford}, {Snow}, {Destree}, {Ross},
  {Ferlet}, {Friedman}, {Gry}, {Jenkins}, {Morton}, {Savage}, {Shull},
  {Sonnentrucker}, {Tumlinson}, {Vidal-Madjar}, {Welty}, \&
  {York}}]{Rachford2009}
{Rachford}, B.~L., {Snow}, T.~P., {Destree}, J.~D., {et~al.} 2009, \apjs, 180,
  125

\bibitem[{{Recchia} {et~al.}(2019){Recchia}, {Phan}, {Biswas}, \&
  {Gabici}}]{Recchia19}
{Recchia}, S., {Phan}, V.~H.~M., {Biswas}, S., \& {Gabici}, S. 2019, \mnras,
  485, 2276

\bibitem[{{Redaelli} {et~al.}(2024){Redaelli}, {Bovino}, {Lupi}, {Grassi},
  {Gaete-Espinoza}, {Sabatini}, \& {Caselli}}]{Redaelli2024}
{Redaelli}, E., {Bovino}, S., {Lupi}, A., {et~al.} 2024, \aap, 685, A67

\bibitem[{{Redaelli} {et~al.}(2022){Redaelli}, {Chac{\'o}n-Tanarro}, {Caselli},
  {Tafalla}, {Pineda}, {Spezzano}, \& {Sipil{\"a}}}]{Redaelli2022}
{Redaelli}, E., {Chac{\'o}n-Tanarro}, A., {Caselli}, P., {et~al.} 2022, \apj,
  941, 168

\bibitem[{{Sabatini} {et~al.}(2023){Sabatini}, {Bovino}, \&
  {Redaelli}}]{Sabatini2023}
{Sabatini}, G., {Bovino}, S., \& {Redaelli}, E. 2023, \apjl, 947, L18

\bibitem[{{Sabatini} {et~al.}(2020){Sabatini}, {Bovino}, {Giannetti},
  {Wyrowski}, {{\'O}rdenes}, {Pascale}, {Pillai}, {Wienen}, {Csengeri}, \&
  {Menten}}]{Sabatini2020}
{Sabatini}, G., {Bovino}, S., {Giannetti}, A., {et~al.} 2020, \aap, 644, A34

\bibitem[{{Savage} {et~al.}(1977){Savage}, {Bohlin}, {Drake}, \&
  {Budich}}]{Savage1977}
{Savage}, B.~D., {Bohlin}, R.~C., {Drake}, J.~F., \& {Budich}, W. 1977, \apj,
  216, 291

\bibitem[{{Sheffer} {et~al.}(2008){Sheffer}, {Rogers}, {Federman}, {Abel},
  {Gredel}, {Lambert}, \& {Shaw}}]{Sheffer2008}
{Sheffer}, Y., {Rogers}, M., {Federman}, S.~R., {et~al.} 2008, \apj, 687, 1075

\bibitem[{{Shingledecker} {et~al.}(2018){Shingledecker}, {Tennis}, {Le Gal}, \&
  {Herbst}}]{Shingledecker2018}
{Shingledecker}, C.~N., {Tennis}, J., {Le Gal}, R., \& {Herbst}, E. 2018, \apj,
  861, 20

\bibitem[{{Shu} {et~al.}(1987){Shu}, {Adams}, \& {Lizano}}]{Shu1987}
{Shu}, F.~H., {Adams}, F.~C., \& {Lizano}, S. 1987, \araa, 25, 23

\bibitem[{{Shull}(2024)}]{Shull2024}
{Shull}, J.~M. 2024, private communication

\bibitem[{{Shull} \& {Danforth}(2019)}]{Shull2019}
{Shull}, J.~M., \& {Danforth}, C.~W. 2019, \apj, 882, 180

\bibitem[{{Shull} {et~al.}(2021){Shull}, {Danforth}, \& {Anderson}}]{Shull2021}
{Shull}, J.~M., {Danforth}, C.~W., \& {Anderson}, K.~L. 2021, \apj, 911, 55

\bibitem[{{Sipil{\"a}} {et~al.}(2021){Sipil{\"a}}, {Silsbee}, \&
  {Caselli}}]{Sipila2021}
{Sipil{\"a}}, O., {Silsbee}, K., \& {Caselli}, P. 2021, \apj, 922, 126

\bibitem[{{Snow} {et~al.}(2000){Snow}, {Rachford}, {Tumlinson}, {Shull},
  {Welty}, {Blair}, {Ferlet}, {Friedman}, {Gry}, {Jenkins}, {Lecavelier des
  Etangs}, {Lemoine}, {Morton}, {Savage}, {Sembach}, {Vidal-Madjar}, {York},
  {Andersson}, {Feldman}, \& {Moos}}]{Snow2000}
{Snow}, T.~P., {Rachford}, B.~L., {Tumlinson}, J., {et~al.} 2000, \apjl, 538,
  L65

\bibitem[{{Sofia} {et~al.}(2004){Sofia}, {Lauroesch}, {Meyer}, \&
  {Cartledge}}]{Sofia2004}
{Sofia}, U.~J., {Lauroesch}, J.~T., {Meyer}, D.~M., \& {Cartledge}, S. I.~B.
  2004, \apj, 605, 272

\bibitem[{{Soler} {et~al.}(2023){Soler}, {Zucker}, {Peek}, {Heyer},
  {Goldsmith}, {Glover}, {Molinari}, {Klessen}, {Hennebelle}, {Testi},
  {Colman}, {Benedettini}, {Elia}, {Mininni}, {Pezzuto}, {Schisano}, \&
  {Traficante}}]{Soler2023}
{Soler}, J.~D., {Zucker}, C., {Peek}, J.~E.~G., {et~al.} 2023, \aap, 675, A206

\bibitem[{{Sonnentrucker} {et~al.}(2007){Sonnentrucker}, {Welty}, {Thorburn},
  \& {York}}]{Sonnentrucker2007}
{Sonnentrucker}, P., {Welty}, D.~E., {Thorburn}, J.~A., \& {York}, D.~G. 2007,
  \apjs, 168, 58

\bibitem[{{Soubiran} {et~al.}(2016){Soubiran}, {Le Campion}, {Brouillet}, \&
  {Chemin}}]{PastelCatalog}
{Soubiran}, C., {Le Campion}, J.-F., {Brouillet}, N., \& {Chemin}, L. 2016,
  \aap, 591, A118

\bibitem[{{Spitzer} \& {Tomasko}(1968)}]{Spitzer1968}
{Spitzer}, Lyman, J., \& {Tomasko}, M.~G. 1968, \apj, 152, 971

\bibitem[{{Stancil} {et~al.}(1999){Stancil}, {Schultz}, {Kimura}, {Gu},
  {Hirsch}, \& {Buenker}}]{Stancil1999}
{Stancil}, P.~C., {Schultz}, D.~R., {Kimura}, M., {et~al.} 1999, \aaps, 140,
  225

\bibitem[{{Stone} {et~al.}(2019){Stone}, {Cummings}, {Heikkila}, \&
  {Lal}}]{Stone2019}
{Stone}, E.~C., {Cummings}, A.~C., {Heikkila}, B.~C., \& {Lal}, N. 2019, Nature
  Astronomy, 3, 1013

\bibitem[{{Stone} {et~al.}(2013){Stone}, {Cummings}, {McDonald}, {Heikkila},
  {Lal}, \& {Webber}}]{Stone2013}
{Stone}, E.~C., {Cummings}, A.~C., {McDonald}, F.~B., {et~al.} 2013, Science,
  341, 150

\bibitem[{{Tango} {et~al.}(2006){Tango}, {Davis}, {Ireland}, {Aerts},
  {Uytterhoeven}, {Jacob}, {Mendez}, {North}, {Seneta}, \&
  {Tuthill}}]{Tango2006}
{Tango}, W.~J., {Davis}, J., {Ireland}, M.~J., {et~al.} 2006, \mnras, 370, 884

\bibitem[{{Tielens}(2005)}]{TielensBook2005}
{Tielens}, A.~G.~G.~M. 2005, {The Physics and Chemistry of the Interstellar
  Medium} (Cambridge: Cambridge Univ. Press)

\bibitem[{{Tielens} \& {Hollenbach}(1985)}]{Tielens_Hollenbach_1985}
{Tielens}, A.~G.~G.~M., \& {Hollenbach}, D. 1985, \apj, 291, 722

\bibitem[{{Turner} {et~al.}(2014){Turner}, {Fromang}, {Gammie}, {Klahr},
  {Lesur}, {Wardle}, \& {Bai}}]{Turner2014}
{Turner}, N.~J., {Fromang}, S., {Gammie}, C., {et~al.} 2014, in Protostars and
  Planets VI, ed. H.~{Beuther}, R.~S. {Klessen}, C.~P. {Dullemond}, \&
  T.~{Henning}, 411

\bibitem[{{Van De Putte} {et~al.}(2023){Van De Putte}, {Cartledge}, {Gordon},
  {Clayton}, \& {Roman-Duval}}]{VanDePutte2023}
{Van De Putte}, D., {Cartledge}, S. I.~B., {Gordon}, K.~D., {Clayton}, G.~C.,
  \& {Roman-Duval}, J. 2023, \apj, 944, 33

\bibitem[{{van Dishoeck} \& {Black}(1982)}]{Vandishoeck1982}
{van Dishoeck}, E.~F., \& {Black}, J.~H. 1982, \apj, 258, 533

\bibitem[{{van Leeuwen}(2007)}]{Hipparcos2}
{van Leeuwen}, F. 2007, \aap, 474, 653

\bibitem[{{Vasyunin} {et~al.}(2017){Vasyunin}, {Caselli}, {Dulieu}, \&
  {Jim{\'e}nez-Serra}}]{Vasyunin2017}
{Vasyunin}, A.~I., {Caselli}, P., {Dulieu}, F., \& {Jim{\'e}nez-Serra}, I.
  2017, \apj, 842, 33

\bibitem[{{Vergely} {et~al.}(2022){Vergely}, {Lallement}, \&
  {Cox}}]{Vergely2022}
{Vergely}, J.~L., {Lallement}, R., \& {Cox}, N.~L.~J. 2022, \aap, 664, A174

\bibitem[{{Wakelam} {et~al.}(2010){Wakelam}, {Smith}, {Herbst}, {Troe},
  {Geppert}, {Linnartz}, {{\"O}berg}, {Roueff}, {Ag{\'u}ndez}, {Pernot},
  {Cuppen}, {Loison}, \& {Talbi}}]{Wakelam2010}
{Wakelam}, V., {Smith}, I.~W.~M., {Herbst}, E., {et~al.} 2010, \ssr, 156, 13

\bibitem[{{Webber}(1998)}]{Webber1998}
{Webber}, W.~R. 1998, \apj, 506, 329

\bibitem[{{Weingartner} \& {Draine}(2001)}]{Weingartner2001}
{Weingartner}, J.~C., \& {Draine}, B.~T. 2001, \apj, 563, 842

\bibitem[{{Weingartner} {et~al.}(2006){Weingartner}, {Draine}, \&
  {Barr}}]{Weingartner2006}
{Weingartner}, J.~C., {Draine}, B.~T., \& {Barr}, D.~K. 2006, \apj, 645, 1188

\bibitem[{{Welty} \& {Hobbs}(2001)}]{Welty2001}
{Welty}, D.~E., \& {Hobbs}, L.~M. 2001, \apjs, 133, 345

\bibitem[{{Wolfire} {et~al.}(2003){Wolfire}, {McKee}, {Hollenbach}, \&
  {Tielens}}]{Wolfire2003}
{Wolfire}, M.~G., {McKee}, C.~F., {Hollenbach}, D., \& {Tielens}, A.~G.~G.~M.
  2003, \apj, 587, 278

\bibitem[{{Wright} {et~al.}(2003){Wright}, {Egan}, {Kraemer}, \&
  {Price}}]{Tycho2_SpectralTypeCatalog}
{Wright}, C.~O., {Egan}, M.~P., {Kraemer}, K.~E., \& {Price}, S.~D. 2003, \aj,
  125, 359

\bibitem[{{Zhang} {et~al.}(2023){Zhang}, {Green}, \& {Rix}}]{Zhang2023}
{Zhang}, X., {Green}, G.~M., \& {Rix}, H.-W. 2023, \mnras, 524, 1855

\bibitem[{Zonca {et~al.}(2019)Zonca, Singer, Lenz, Reinecke, Rosset, Hivon, \&
  Gorski}]{HEALPIX_1}
Zonca, A., Singer, L., Lenz, D., {et~al.} 2019, Journal of Open Source
  Software, 4, 1298

\bibitem[{{Zucker} {et~al.}(2022){Zucker}, {Goodman}, {Alves}, {Bialy},
  {Foley}, {Speagle}, {Gro{\^I}{\texttwosuperior}schedl}, {Finkbeiner},
  {Burkert}, {Khimey}, \& {Swiggum}}]{Zucker2022}
{Zucker}, C., {Goodman}, A.~A., {Alves}, J., {et~al.} 2022, \nat, 601, 334

\end{thebibliography}


\begin{appendix} 
\restartappendixnumbering

\section{Distribution of gas density for selected sight lines}
\label{app1}

\begin{figure*}
\begin{center}
	\includegraphics[width=0.91\textwidth]{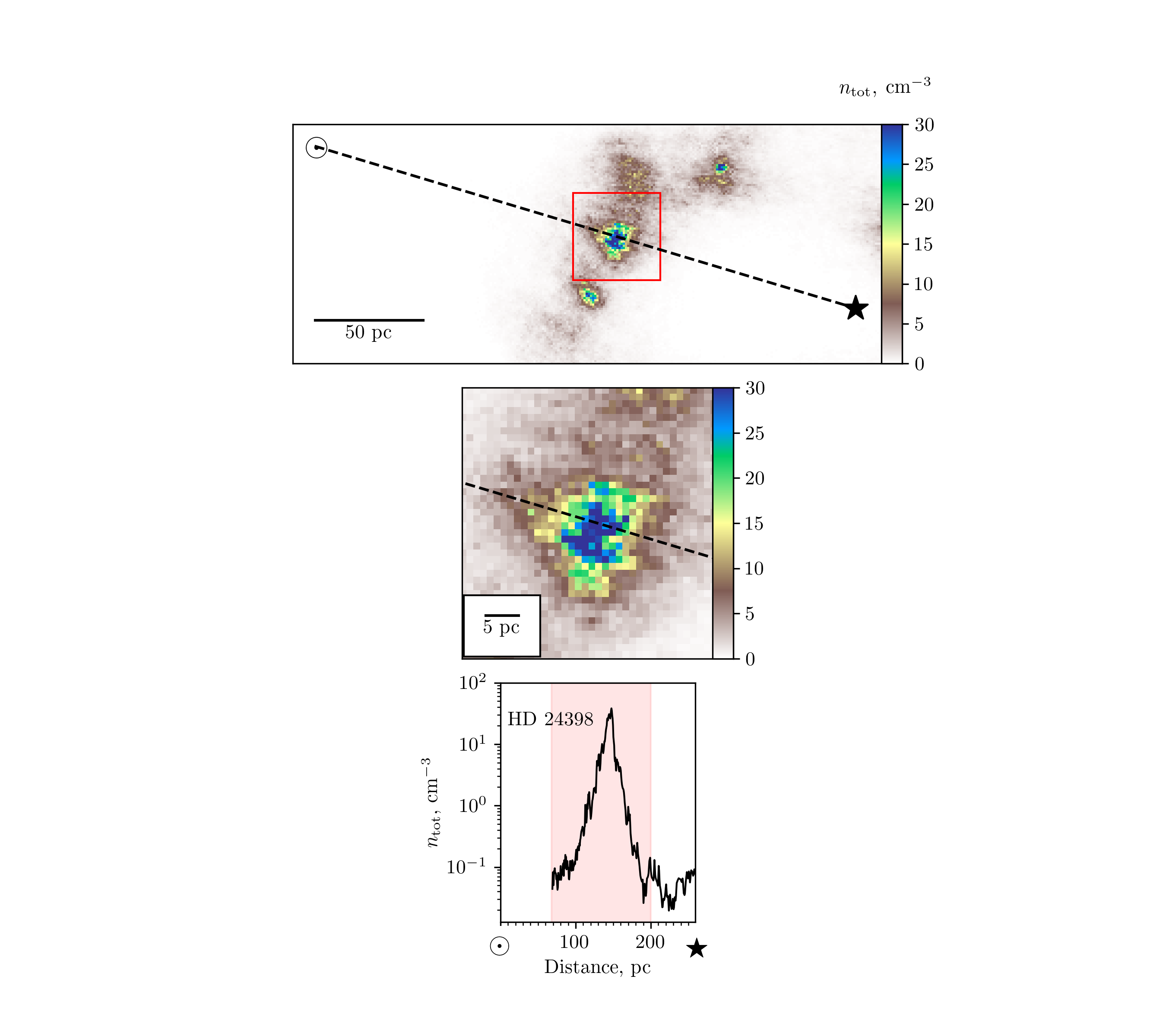}
    \caption{Line of sight to HD~24398 (see description in Appendix~\ref{app1}).} 
    \label{f_app1_1}
\end{center}
\end{figure*}

\begin{figure*}
\begin{center}
	\includegraphics[width=0.91\textwidth]{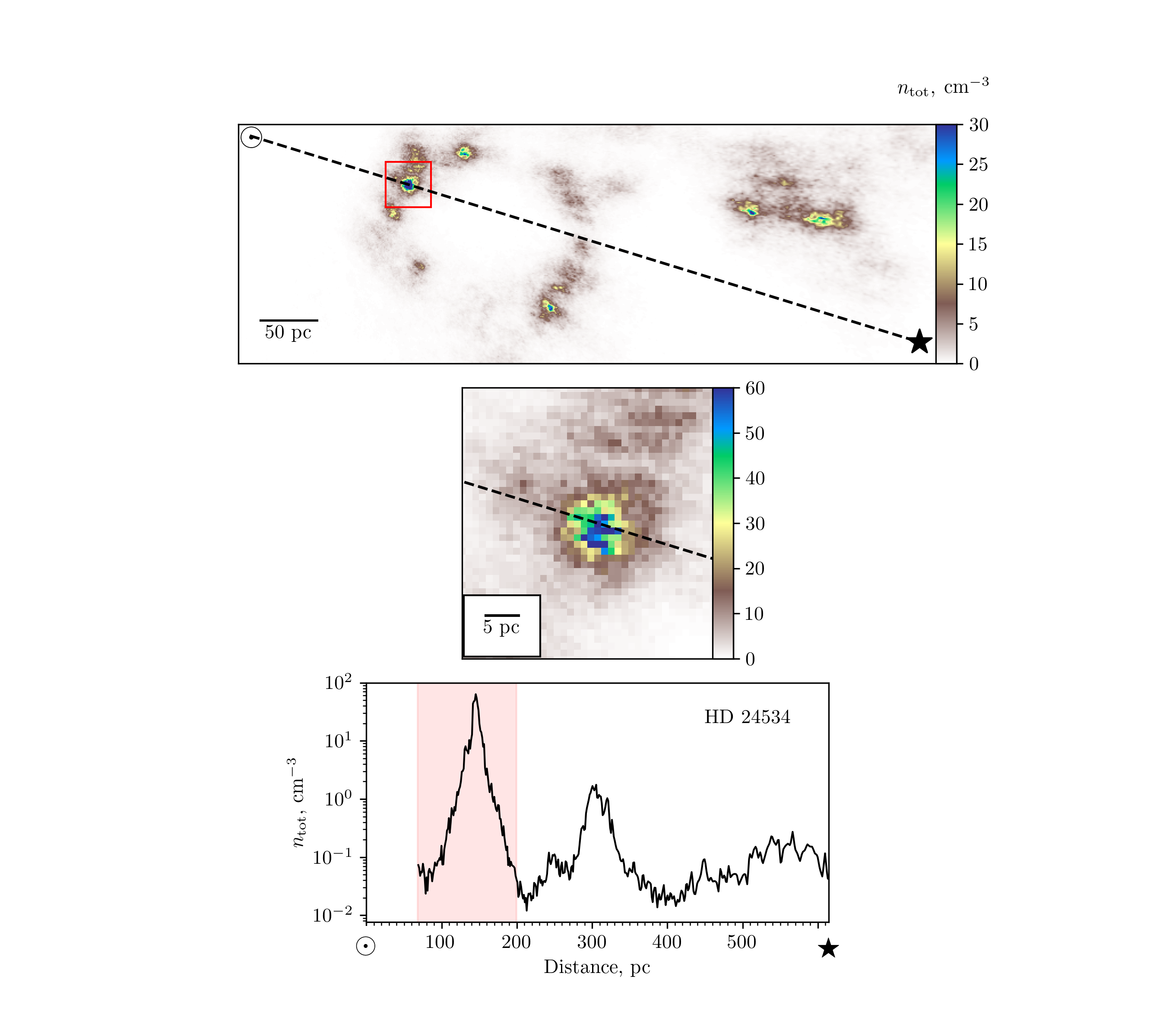}
    \caption{Line of sight to HD~24534.} 
    \label{f_app1_2}
\end{center}
\end{figure*}

\begin{figure*}
\begin{center}
	\includegraphics[width=\textwidth]{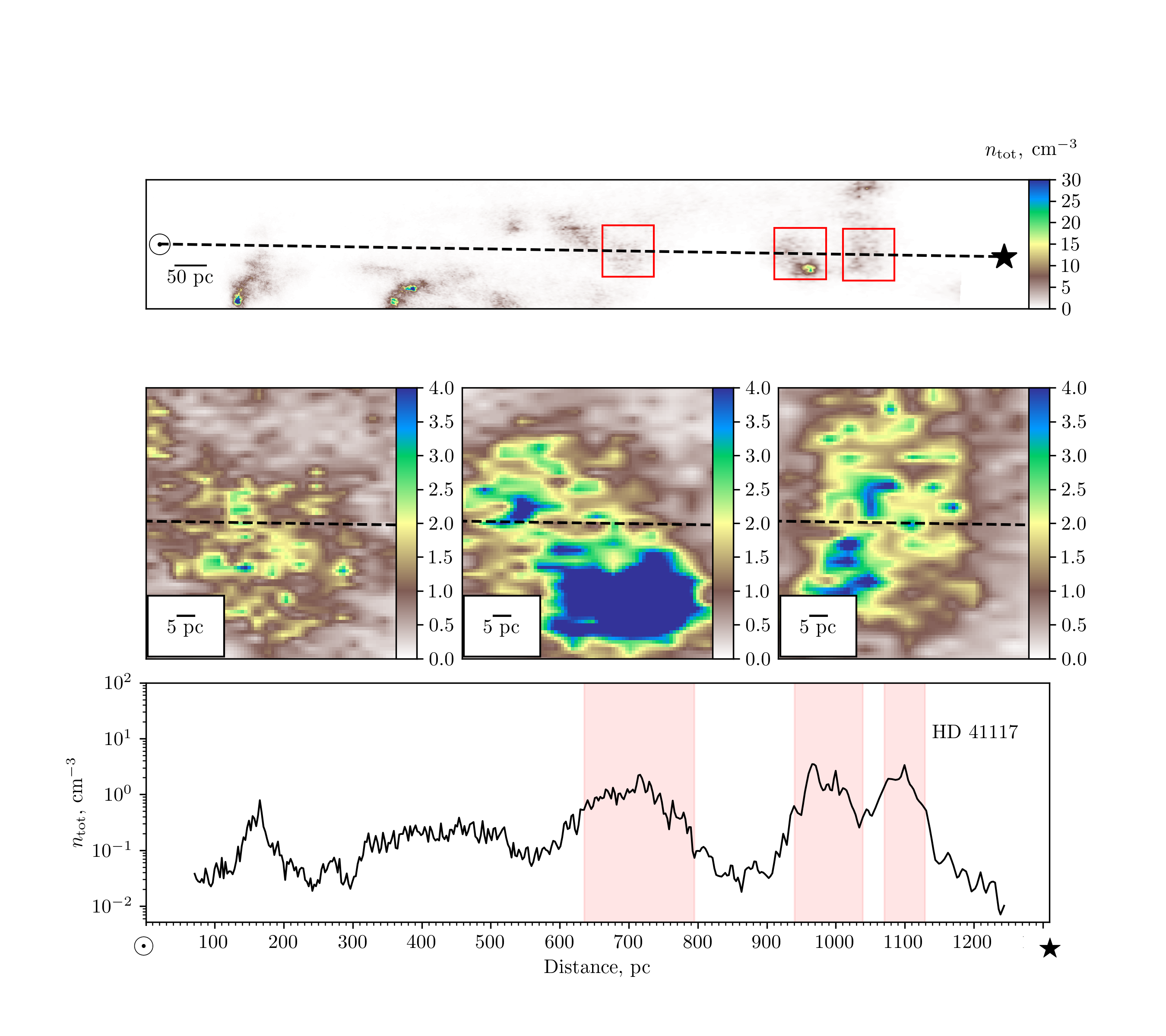}
    \caption{Line of sight to HD~41117.} 
    \label{f_app1_3}
\end{center}
\end{figure*}

\begin{figure*}
\begin{center}
	\includegraphics[width=0.91\textwidth]{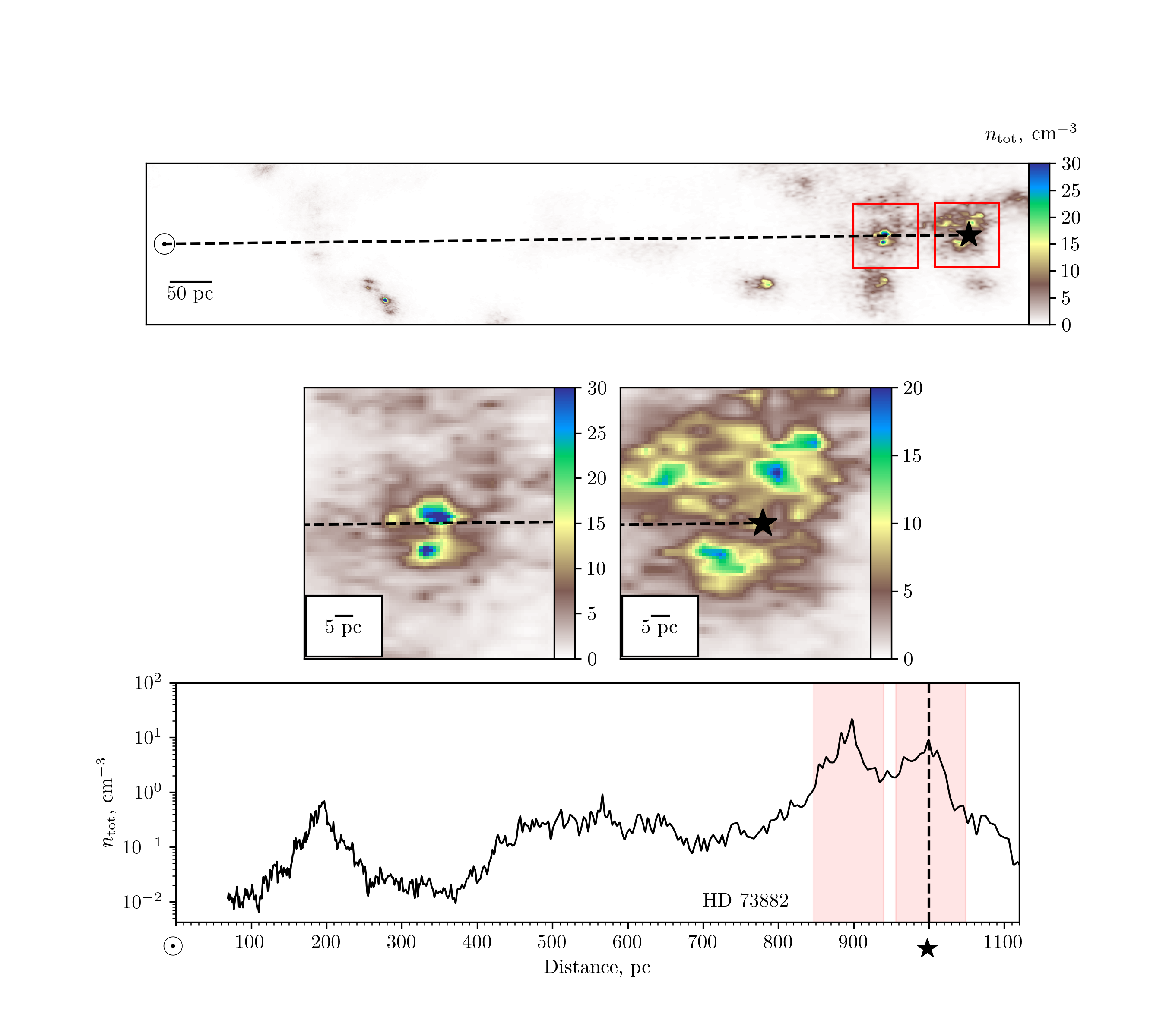}
    \caption{Line of sight to HD~73882.} 
    \label{f_app1_4}
\end{center}
\end{figure*}

\begin{figure*}
\begin{center}
	\includegraphics[width=0.91\textwidth]{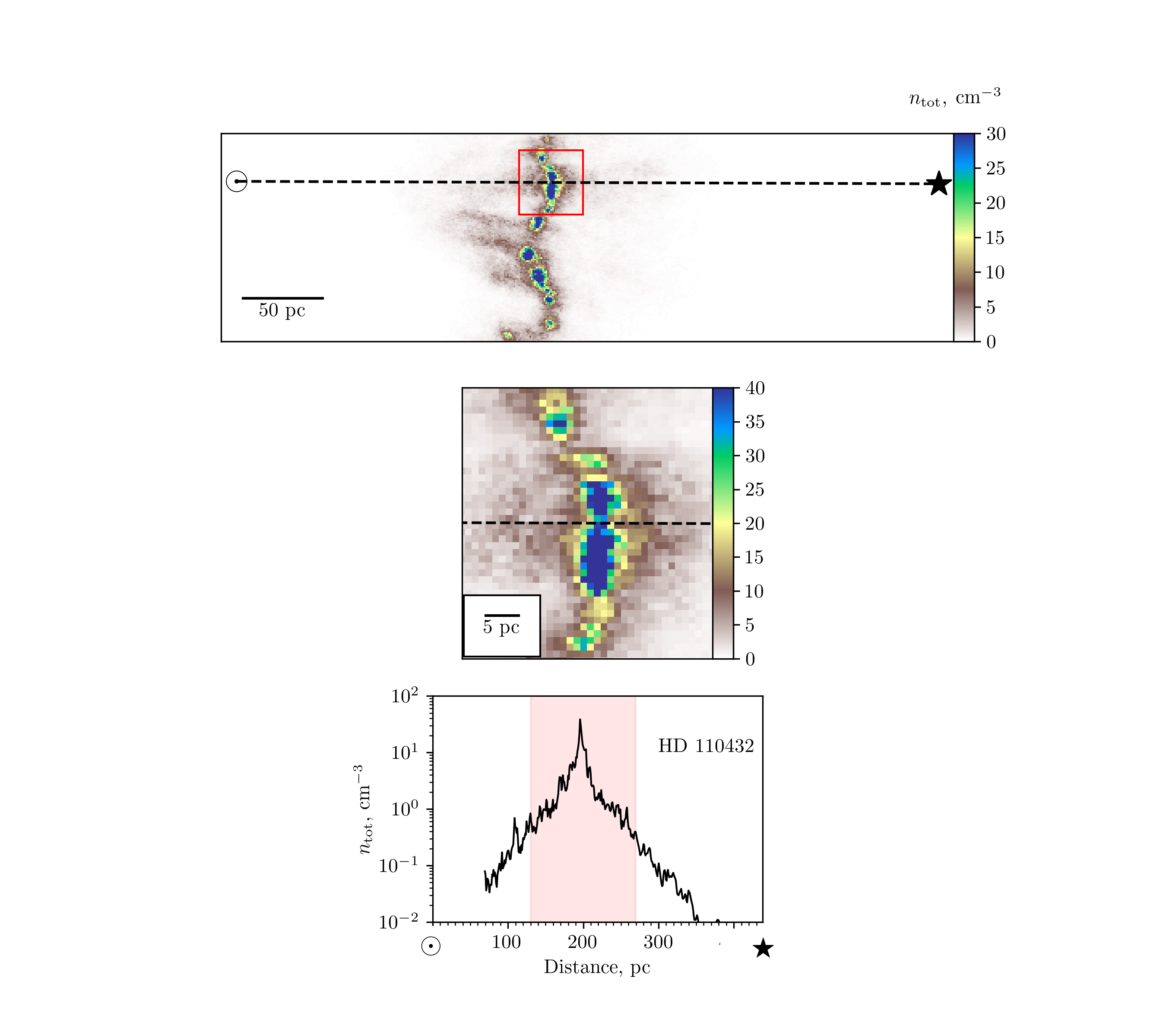}
    \caption{Line of sight to HD~110432.} 
    \label{f_app1_5}
\end{center}
\end{figure*}

\begin{figure*}
\begin{center}
	\includegraphics[width=0.91\textwidth]{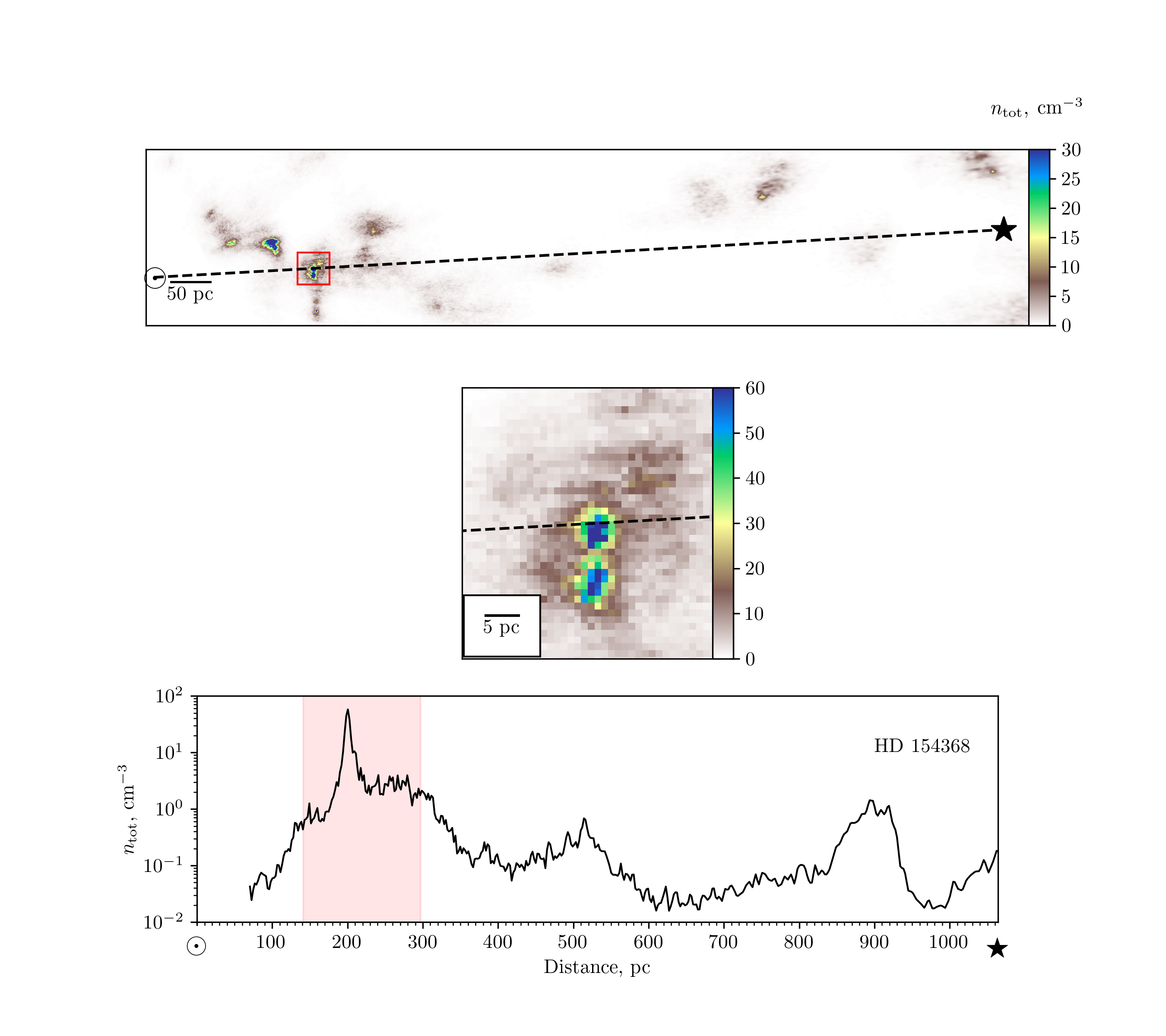}
    \caption{Line of sight to HD~154368.} 
    \label{f_app1_6}
\end{center}
\end{figure*}

\begin{figure*}
\begin{center}
	\includegraphics[width=0.91\textwidth]{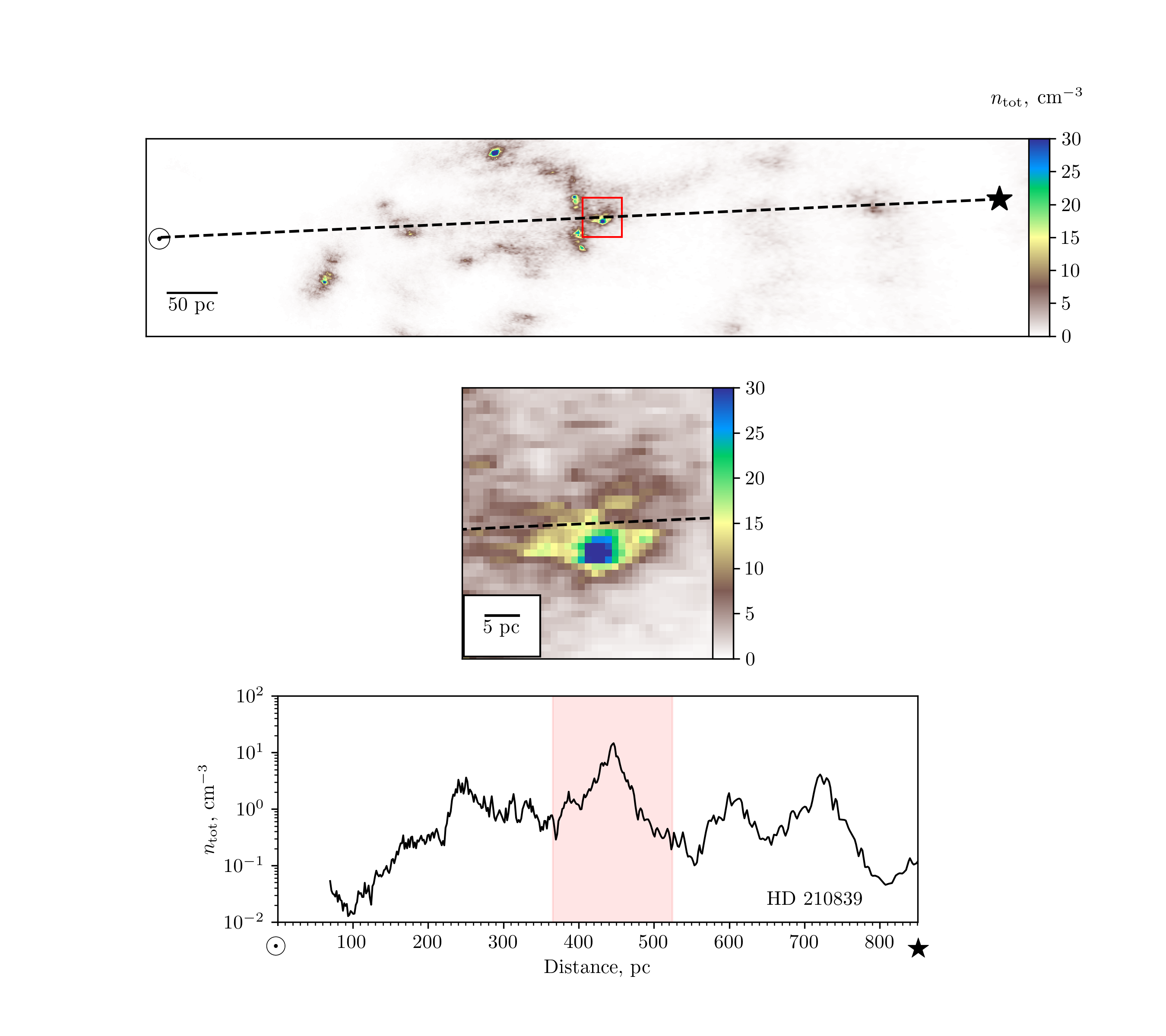}
    \caption{Line of sight to HD~210839.} 
    \label{f_app1_7}
\end{center}
\end{figure*}

\begin{figure*}
\begin{center}
	\includegraphics[width=0.91\textwidth]{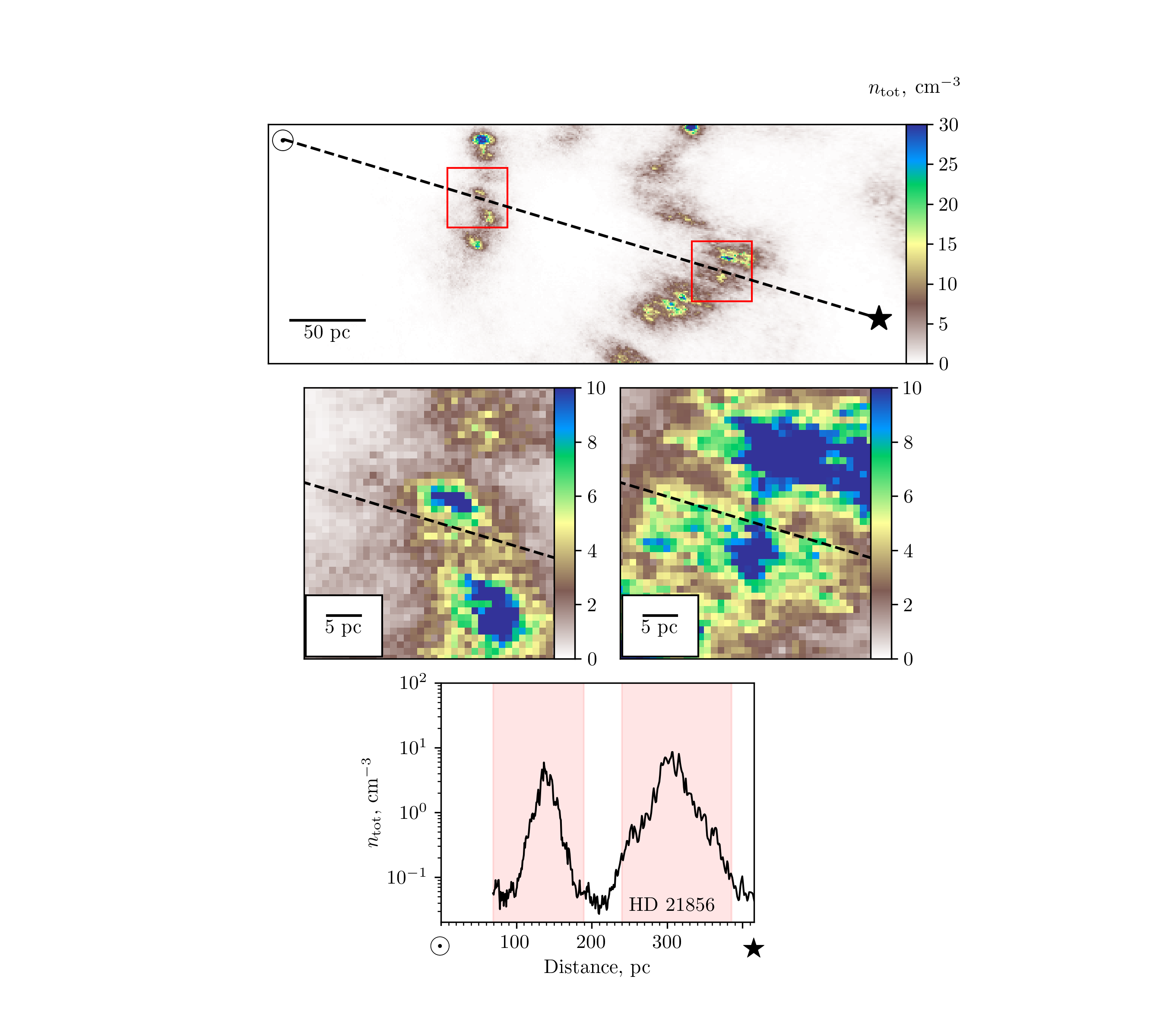}
    \caption{Line of sight to HD~21856.} 
    \label{f_app1_8}
\end{center}
\end{figure*}

\begin{figure*}
\begin{center}
	\includegraphics[width=0.91\textwidth]{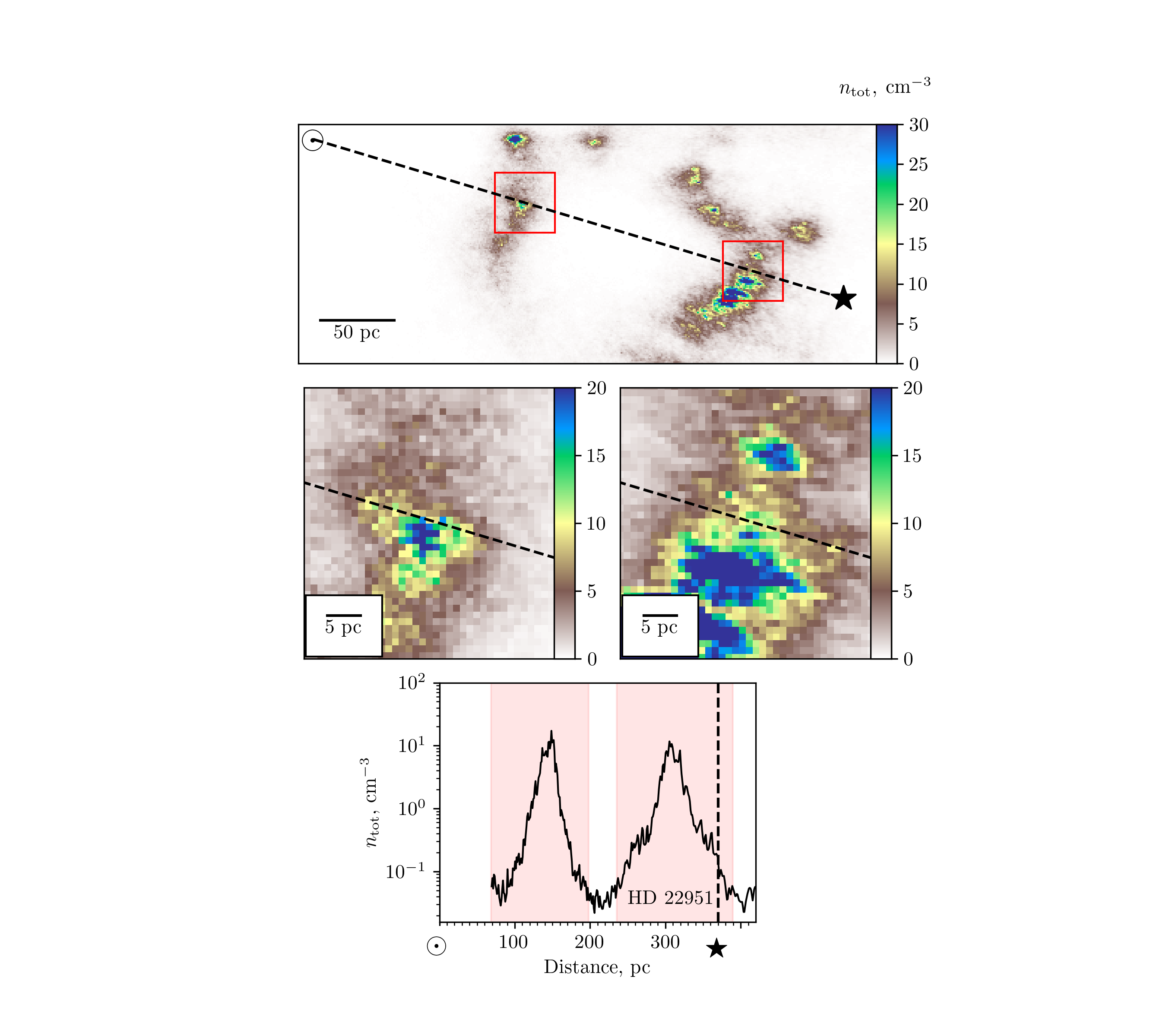}
    \caption{Line of sight to HD~22951.} 
    \label{f_app1_9}
\end{center}
\end{figure*}

\begin{figure*}
\begin{center}
	\includegraphics[width=0.91\textwidth]{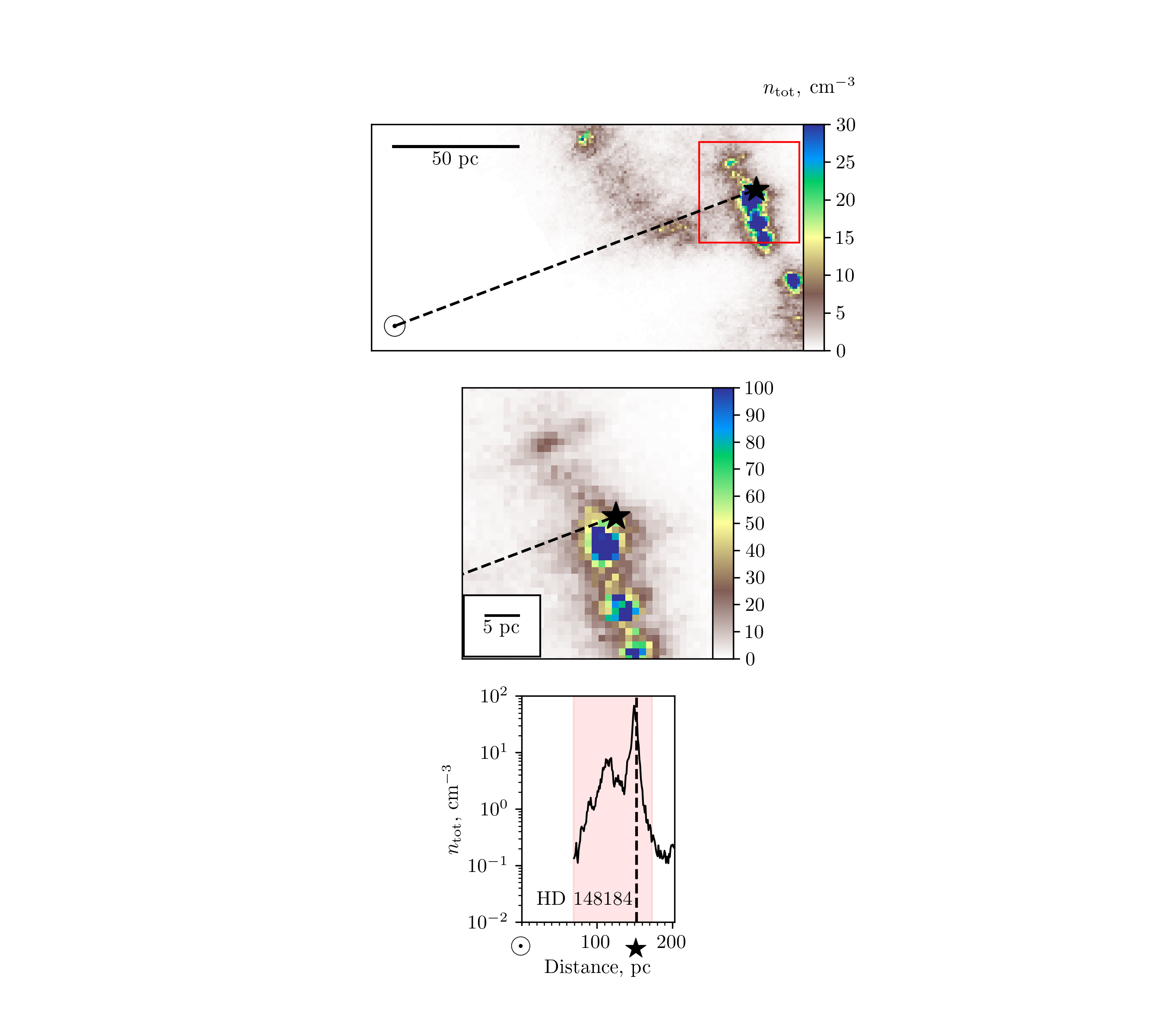}
    \caption{Line of sight to HD~148184.} 
    \label{f_app1_10}
\end{center}
\end{figure*}

\begin{figure*}
\begin{center}
	\includegraphics[width=\textwidth]{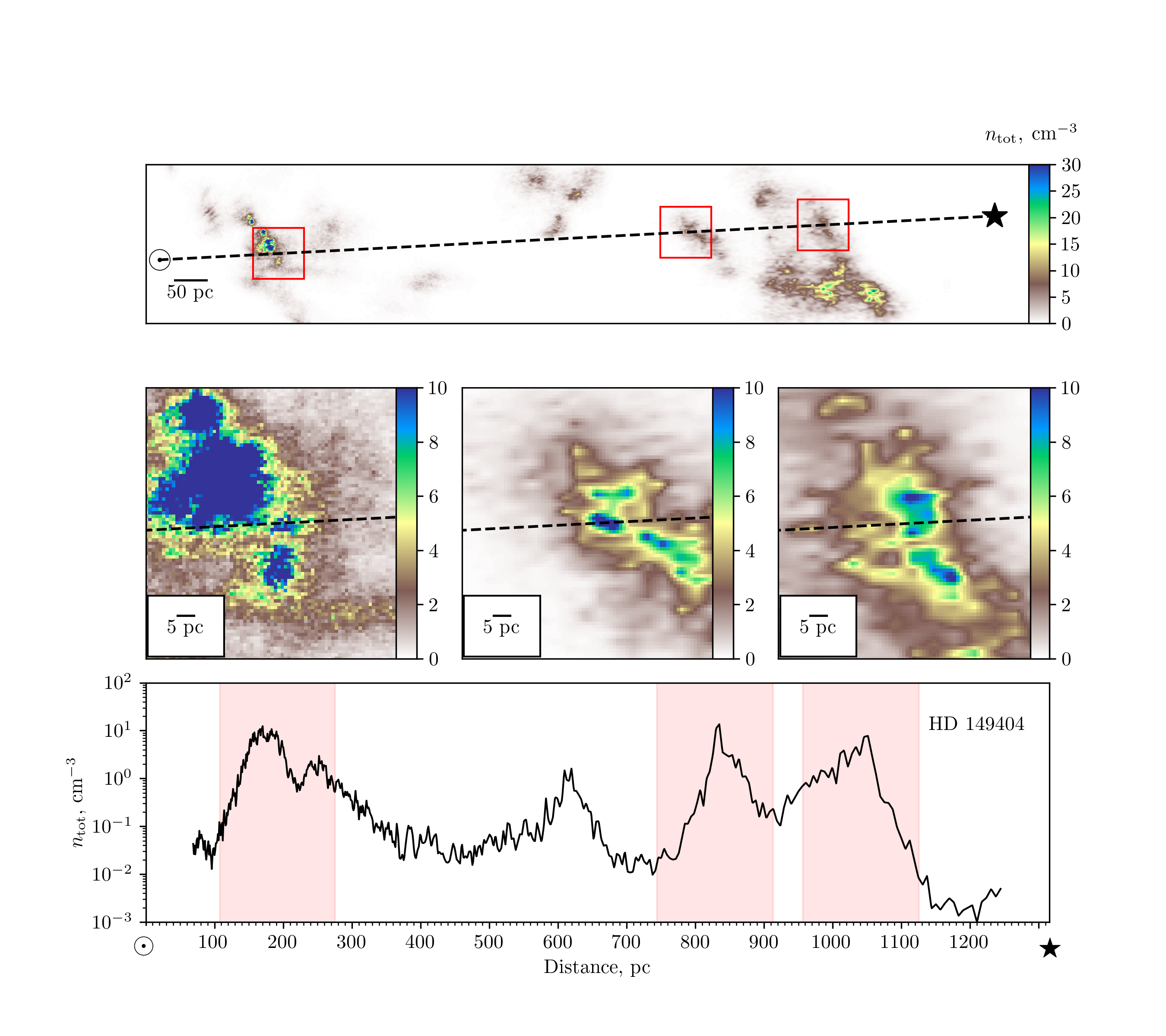}
    \caption{Line of sight to HD~149404.} 
    \label{f_app1_11}
\end{center}
\end{figure*}

\begin{figure*}
\begin{center}
	\includegraphics[width=0.91\textwidth]{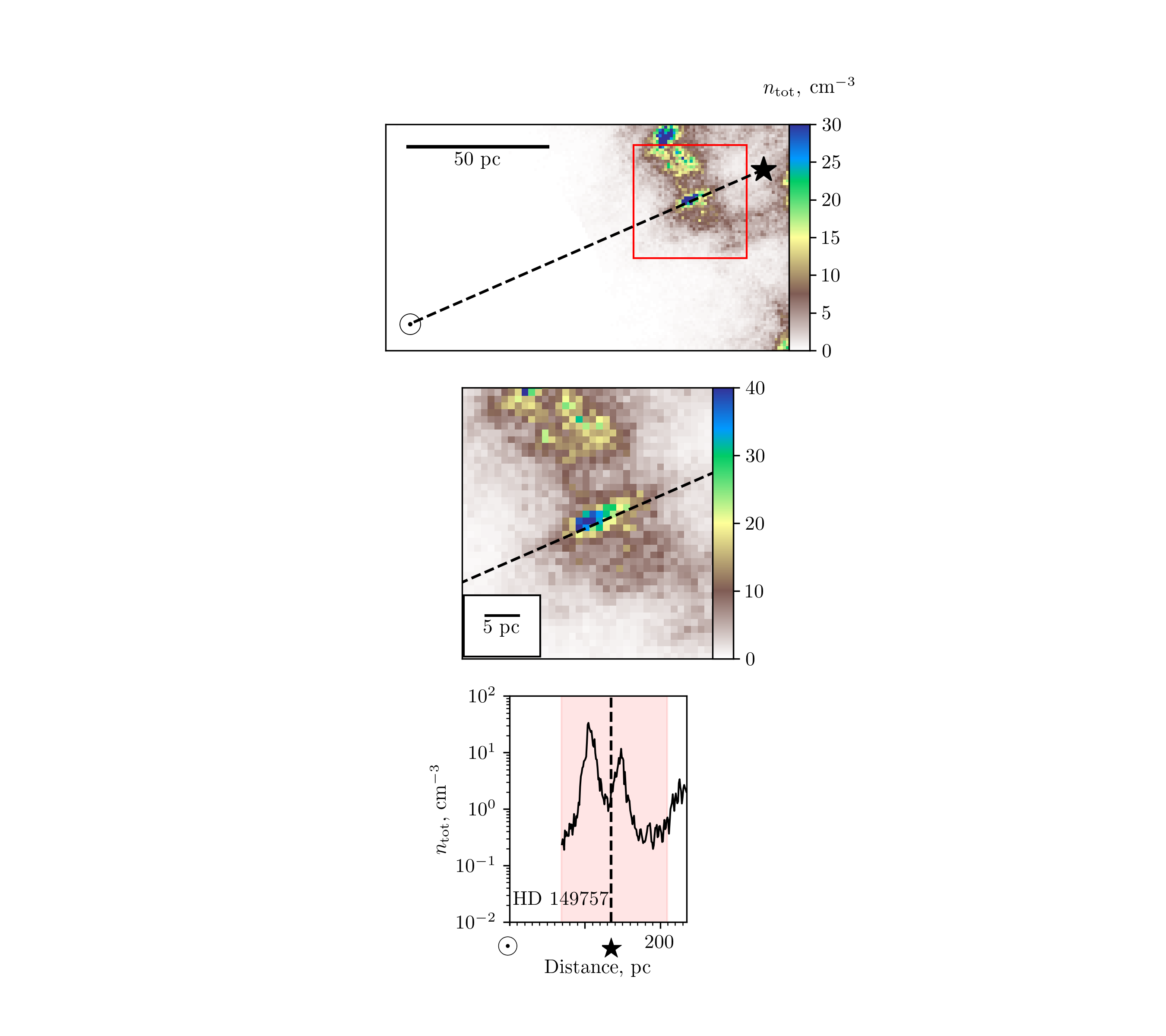}
    \caption{Line of sight to HD~149757.} 
    \label{f_app1_12}
\end{center}
\end{figure*}

The upper panel of each figure (\ref{f_app1_1}--\ref{f_app1_12}) shows distribution of the total volume gas density, $n_{\rm tot}$, derived from the dust extinction map \citep{Edenhofer2024}. Depicted is a vertical plane containing the line of sight to the corresponding target star ($\star$). The middle panel shows zoom-in view of the gas clump(s) simulated in Sections~\ref{detections} and \ref{non-detections}. In the bottom panel, $n_{\rm tot}$ is plotted versus the distance along the sight line; the shading indicates the range of distances included in the simulations, where the map density is rescaled by the factor of $N({\rm H}_{\rm tot})/N({\rm H}_{\rm tot}^{\rm map})$ (see Section~\ref{3D_PDR} for details). To facilitate comparison of the gas distribution for different sight lines, we use the same distance scale in the lower panels. Note that the extinction map and thus the derived gas density starts at a distance of 69~pc from the Sun (see Section~\ref{3D_maps}).

Figures~\ref{f_app1_1}--\ref{f_app1_7} show seven sight lines with H$_3^+$ detection, and Figures~\ref{f_app1_8}--\ref{f_app1_12} show five sight lines without detection. For HD~41117 (Figure~\ref{f_app1_3}) and HD~149404 (Figure~\ref{f_app1_11}), where the target stars are beyond the outer border of the extinction map by \citet{Edenhofer2024}, we used lower-resolution maps by \citet{Lallement2022} and \citet{Vergely2022} to verify that the remaining amount of gas is negligible. 

\section{Corrections and updates to {\sc 3d-pdr}}
\label{app2}

An erroneous factor of 1/2 in front of all photoreaction rates is removed, including the rates of H$_{2}$ and CO photodissociation as well as of carbon photoionization.

Electron-ion recombination on dust grains (including a population of PAH grains) is added. We follow the approach of \citet{Weingartner2001}, computing the recombination rates from their Equation~(8).

We apply $A_{V}/N_{\rm tot}=5.3 \times 10^{-22}$~mag~cm$^{2}$ for the visual extinction, according to \citet{DraineBook2011}. 

We use an analytical approximation for the H$_{2}$ self-shielding function, given by Equation~(37) from \citet{Draine_self_shielding}.

For the H$_{2}$ photodissociation rate in the absence of dust extinction and self-shielding effects, we adopt $5.7\times10^{-11}$~s$^{-1}$ \citep{Heays_2017}. 

The carbon photoionization rate in {\sc 3d-pdr} is being computed using Equation~(23) from \citet{Bisbas2012}:
\begin{equation}
    r = \alpha G_{\rm D}e^{-\tau},
\end{equation}
where $\alpha$ is the unshielded ionization rate, $G_{\rm D}$ is the FUV field in Draine's units, and $\tau$ is the optical depth accounting for the dust extinction, self-shielding by neutral carbon, and H$_{2}$ absorption. Following \citet{Tielens_Hollenbach_1985}, in the present paper we use their Equation~(A6) for the optical depth,  
\begin{equation}
    \tau=k_{\rm dust}A_V+ k_{\rm C}N({\rm C}) + k_{{\rm H}_2}N({\rm H}_2),
\end{equation}
with $k_{\rm C}=10^{-17}$~cm$^2$ and $k_{{\rm H}_2}=1.4\times10^{-22}$~cm$^2$, but utilize the updated values of $\alpha=3.5\times 10^{-10}$~s$^{-1}$ and $k_{\rm dust}=2.77$ from \citet{Heays_2017} (instead of $k_{\rm dust}=2.4$ from \citealt{Tielens_Hollenbach_1985}, or $k_{\rm dust}=2.41$ from \citealt{Black1977}).

For the photoelectric heating rate as well as for the rate of photoelectric grain charging, the accurate value of the $A_{V}$ prefactor $k_{\rm ph}$ is uncertain. Following Equation~(117) in \citet{Black1977}, we employ the value of $k_{\rm ph}=2.3$, which corresponds to FUV photons with energies above 10~eV \citep[as the photoelectic yield sharply decreases at lower energies, see][]{Weingartner2006}. Similar to the updated value of $k_{\rm dust}$ for carbon photoionization, the actual value of $k_{\rm ph}$ may also be slightly higher (around 2.6). However, for the relatively diffuse gas clumps studied in this paper, the effect of such uncertainty is verified to be less than a few percent in gas temperature.

For the reaction of dissociative recombination ${\rm H}_{3}^{+} + e^{-} \to {\rm H}_{2} + {\rm H}$ or 3H, we adopt the rate of $6.8\times10^{-8}(T/300)^{-0.5}$~cm$^3$~s$^{-1}$ from \citet{McCall2004}.

We also update several rates of thermal reactions relevant to OH$^{+}$ chemistry. In {\sc 3d-pdr}, the rates are being approximated by
\begin{equation}
    R = \alpha (T/300)^{\beta}e^{-\gamma/T},
\end{equation}
where $\alpha$, $\beta$, and $\gamma$ are the fitting parameters. For the reactions ${\rm O}^{+} + {\rm H}_{2} \to {\rm OH}^{+} + {\rm H}$ and ${\rm OH}^{+}+ {\rm H}_{2} \to {\rm H}_{2}\rm{O}^{+} + {\rm H}$ we use $\alpha=1.4\times10^{-9}$~cm$^3$~s$^{-1}$ and $1.2\times10^{-9}$~cm$^3$~s$^{-1}$, respectively, while $\beta=\gamma=0$ \citep{Kovalenko2018}. For the charge-transfer reaction ${\rm H}^{+} + {\rm O}(^3{\rm P}_2) \to {\rm O}^{+} + {\rm H}$ we adopt $\alpha = 5.52\times10^{-10}$~cm$^3$~s$^{-1}$, $\beta=0.223$, and $\gamma=228$~K based on \citet{Stancil1999}. For the dissociative recombination ${\rm OH}^{+} + e^{-} \to {\rm O} + {\rm H}$ we employ the parameters from Table~2 of \citet{Kalosi2023}. 

\section{Results of simulations} 
\label{app3}

Values of $N({\rm H}_2)$ and $N({\rm H}_3^+)$ computed for each selected sight line are summarized in Tables~\ref{table3} (detections) and \ref{table4} (non-detections); $\zeta_{\rm H_2}$ in units of $10^{-17}$~s$^{-1}$, $N({\rm H}_{\rm tot})$ and $N({\rm H}_2)$ in units of $10^{20}$~cm$^{-2}$, and $N({\rm H}_3^+)$ in units of $10^{13}$~cm$^{-2}$. The three values of $N({\rm H}_{\rm tot})$ for each sight line correspond to the measured mean and 68\% confidence limits listed in Table~\ref{table1}. 

\begin{table*}
  \caption{Detections: values of $N({\rm H}_2)$ and $N({\rm H}_3^+)$ computed for different combinations of $N({\rm H}_{\rm tot})$ and $\zeta_{\rm H_2}$$^{\rm a}$}
  \vspace{-.5cm}
     \begin{center}
         \begin{tabular}{c|l||C{\mycolumnwidth}C{2.8mm}C{\mycolumnwidth}|C{\mycolumnwidth}C{2.8mm}C{\mycolumnwidth}|C{\mycolumnwidth}C{2.8mm}C{\mycolumnwidth}|C{\mycolumnwidth}C{2.8mm}C{\mycolumnwidth}|}
             \hhline{~-------------}
              & 
             \diagbox[width=2.8cm,  height=1.3cm, outerleftsep=-5pt, outerrightsep=24pt,innerleftsep=-0.5cm]{$N({\rm H}_{\rm tot})$}{$\zeta_{\rm H_2}$} &\multicolumn{3}{c|}{1.0}  & \multicolumn{3}{c|}{5.0} & \multicolumn{3}{c|}{10.0} & \multicolumn{3}{c|}{50.0}\\ \hhline{~-------------}

             \noalign{\myspace}
             
             \hhline{~-------------} 
             \multirow{3}{*}{\rotatebox[origin=c]{90}{HD~24398}} 

             &\hspace{.6cm}$+6.3$& 8.46 & \& & 1.31 & 8.29 &\& & 5.93 & 8.09 & \&  & 10.64  & 6.91 & \& & 31.62\\ \hhline{~-------------}
             & 16.6 & 5.42 & \& & 1.03 & 5.29& \& & 4.57 & 5.14 & \& & 8.05 & 4.26 & \& & 21.40\\ \hhline{~-------------}
             &\hspace{.6cm}$-3.2$ & 3.88 & \& & 0.83 & 3.77 & \& & 3.64 & 3.65& \& & 6.29 & 2.96 & \& & 15.16\\ \hhline{~-------------}
             
             \noalign{\myspace}
             
             \hhline{~-------------}
            \multirow{3}{*}{\rotatebox[origin=c]{90}{HD~24534}} 
            & \hspace{.6cm}$+1.8$& 7.54 & \& & 1.05 & 7.40 & \& & 4.72 & 7.24 & \& & 8.41 & 6.27 & \& & 25.69\\
            \hhline{~-------------}
            & 22.1 & 6.76 & \& & 0.98 &  6.63 & \& & 4.38 & 6.48 & \& & 7.81 & 5.58 & \& & 23.50\\ \hhline{~-------------}
            & \hspace{.6cm}$-1.6$ & 6.07 & \&  & 0.90 & 5.95 & \& & 4.07 & 5.81 & \& &  7.26& 4.97 &  \&  & 21.40\\ \hhline{~-------------}

             \noalign{\myspace}

             \hhline{~-------------}
            \multirow{3}{*}{\rotatebox[origin=c]{90}{HD~41117}} 
            & \hspace{.6cm}$+12.8$& 7.52 & \& & 4.55 & 6.81 & \& & 14.55 & 6.18 & \& & 17.82 & 3.91 & \& & 18.24\\
            \hhline{~-------------}
            & 36.6 & 3.22 & \& & 1.59 & 2.84 & \& & 4.41 & 2.57 & \& & 5.01 & 1.60 & \& & 4.88\\ \hhline{~-------------}
            & \hspace{.6cm}$-7.6$ & 1.37 & \&  & 0.46 & 1.19 &  \& & 1.14 & 1.08 & \&  & 1.25 & 0.68 & \&  & 1.27\\ \hhline{~-------------}

             \noalign{\myspace}

             \hhline{~-------------}
            \multirow{3}{*}{\rotatebox[origin=c]{90}{HD~73882}} 
            & \hspace{.6cm}$+8.0$& 10.61 & \& & 3.34 & 10.05 & \& & 13.03 & 9.48 & \& & 19.82 & 6.91 & \& & 36.17\\
            \hhline{~-------------}
            & 39.9 & 7.22 & \& & 2.22 & 6.79 & \& & 8.56 & 6.38 & \& & 13.01 & 4.59 & \& & 23.37\\ \hhline{~-------------}
            & \hspace{.6cm}$-5.9$ & 4.94 & \&  & 1.46 & 4.62 &  \& & 5.61 & 4.33 & \&  & 8.58 & 3.07 & \&  & 14.87\\ \hhline{~-------------}

             \noalign{\myspace}

             \hhline{~-------------}
            \multirow{3}{*}{\rotatebox[origin=c]{90}{HD~110432}} 
            & \hspace{.6cm}$+3.6$& 3.96 & \& & 0.68 & 3.85 & \& & 2.91 & 3.73 & \& & 4.95 & 3.10 & \& & 12.39\\
            \hhline{~-------------}
            & 16.3 & 2.57 & \& & 0.42 & 2.52 & \& & 1.86 & 2.43 & \& & 3.11 & 2.01 & \& & 7.67 \\ \hhline{~-------------}
            & \hspace{.6cm}$-2.2$ & 1.85 & \&  & 0.30 & 1.79 &  \& & 1.27 & 1.73 & \&  & 2.13 & 1.42 & \&  & 5.14\\ \hhline{~-------------}

             \noalign{\myspace}

             \hhline{~-------------}
            \multirow{3}{*}{\rotatebox[origin=c]{90}{HD~154368}} 
            & \hspace{.6cm}$+5.5$& 10.51 & \& & 1.21 & 10.31 & \& & 5.46 & 10.10 & \& & 9.83 & 8.91 & \& & 30.31\\
            \hhline{~-------------}
            & 39.3 & 8.71 & \& & 1.05 & 8.54 & \& & 4.77 & 8.36 & \& & 8.50 & 7.33 & \& & 25.41\\ \hhline{~-------------}
            & \hspace{.6cm}$-4.3$ & 7.34 & \&  & 0.94 & 7.19 &  \& & 4.20 & 7.03 & \&  & 7.41 & 6.14 & \&  & 21.67\\ \hhline{~-------------}

             \noalign{\myspace}

             \hhline{~-------------}
            \multirow{3}{*}{\rotatebox[origin=c]{90}{HD~210839}} 
            & \hspace{.6cm}$+2.7$& 4.33 & \& & 1.80 & 4.08 & \& & 6.87 & 3.84 & \& & 10.28 & 2.72 & \& & 16.12\\
            \hhline{~-------------}
            & 30.2 & 3.69 & \& & 1.56 & 3.47 & \& & 5.88 & 3.25 & \& & 8.66 & 2.28 & \& & 13.01\\ \hhline{~-------------}
            & \hspace{.6cm}$-2.3$ & 3.15 & \&  & 1.34 & 2.95 &  \& & 5.00 & 2.76 & \&  & 7.24 & 1.91 & \&  & 10.45\\ \hhline{~-------------}
                        
            \end{tabular}
    \end{center}

$^{\rm a}\,$The optimum values of $N({\rm H}_2)$, $N({\rm H}_{\rm tot})$, and $\zeta_{\rm H_2}$, derived from the bicubic interpolation between the computed values, are listed in Table~\ref{table2}.
    
  
    


    \begin{tikzpicture}[overlay, remember picture]
    \node[draw,fill=white] at (4.07cm,15.62cm) {$N({\rm H}_2)$ \& $N({\rm H}_3^+)$};
    \end{tikzpicture}

\label{table3}
\end{table*}

\begin{table*}
  \caption{Non-detections: values of $N({\rm H}_2)$ and $N({\rm H}_3^+)$ computed for different combinations of $N({\rm H}_{\rm tot})$ and $\zeta_{\rm H_2}$$^{\rm a}$}
  \vspace{-0.5cm}
     \begin{center}
         \begin{tabular}{c|l||C{\mycolumnwidth}C{2.8mm}C{\mycolumnwidth}|C{\mycolumnwidth}C{2.8mm}C{\mycolumnwidth}|C{\mycolumnwidth}C{2.8mm}C{\mycolumnwidth}|C{\mycolumnwidth}C{2.8mm}C{\mycolumnwidth}|}
             \hhline{~-------------}
             & 
             \diagbox[width=2.8cm,  height=1.3cm, outerleftsep=-5pt, outerrightsep=24pt,innerleftsep=-0.5cm]{$N({\rm H}_{\rm tot})$}{$\zeta_{\rm H_2}$} &\multicolumn{3}{c|}{1.0}  & \multicolumn{3}{c|}{5.0} & \multicolumn{3}{c|}{10.0} & \multicolumn{3}{c|}{50.0}\\ \hhline{~-------------}

             \noalign{\myspace}

             \hhline{~-------------}
            \multirow{3}{*}{\rotatebox[origin=c]{90}{HD~21856}}
            & \hspace{.6cm}$+2.6$& 1.25 & \& & 0.38 & 1.17 & \& & 1.15 & 1.08  & \& & 1.50 & 0.75 & \& & 1.90\\
            \hhline{~-------------}
            & 13.5 & 0.62 & \& & 0.13 & 0.57 & \& & 0.39 & 0.52 & \& & 0.48 & 0.36 & \& & 0.60\\ \hhline{~-------------}
            & \hspace{.6cm}$-2.0$ & 0.30 & \&  & 0.04  & 0.27 &  \& & 0.11 & 0.25 & \&  & 0.14 & 0.16 & \&  & 0.17\\ \hhline{~-------------}

             \noalign{\myspace}

             \hhline{~-------------}
            \multirow{3}{*}{\rotatebox[origin=c]{90}{HD~22951}} 
            & \hspace{.6cm}$+4.9$& 4.16 & \& & 1.22  & 3.96 & \& & 4.69 & 3.75 & \& & 7.07 & 2.79 & \& & 12.05\\
            \hhline{~-------------}
            & 17.7 & 2.13 & \& & 0.56 & 2.01 & \& & 2.04 & 1.90 & \& & 2.99 & 1.41 & \& & 4.91\\ \hhline{~-------------}
            & \hspace{.6cm}$-3.3$ & 1.10 & \&  & 0.24 & 1.04 &  \& & 0.87 & 0.98 & \&  & 1.24 & 0.73 & \&  & 2.03\\ \hhline{~-------------}

             \noalign{\myspace}

             \hhline{~-------------}
            \multirow{3}{*}{\rotatebox[origin=c]{90}{HD~148184}}
            & \hspace{.6cm}$+5.9$& 6.86 & \& & 0.46 & 6.78 & \& & 2.18 & 6.68 & \& & 4.10 & 6.06 & \& & 15.14\\
            \hhline{~-------------}
            & 23.6 & 4.63 & \& & 0.37 & 4.57 & \& & 1.73 & 4.49 & \& & 3.25 & 4.02 & \& & 11.55\\ \hhline{~-------------}
            & \hspace{.6cm}$-3.9$ & 3.20 & \&  & 0.29 & 3.15 &  \& & 1.34 & 3.09 & \&  & 2.49 & 2.70 & \&  & 8.18\\ \hhline{~-------------}

             \noalign{\myspace}

             \hhline{~-------------}
            \multirow{3}{*}{\rotatebox[origin=c]{90}{HD~149404}}
            & \hspace{.6cm}$+11.5$ & 3.18 & \& & 0.65 & 2.99 & \& & 2.40 & 2.82 & \& & 3.55 & 2.07 & \& & 5.85\\
            \hhline{~-------------}
            & 39.1 & 1.16 & \& & 0.18 & 1.08 & \& & 0.63 & 1.01 & \& & 0.88 & 0.73 & \& & 1.34\\
            \hhline{~-------------}
            & \hspace{.6cm}$-7.1$ & 0.48 & \& & 0.05 & 0.43 & \& & 0.17 & 0.39 & \& & 0.21 & 0.29 & \& & 0.33\\
            \hhline{~-------------}

             \noalign{\myspace}

             \hhline{~-------------}
            \multirow{3}{*}{\rotatebox[origin=c]{90}{HD~149757}} 
            & \hspace{.6cm}$+2.0$& 4.03 & \& & 0.59 & 3.93 & \& & 2.63 & 3.83 & \& & 4.58 & 3.22 & \& & 12.19\\
            \hhline{~-------------}
            & 14.4 & 3.00 & \& & 0.43 & 2.91 & \& & 1.86 & 2.86 & \& & 3.31  & 2.38 & \& & 8.33 \\ \hhline{~-------------}
            & \hspace{.6cm}$-1.5$ & 2.26 & \&  & 0.32 & 2.20 &  \& & 1.34 & 2.08 & \&  & 2.15 & 1.76 & \&  & 5.46\\ \hhline{~-------------}
                        
            \end{tabular}
    
    \end{center}

$^{\rm a}\,$The optimum values of $N({\rm H}_2)$ and $N({\rm H}_{\rm tot})$ with the corresponding upper limit for $\zeta_{\rm H_2}$, derived from the bicubic interpolation between the computed values, are listed in Table~\ref{table2}. No constraints can be put on $\zeta_{\rm H_2}$ for HD~21856 and HD~149404 due to the high noise level in $N({\rm H}_3^+)$ measurements.
    
    \begin{tikzpicture}[overlay, remember picture]
    \node[draw,fill=white] at (4.03cm,12.37cm) {$N({\rm H}_2)$ \& $N({\rm H}_3^+)$};
    \end{tikzpicture}

\label{table4}

\end{table*}

\end{appendix}

\end{document}